\documentstyle[preprint,aps,graphicx,floats]{revtex}

\begin{document}

\preprint{$
\begin{array}{l}
\mbox{UB-HET-07-02}\\[-3mm]
\mbox{July~2007} \\ [3mm]
\end{array}
$}

\title{High $p_T$ Top Quarks at the Large Hadron Collider}

\author{U.~Baur\footnote{baur@ubhex.physics.buffalo.edu}}
\address{Department of Physics,
State University of New York, \\
Buffalo, NY 14260, USA}
\author{L.H.~Orr\footnote{orr@pas.rochester.edu}}
\address{Dept. of Physics and Astronomy, University of Rochester,\\
Rochester, NY 14627, USA}

\maketitle 

\begin{abstract}
\baselineskip13.pt  
Many new physics models predict resonances with masses in the TeV
range which decay into a pair of top quarks. With its large cross
section, $t\bar t$ production at the Large Hadron Collider (LHC) offers
an excellent opportunity to search for such particles. The
identification of very 
energetic top quarks is crucial in such an analysis. We consider
in detail the $t\bar t\to\ell^\pm\nu b\bar bq\bar q'$ ($\ell=e,\,\mu$)
final state for 
high $p_T$ top quarks. In this phase space region, two or more of the
final state quarks can merge into a single jet due to the large Lorentz boost
of the parent top quark. As a result, a large fraction of 
$t\bar t\to\ell^\pm\nu b\bar bq\bar q'$ events with an invariant mass in
the TeV region contains less than four observable jets. Requiring
one or two tagged $b$-quarks, we calculate the $W+$jets, $Wb+$jets,
$Wb\bar b+$jets, $Wbt$, and single top plus jets backgrounds for these final
states, and identify cuts which help to suppress them. In
particular, we discuss whether a cut on the jet invariant mass may be
useful in reducing the background in the $\ell\nu+2$~jets channel. We
also investigate how next-to-leading order QCD 
corrections affect high $p_T$ top quark production at the LHC.
We find that the $\ell\nu+2$~jets and $\ell\nu+3$~jets final states with
one or two $b$-tags will significantly improve the chances for
discovering new heavy particles in the $t\bar t$ channel at the LHC.
\end{abstract}

\newpage


\tightenlines

\section{Introduction}
\label{sec:one}
The Large Hadron Collider (LHC) is scheduled to have its first physics
run in 2008. Investigating jet, weak boson and top quark
production are some of the goals of the 2008 run. Top pair production at
the LHC, with a cross section which is about a factor~100 larger than at
the Fermilab Tevatron, will make it possible to
precisely determine the top quark properties~\cite{Beneke:2000hk}. It also 
offers an excellent chance to search for new physics in the early
operational phase of the LHC. Once the LHC reaches design luminosity,
$t\bar t$ production will provide access to new phenomena in the
multi-TeV region. 
Many extensions of the Standard Model (SM) predict particles which decay
into $t\bar t$ pairs, and thus show up as resonances in 
the $t\bar t$ invariant mass, $m(t\bar t)$, distribution. The masses of
these particles are typically in the TeV range. For example,
topcolor~\cite{topcolor,Hill:2002ap} and  Little
Higgs~\cite{Arkani-Hamed:2001nc,Schmaltz:2005ky,atlaslh} models predict
weakly coupled new vector bosons, models with extra
dimensions~\cite{Arkani-Hamed:1998rs,Randall:1999ee,anton} can have 
Kaluza-Klein (KK) excitations of the
graviton~\cite{Fitzpatrick:2007qr,Arai:2007ts} the
weak~\cite{anton,McMullen:2001zj} and the strong gauge
bosons~\cite{Agashe:2006hk,Lillie:2007yh,lillie2,abdel,Ghavri:2006kc,nandi,atlasglu,Burdman:2006gy}
which couple to top quarks, while massive axial vector bosons appear in
torsion gravity models~\cite{Belyaev:2007fn}. Resonances in the $t\bar
t$ channel also occur in technicolor~\cite{Eichten:1994nc,pallin},
chiral color~\cite{chicol} and models with a strong $SU(3)\times SU(3)$
gauge symmetry~\cite{Simmons:1996fz,Choudhury:2007ux}. In some
models~\cite{Fitzpatrick:2007qr,Agashe:2006hk,Lillie:2007yh,lillie2,abdel},
the couplings 
of the new particles to light quarks and gluons is suppressed, and the
$t\bar t$ final state becomes their main discovery channel. 

Top quarks decay either hadronically, $t\to Wb\to bq\bar q'$
($q,\,q'=u,\,d,\,s,\,c$), or 
semileptonically, $t\to Wb\to b\ell\nu$ ($\ell=e,\,\mu$; decays with
$\tau$ leptons in the final state are ignored here). Pair
production of top quarks thus results in so-called ``di-lepton+jets''
events, $t\bar t\to\ell^\pm\nu_\ell{\ell'}^\mp\nu_{\ell'}b\bar b$,
``lepton+jets'' events, $t\bar t\to\ell^\pm\nu b\bar bq\bar q'$, or
the ``all-hadronic'', $t\bar t\to b\bar b+4$~quarks, final state. For
sufficiently small top quark transverse momenta, $p_T(t)$, a substantial
fraction of $t\bar t$ 
events has a number of isolated jets within the $p_T$ and rapidity range
covered by the detector which is equal to or greater (when large angle,
hard QCD radiation is included) than the number of
quarks in the final 
state. This is reflected in the standard $t\bar t$ selection criteria of
the LHC experiments. For example, to identify lepton+jets events,
ATLAS and CMS require an isolated charged lepton, missing transverse
momentum, and at least four isolated hadronic jets. For events with more
than four
jets, the four leading (highest transverse momentum) jets  are selected. 
Of these four jets two have to be tagged as a 
$b$-quarks~\cite{atlastdr,cmstdr}. The main background in this case
originates from $Wb\bar b+2$~jets and $W+4$~jets production, and is quite
small~\cite{atlastdr,cmstdr}. 

While the standard top quark selection criteria work well for top quark
transverse momenta less 
than a few hundred GeV and $t\bar t$ invariant masses below 1~TeV, 
they are not adequate in the TeV region where signatures from new $t\bar t$
resonances are 
expected. In this region the top quark decay products are highly
boosted and thus almost collinear. This frequently results in
non-isolated leptons and/or merged or overlapping jets for lepton+jets
and all-hadronic $t\bar t$ events, {\it ie.} the
number of jets may be smaller than the number of final state
quarks. Furthermore, the $b$-tagging efficiency in the TeV region may be
significantly smaller than at low
energies~\cite{atlaslh,Lillie:2007yh,atlasglu}. Imposing standard $t\bar t$ 
selection criteria for very energetic top quarks therefore may dramatically
reduce the observable cross section for $t\bar t$ invariant masses in
the TeV range, and severely limit the sensitivity of the LHC experiments
to $t\bar t$ resonances in this range. 

Extending the selection criteria to include topologies with fewer jets is 
an obvious strategy for improving the selection efficiency for very
energetic top quarks. On the other hand, this may significantly increase the
background. For example, for 
lepton+jets events where all three quarks originating from the
hadronically decaying top quark merge into one jet, $Wb\bar b$ and
$Wjj$ production contribute to the background. These processes occur
at a lower order in perturbation theory than $pp\to Wb\bar b+2$~jets and
$pp\to W+4$~jets and therefore are potentially more dangerous. 
Relaxing the selection criteria further by requiring 
only one tagged $b$-quark in $t\bar t$ events may partially
compensate for the reduced $b$-tagging efficiency at very high
energies. However, this will also increase the reducible background where
a light quark or gluon jet is misidentified as a $b$-quark. Since the
mistagging probability worsens significantly with
energy~\cite{atlaslh,atlasglu}, the background for final states with
only one $b$-tag is potentially much larger than if two $b$-tags are
required. 

The importance of modifying the selection criteria for very energetic
top quarks to optimize the search for KK excitations of the
gluon in bulk Randall-Sundrum models has been discussed in
Refs.~\cite{Agashe:2006hk} and~\cite{Lillie:2007yh}. In this paper we
follow a more general approach and investigate whether it is
feasible to improve the $t\bar t$ selection efficiency for very
energetic top quarks. In Sec.~\ref{sec:two} we discuss the signatures
and the selection of $t\bar t$ events with high $p_T$ top quarks in the
lepton+jets, di-lepton+jets, and the all-hadronic decay modes. We also
investigate how next-to-leading order (NLO) QCD corrections affect the cross
section for the lepton+jets channel in the phase space region of
interest. The main result 
of Sec.~\ref{sec:two} is that the $\ell\nu+2$~jets and
$\ell\nu+3$~jets final state topologies with one or two tagged $b$-jets
offer the best chances to improve the $t\bar t$ selection efficiency. In
Sec.~\ref{sec:three} we calculate the differential cross sections of the
SM background processes as a function of the $t\bar t$ invariant mass
and the top quark transverse momentum for these final 
states. More precisely, we consider the $Wb\bar b+$jets, $(Wb+W\bar
b)+$jets, $W+$jets, $(t+\bar t)+$~jets, $(t\bar b+\bar tb)+$~jets,
$Wbt$, and $Wt(j)$ 
backgrounds. We also show that cluster transverse mass and invariant
mass cuts are sufficient to control the background at large values of
$m(t\bar t)$ and 
$p_T(t)$. Of particular interest for suppressing the background is a cut
on the jet invariant mass in $Wjj$ and $(t+\bar t)j$ production. In
Sec.~\ref{sec:four} we investigate the efficiency of such a cut. Our
conclusions are presented in Sec.~\ref{sec:five}.

Considering the $\ell\nu+2$~jets and $\ell\nu+3$~jets final state
topologies with one or two tagged $b$-jets in $t\bar t$ production is,
of course, not new. These final states have been successfully used to
search for the top quark in Run~1 of the Fermilab
Tevatron~\cite{topcdf,topd0} where it was 
essential to maximize the number of signal events. This is also the case
at the LHC in the high invariant mass and $p_T$ region when searching
for signals of new physics. However, there is an important difference
between the top quark search at the Tevatron and the search for new
physics in the $t\bar t$ channel at the LHC. At the Tevatron, most
$\ell\nu+2$~jets and $\ell\nu+3$~jets $t\bar t$ events are the result of
one or two jets which do not pass the $p_T$ and rapidity cuts
imposed. For very energetic top quarks at the LHC, jet merging is the
main source of such events. 

All tree level (NLO QCD) cross sections in this paper are computed using
CTEQ6L1 (CTEQ6M)~\cite{Pumplin:2002vw} parton distribution functions
(PDFs). For the CTEQ6L1 PDF's, the 
strong coupling constant is evaluated at leading order with
$\alpha_s(M_Z^2)=0.130$. The factorization and renormalization scales
for the calculation of the $t\bar t$ signal are set equal to
$\sqrt{m_t^2+p_T^2(t)}$, where $m_t=173$~GeV is the top quark mass. The
value of the top quark mass chosen is consistent with the most recent
experimental data~\cite{:2007bx}. The choice of factorization and
renormalization scales of the background processes is discussed in
Sec.~\ref{sec:three}. The Standard Model (SM) 
parameters used in all tree-level calculations are~\cite{Mangano:2002ea}
\begin{eqnarray}
\label{eq:input1}
G_{\mu} = 1.16639\times 10^{-5} \; {\rm GeV}^{-2}, & \quad & \\
M_Z = 91.188 \; {\rm GeV}, & \quad & M_W=80.419  \; {\rm GeV}, \\
\label{eq:input2} 
\sin^2\theta_W=1-\left({M^2_W\over M_Z^2}\right), & \quad &
\alpha_{G_\mu} = {\sqrt{2}\over\pi}\,G_F \sin^2\theta_W M_W^2,
\end{eqnarray}
where $G_F$ is the Fermi constant, $M_W$ and $M_Z$ are the $W$ and
$Z$ boson masses, $\theta_W$ is the weak mixing angle, and
$\alpha_{G_\mu}$ is the electromagnetic coupling constant in the $G_\mu$
scheme. 

\section{Detecting very energetic top quarks}
\label{sec:two}

\subsection{The lepton+jets final state at leading order}
\label{sec:twoa}

We begin our discussion by examining the lepton+jets final state in more
detail. The di-lepton+jets and the all-hadronic final states will be
discussed in Sec.~\ref{sec:twod}.
Approximately 30\% of all top quark pairs yield lepton+jets
events. We calculate the SM $t\bar t\to\ell\nu b\bar bq\bar q'$ cross
section at leading-order (LO), including all decay correlations. Top
quark and $W$ decays are treated in the narrow width approximation.
We require that both $b$-quarks are tagged
with a constant efficiency of $\epsilon_b=0.6$ and that there are two
additional jets in the event which are not tagged. We sum over electron
and muon final states and impose the
following acceptance cuts on $\ell\nu b\bar{b}jj$ events at the
LHC ($pp$ collisions at $\sqrt{s}=14$~TeV):
\begin{eqnarray}\label{eq:cuts1}
p_T(\ell)>20~{\rm GeV}, & \qquad & |\eta(\ell)|<2.5, \\ \label{eq:cuts2}
p_T(j)>30~{\rm GeV}, & \qquad & |\eta(j)|<2.5, \\ \label{eq:cuts3}
p_T(b)>30~{\rm GeV}, & \qquad & |y(b)|<2.5,\\ \label{eq:cuts4}
p\llap/_T>40~{\rm GeV}. & \qquad & 
\end{eqnarray}
Here, $\eta$ ($y$) is the pseudo-rapidity (rapidity), $\ell=e,\,\mu$, 
and $p\llap/_T$ is 
the missing transverse momentum originating from the neutrino in
$t\to b\ell\nu$ which escapes undetected. We include minimal
detector effects via Gaussian smearing of parton momenta according to
ATLAS~\cite{atlastdr} expectations, and take into account the $b$-jet energy
loss via a parametrized function. Charged leptons are assumed to be
detected with an efficiency of $\epsilon_\ell=0.85$. All numerical
results presented in this paper include the appropriate combination of
$b$-tagging and lepton detection efficiencies unless specified
otherwise. In addition to the cuts listed in Eqs.~(\ref{eq:cuts1})
--~(\ref{eq:cuts4}), the LHC experiments will
also impose isolation cuts on all final state objects except the
missing transverse momentum by requiring the separation in
pseudo-rapidity -- azimuth space to be larger than a minimum value,
$R_{min}$: 
\begin{equation}
\Delta R=[(\Delta\eta)^2+(\Delta\Phi)^2]^{1/2}>R_{min}. \label{eq:cuts5}
\end{equation}
The minimum separation usually is in the range $R_{min}=0.4-0.7$.

New particles which decay into a pair of top quarks lead to resonances
in the $t\bar t$ invariant mass distribution and to a Jacobian peak in
the top quark transverse momentum distribution. In the following we
therefore concentrate on these observables. Since the neutrino escapes
undetected, $m(t\bar t)$ cannot be directly reconstructed. However,
assuming that the charged lepton and the missing transverse momentum
come from a $W$ boson with a fixed invariant mass $m(\ell\nu)=M_W$, it
is possible to reconstruct the longitudinal momentum of the neutrino,
$p_L(\nu)$, albeit with a twofold ambiguity. In our calculations of the
$m(t\bar t)$ distribution in the lepton+jets final state, we
reconstruct the $t\bar t$ invariant mass using both solutions for
$p_L(\nu)$ with equal weight. The energy loss of the $b$-quarks slightly
distorts the $p\llap/_T$ distribution. As a result, the quadratic
equation for $p_L(\nu)$ does not always have a solution. Events for
which this is the case are discarded in our analysis. This results in a
$\approx 10\%$ reduction of the $t\bar t$ cross section in the $m(t\bar
t)$ distribution. More advanced
algorithms~\cite{Barger:2006hm} improve the reconstruction of the mass
of the new physics signal, however, they have little effect on the shape
of the SM $m(t\bar t)$ distribution.

In order to reconstruct the $t$ or $\bar t$ transverse momentum one has
to correctly assign the $b$ and $\bar b$ momenta to the parent top or
anti-top quark. Since it is impossible to determine the $b$-charge on an
event-by-event basis, we combine $p\llap/_T$, $p_T(\ell)$, and the
transverse momentum of the tagged $b$-jet with the smaller
separation from the charged lepton to form the transverse momentum of
the semileptonically decaying top quark. The $p_T$'s of the two
non-tagged jets and the other $b$-jet form the transverse momentum of
the hadronically decaying top\footnote{Alternatively, one could select
the $bjj$ combination which minimizes
$|m(bjj)-m_t|$~\cite{Affolder:2000dt}.}. We find that the reconstructed
 and true top quark transverse momentum distributions are virtually
identical except for transverse momenta below 50~GeV
where deviations at the few percent level are observed. 

\begin{figure}[th!] 
\begin{center}
\includegraphics[width=14.6cm]{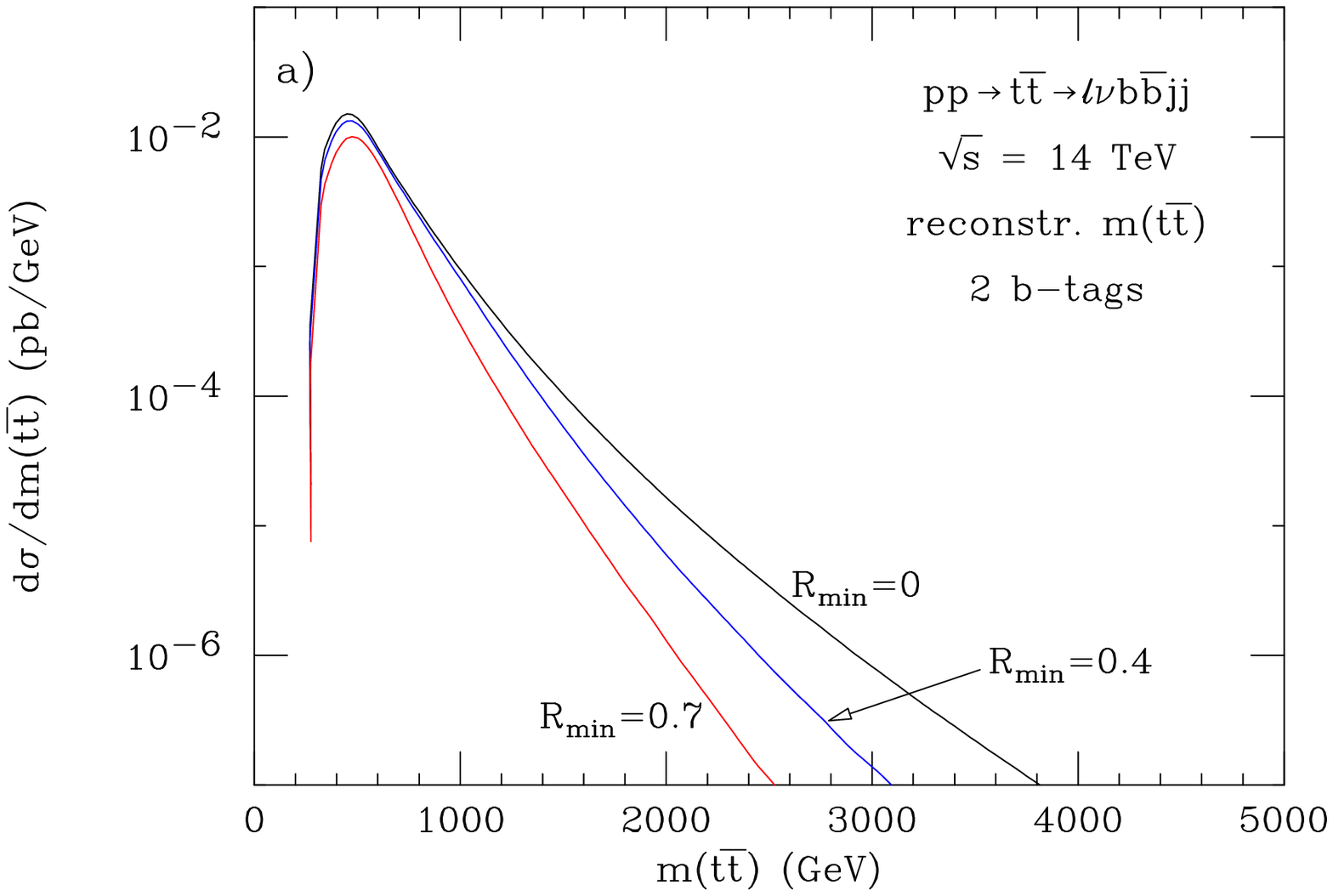} \\[3mm]
\includegraphics[width=14.6cm]{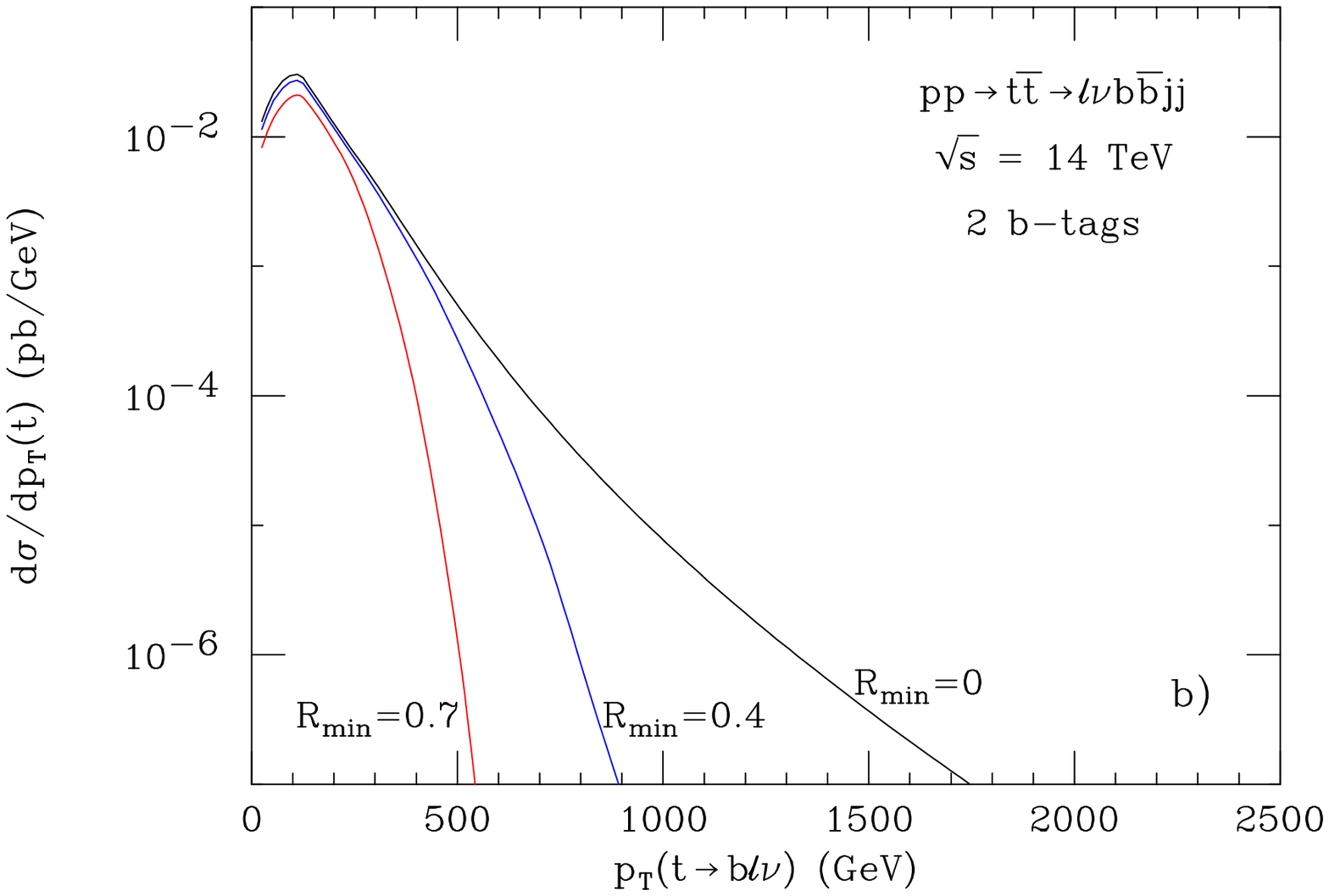}
\vspace*{2mm}
\caption[]{\label{fig:one} 
The LO $pp\to t\bar t\to\ell\nu b\bar bjj$ differential cross section
at the LHC for three choices of $R_{min}$ as a function of a) the
reconstructed $t\bar t$ invariant mass, and b) the reconstructed $p_T$
of the semileptonically decaying top quark. The cuts imposed are listed
in Eqs.~(\ref{eq:cuts1}) --~(\ref{eq:cuts4}).}
\vspace{-7mm}
\end{center}
\end{figure}

The angle between the momentum vector of a top quark decay particle and the 
flight direction of the parent $t$ quark tends to be small for very
energetic top quarks, due to the large Lorentz boost. Imposing a
standard isolation cut on the charged lepton and the jets in $t\bar
t\to\ell\nu b\bar bjj$ events thus significantly reduces the $t\bar t$
cross section at large values of $m(t\bar t)$ and $p_T(t)$. This is seen
in Fig.~\ref{fig:one}, where we show the $t\bar t$ invariant mass and
the $p_T(t\to b\ell\nu)$ distribution for the $\ell\nu b\bar bjj$ final
state and three choices of $R_{min}$, with $R_{min}=0$ corresponding to
no isolation cut being imposed.
The smallest differential cross section shown in Fig.~\ref{fig:one},
$10^{-7}$~pb/GeV, corresponds to 1~event in a 100~GeV bin for an
integrated luminosity of 100~fb$^{-1}$. It can be viewed as a crude
measure of the LHC reach once it operates at design luminosity. At LO,
the transverse momentum distributions 
of the $t$ and the $\bar t$ quark are identical. We therefore do not
show the transverse momentum distribution of the hadronically decaying
top quark in Fig.~\ref{fig:one}. The figure demonstrates that the
isolation cut greatly reduces the cross section in the TeV region, in
particular in the $p_T(t)$ distribution. 

Events which fail the isolation
cut either have a charged lepton which is embedded inside a jet, or some
of the final state quark jets merge and one observes lepton+jets events
with fewer than four jets. Events with non-isolated leptons are
difficult to utilize. 
In most $t\bar t$ lepton+jets events with a non-isolated
lepton, the lepton is embedded in the $b$-jet which originates from the
same parent top quark. Such a lepton can easily be confused with a
charged lepton originating from semileptonic $b$-decay. Furthermore,
such events look similar to QCD $b\bar b+$jets events where one or more
badly mismeasured jets result in a significant amount of missing
transverse momentum. Finally, since the neutrino is not required to be
isolated, the $\Delta R$ cut affects the $t\to b\ell\nu$ decay much less than
$t\to bjj$. Trying to utilize lepton+jets events with fewer than
four jets thus is potentially more beneficial than attempting to use
events where the lepton is not isolated. 

In the following we therefore focus on $\ell\nu+n$~jets events with
$n<4$. If we require two $b$-tagged jets, the $n=1$ final state does not
contribute. This leaves the $\ell\nu+2$~jets and $\ell\nu+3$~jets final
states. 

\begin{figure}[th!] 
\begin{center}
\includegraphics[width=13.9cm]{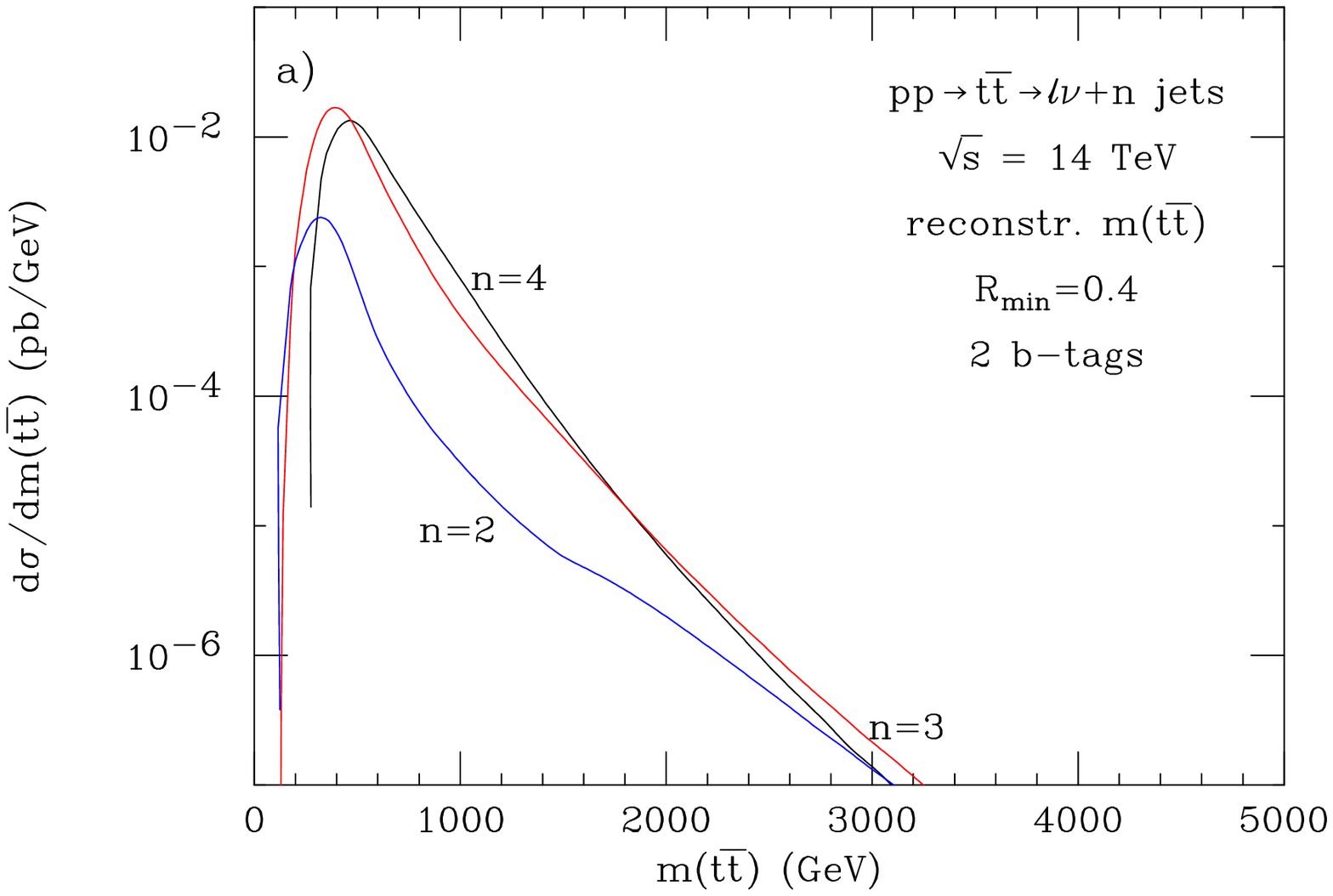} \\[3mm]
\includegraphics[width=13.9cm]{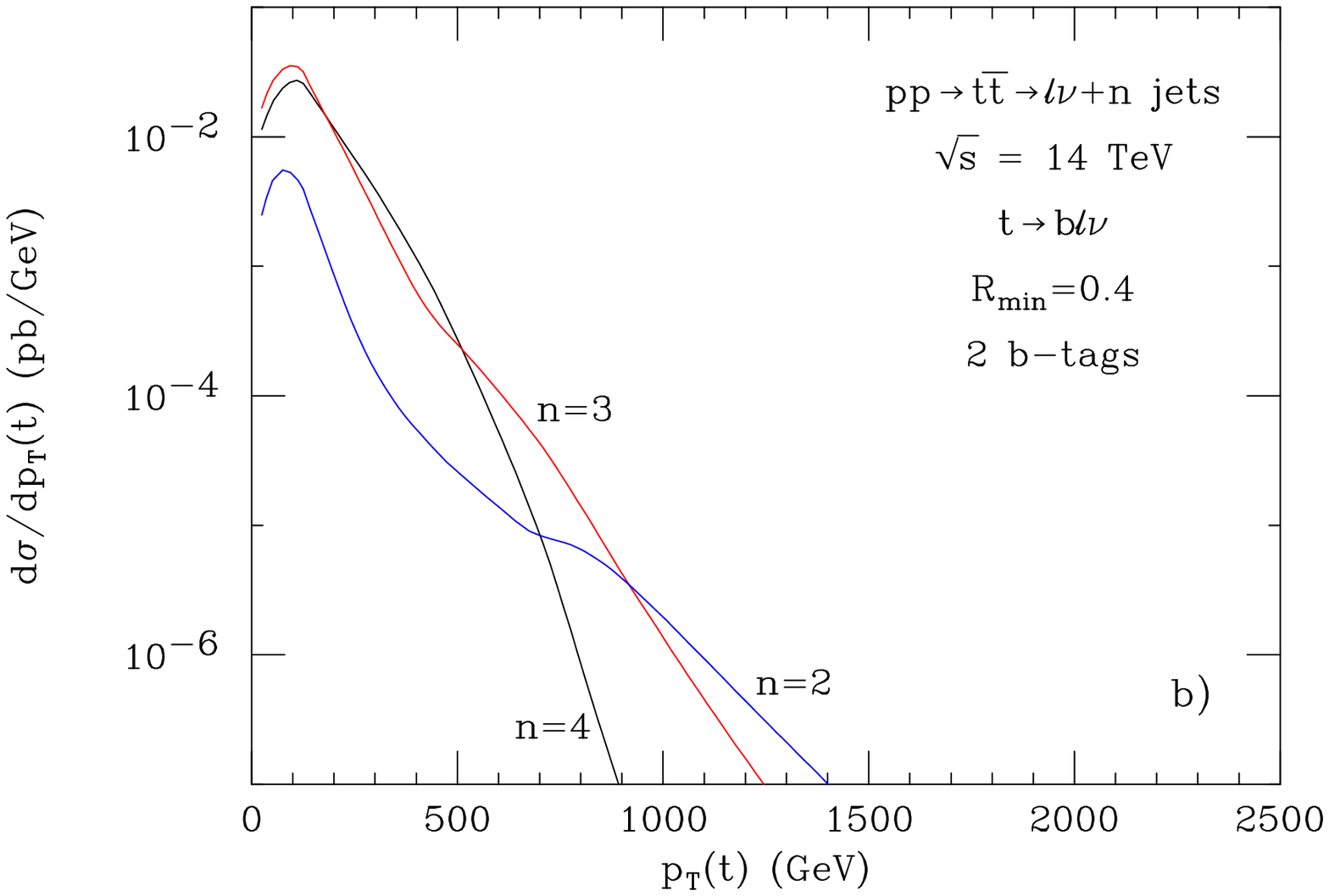}
\vspace*{2mm}
\caption[]{\label{fig:two} 
The LO $pp\to t\bar t\to\ell\nu +n$~jets differential cross section
at the LHC as a function of a) the
reconstructed $t\bar t$ invariant mass, and b) the reconstructed $p_T$
of the semileptonically decaying top quark. Shown are the cross sections
for two (blue), three (red) and four (black) jets in the final
state. Two of the jets are assumed to be $b$-tagged. The cuts imposed
are listed in Eqs.~(\ref{eq:cuts1}) 
--~(\ref{eq:cuts4}). In addition an isolation cut (see
Eq.~(\ref{eq:cuts5})) with $R_{min}=0.4$ is imposed and jets with
$\Delta R<0.4$ are merged.} 
\vspace{-7mm}
\end{center}
\end{figure}

We calculate LO $pp\to t\bar t\to\ell\nu+n$~jets production by merging
light quark jets from $W\to q\bar q'$ and $b$-quark jets if 
\begin{equation}
\Delta R(i,j)<0.4,
\end{equation}
$i,\,j=q,\,q',\,b$. If a $b$-quark jet and a light quark jet merge, their
momenta are combined into a $b$-jet. Jets are counted and used in the
reconstruction of $m(t\bar t)$ if they 
satisfy Eqs.~(\ref{eq:cuts2}) and~(\ref{eq:cuts3}) after merging. The
$m(t\bar t)$ and $p_T(t\to b\ell\nu)$ differential cross sections for
$pp\to t\bar t\to\ell\nu+n$~jets with two $b$-tags and $n=2,\,3$ are shown in
Fig.~\ref{fig:two}a and~\ref{fig:two}b, respectively. For comparison, we
also show the $\ell\nu+4$~jets distributions.
Taking into account the $\ell\nu+2$~jets and $\ell\nu+3$~jets final states
is seen to increase the cross section for very large values of $m(t\bar
t)$ by more than a factor of~3. The effect is even more pronounced in the
$p_T(t)$ distribution where the 2~jet and 3~jet final states extend the
range which is accessible from 900~GeV to about 1.4~TeV. The LO results
shown in Fig.~\ref{fig:two} are, of course, expected to be somewhat
modified by QCD corrections. The QCD corrections for the
lepton+jets final state topologies will be discussed in more detail in
Sec.~\ref{sec:twob}. They do not qualitatively change the results
presented in Fig.~\ref{fig:two}.

At small transverse momenta and invariant masses, most of the 2~jet and
3~jet final states originate from 4~jet events where one or both light
quark jets do not pass the $p_T$ and rapidity cuts of
Eq.~(\ref{eq:cuts2}). With increasing energies, more and more
$\ell\nu+n$~jet events with $n=2,\,3$ contain jets which originate from
jet merging. This is most pronounced in the $n=2$ case where it leads to
a shoulder in the differential cross section at $m(t\bar t)\approx
1.5$~TeV and $p_T(t)\approx 700$~GeV. At large invariant masses or
$p_T$'s, $t\bar t\to\ell\nu+2$~jet events originate almost exclusively
from the merging of all three quarks in $t\to bq\bar q'$ into one
``$t$-jet'' which, however, is tagged as a $b$-jet. The invariant mass
of such jets, which is close to $m_t$, and their shower
profile~\cite{vachon} are potential tools for discriminating signal and
QCD $Wjj$/$Wb\bar b$ background
events~\cite{Agashe:2006hk,Lillie:2007yh}. The jet 
invariant mass cut will be discussed in more detail in Sec.~\ref{sec:four}. 
Alternatively, one can pursue a strategy similar to that discussed in
Refs.~\cite{Butterworth:2007ke} and~\cite{Butterworth:2002tt}. 
$t\bar t\to\ell\nu+3$jet events at high energies originate either from
the merging of the two light quark jets from $W\to q\bar q'$, or from the
merging of one light quark jet with a $b$-jet.

\begin{figure}[th!] 
\begin{center}
\begin{tabular}{cc}
\includegraphics[width=8.2cm]{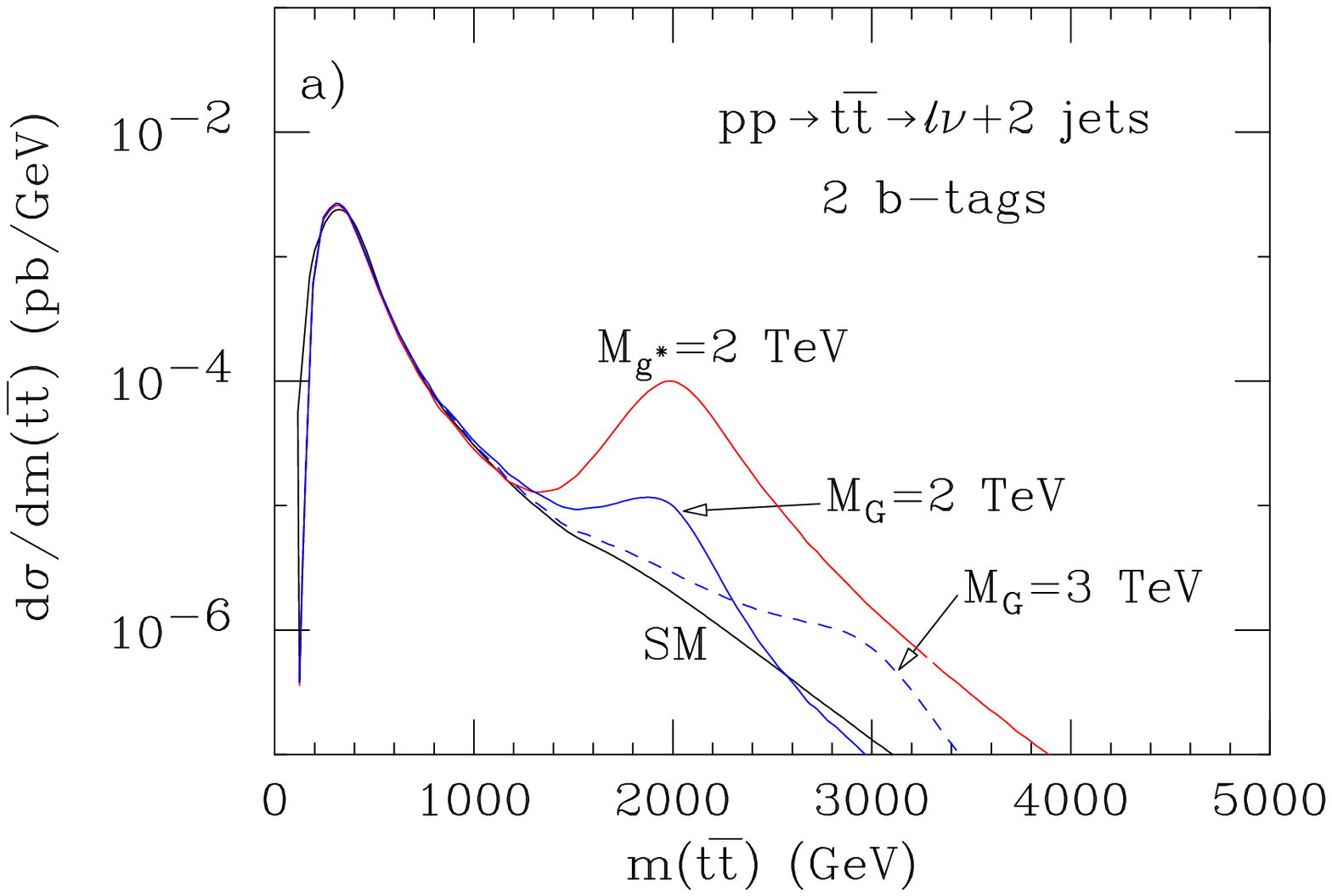} & 
\includegraphics[width=8.2cm]{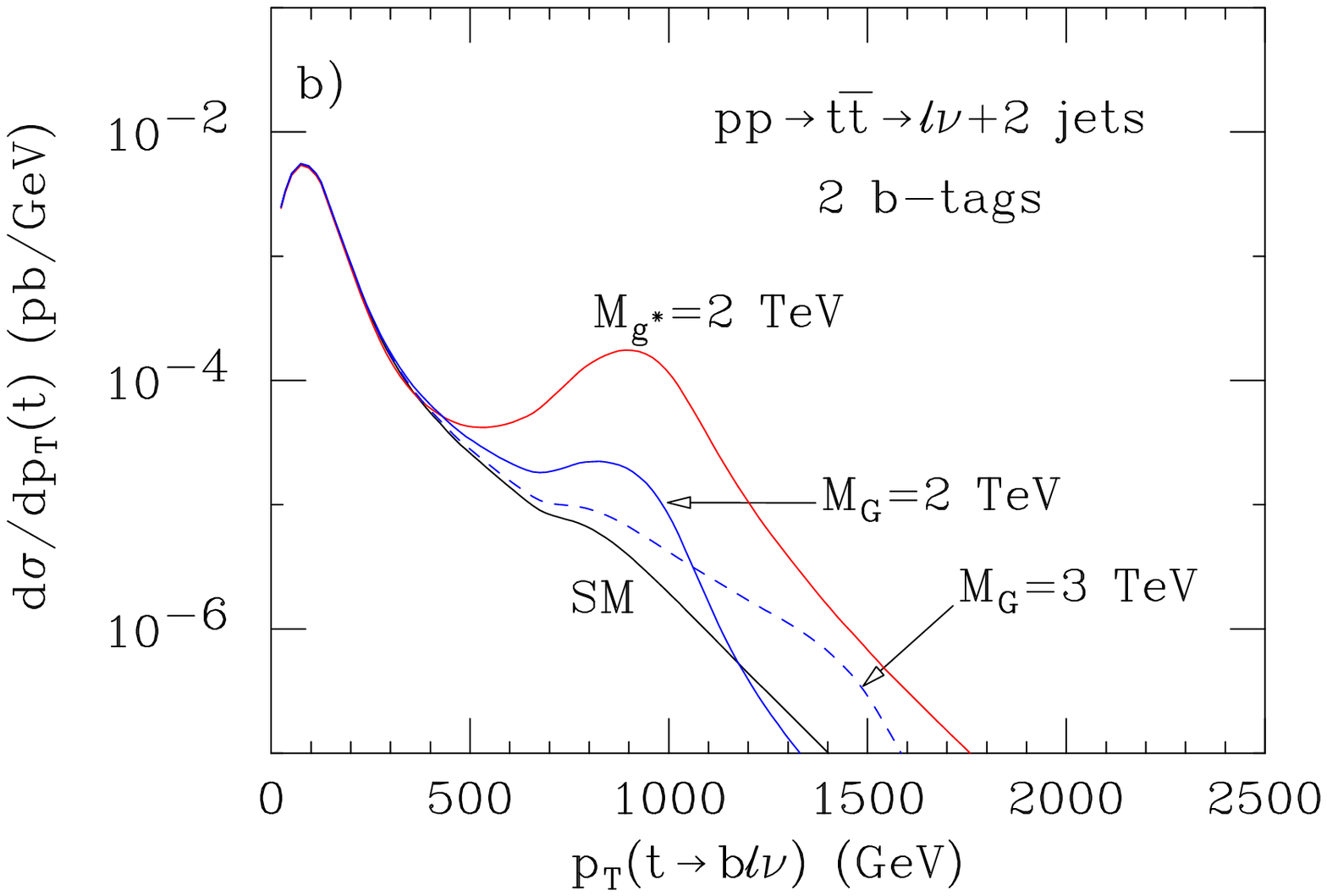} \\[4mm]
\includegraphics[width=8.2cm]{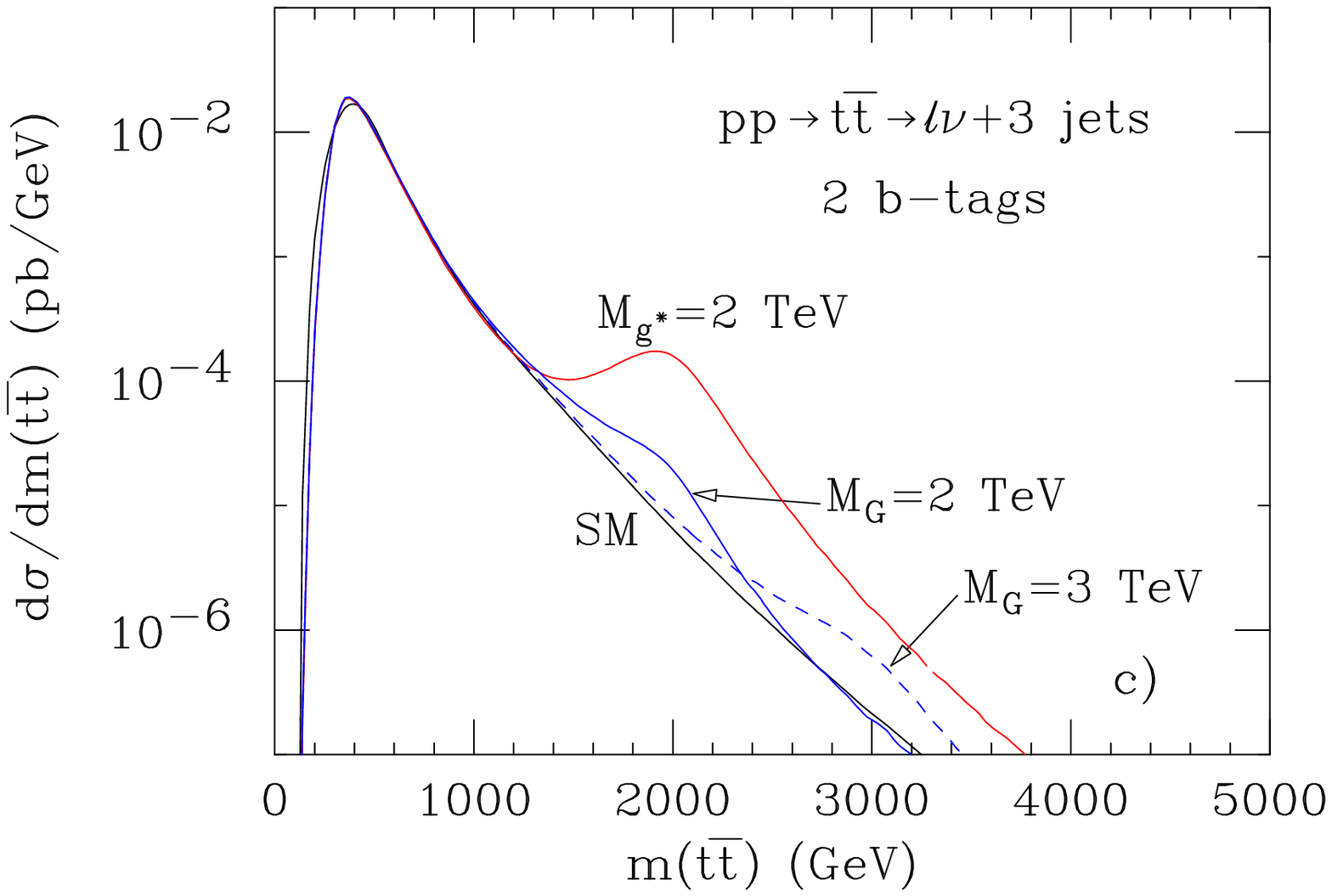} & 
\includegraphics[width=8.2cm]{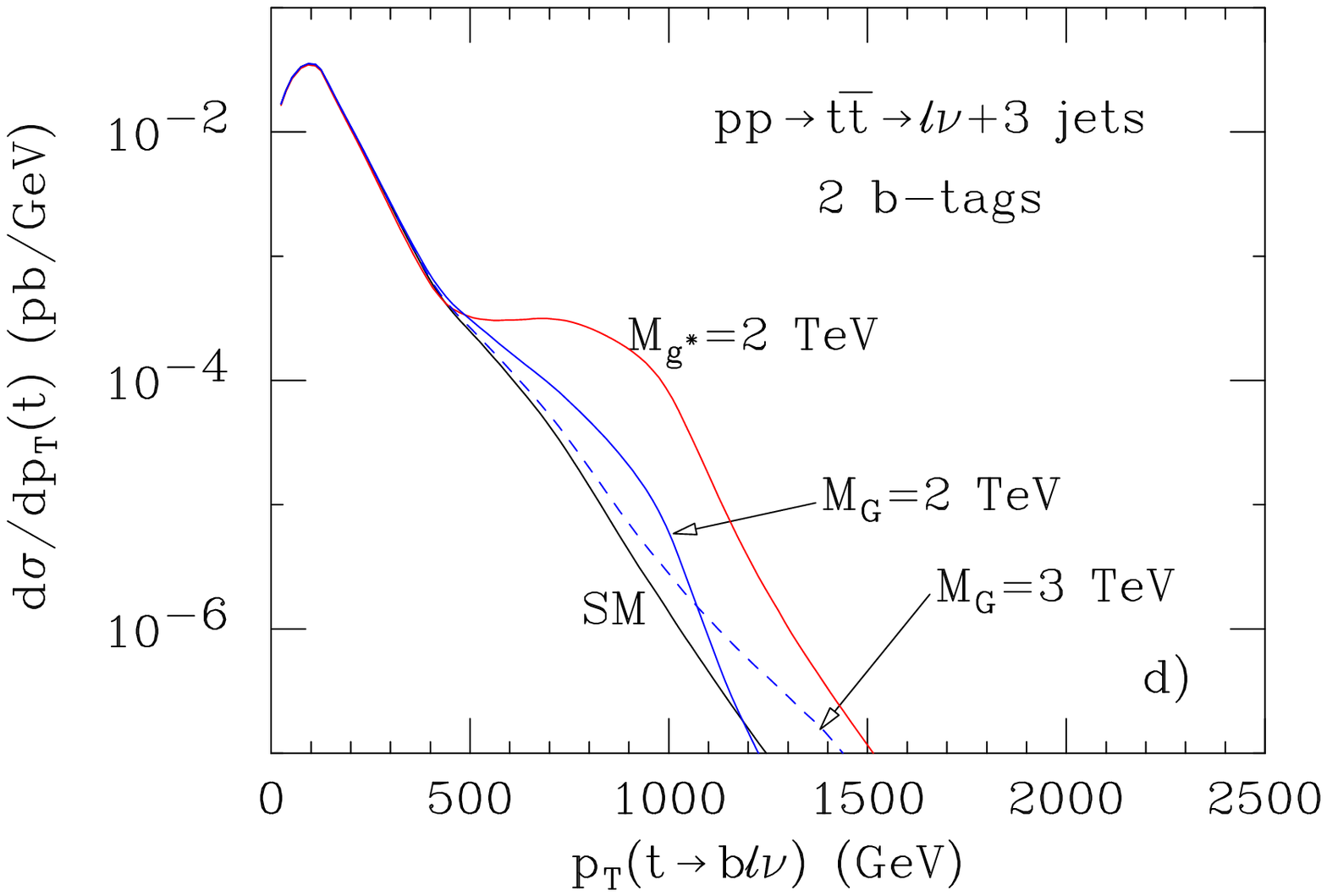} \\[4mm]
\includegraphics[width=8.2cm]{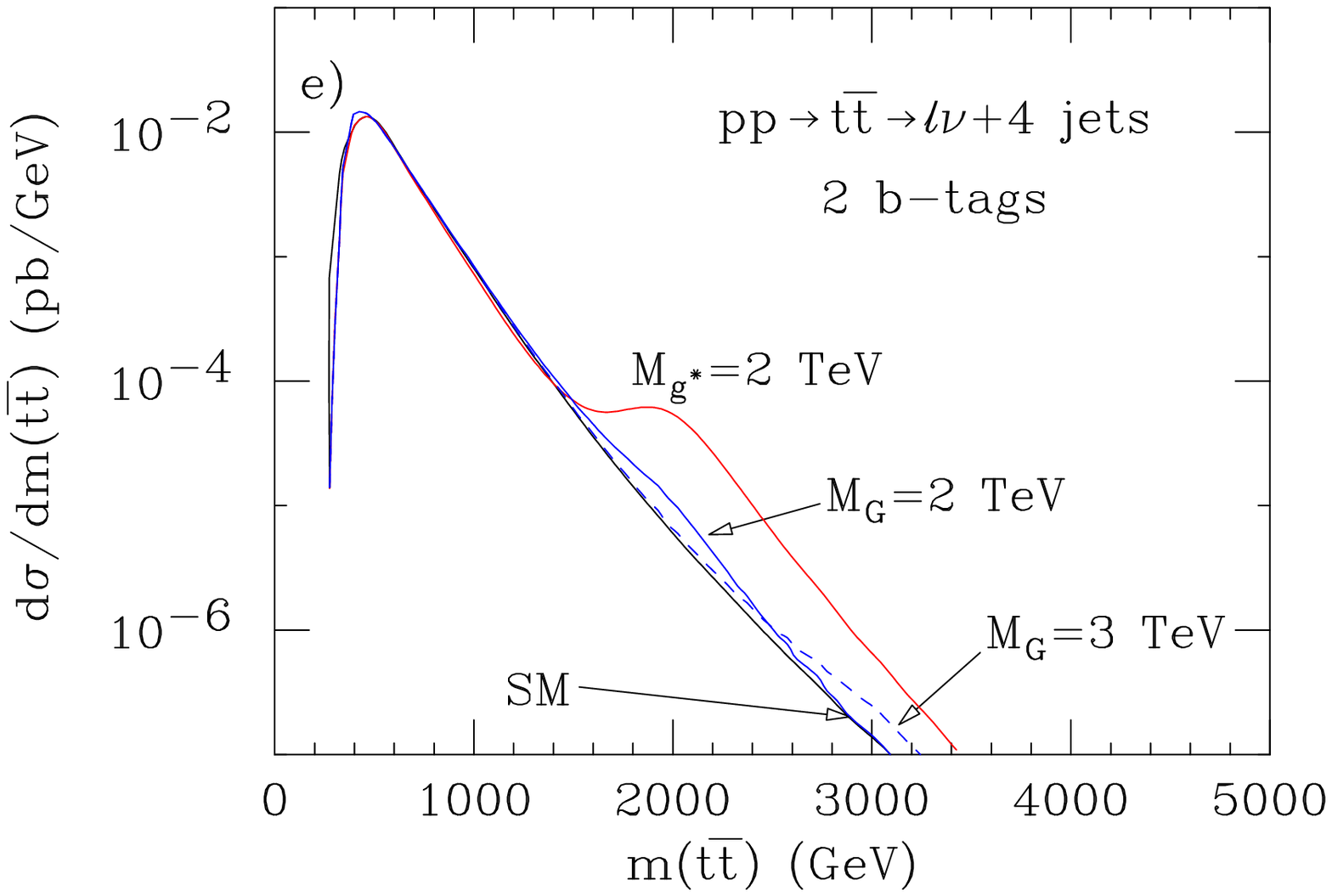} & 
\includegraphics[width=8.2cm]{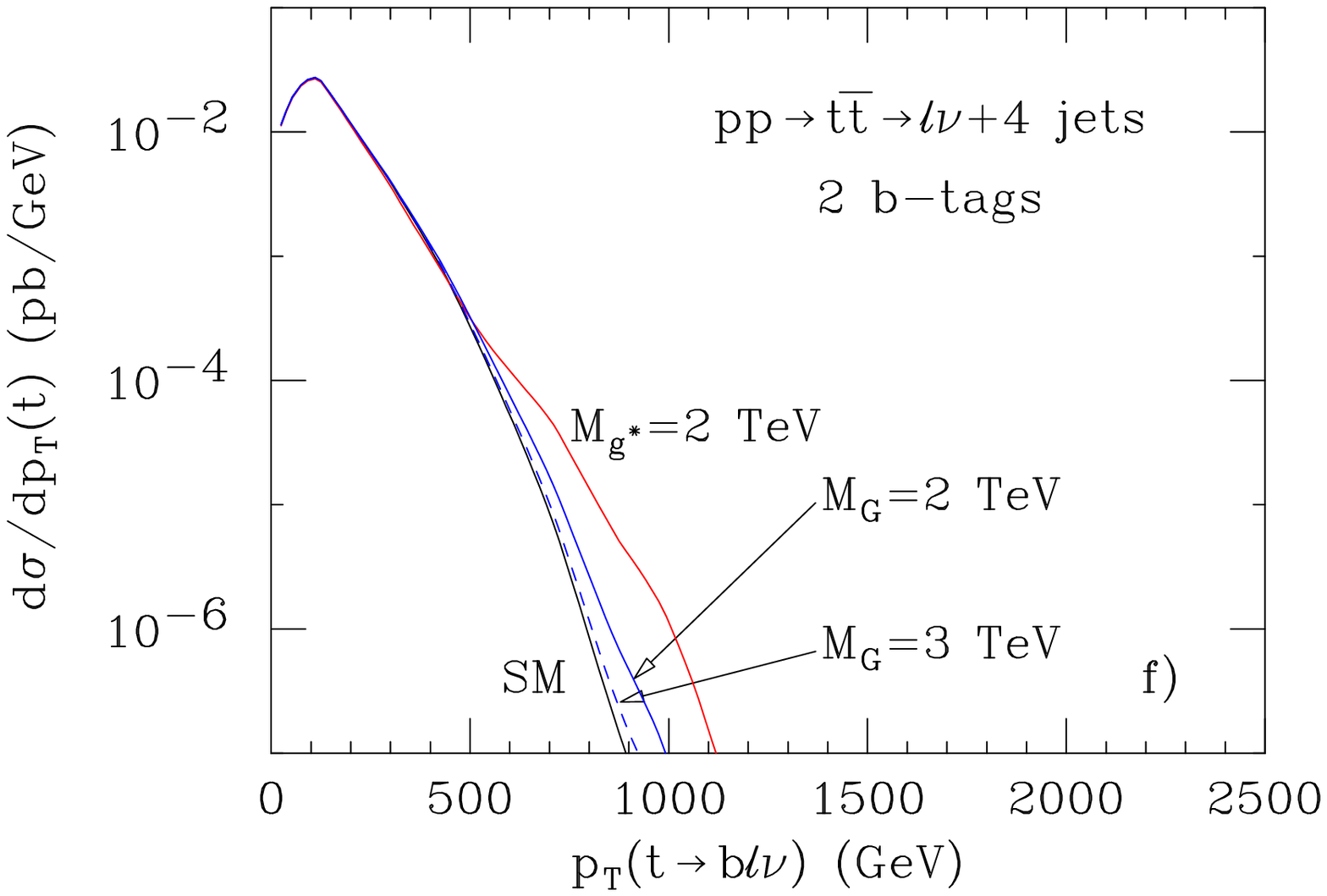}
\end{tabular}
\vspace*{2mm}
\caption[]{\label{fig:three} 
The LO $pp\to t\bar t\to\ell\nu +n$~jets, $n=2,\,3,\,4$, differential
cross section at the LHC as a function of the
reconstructed $t\bar t$ invariant mass, and the reconstructed $p_T$
of the semileptonically decaying top quark. Two of the jets are assumed
to be $b$-tagged. Shown are the cross sections 
for two, three and four jets in the final
state in the SM (black lines) and for two types of KK gluon
excitations (see text). The cuts imposed are listed
in Eqs.~(\ref{eq:cuts1}) --~(\ref{eq:cuts4}). In addition an isolation
cut (see Eq.~(\ref{eq:cuts5})) with $R_{min}=0.4$ is imposed and jets with
$\Delta R<0.4$ are merged.} 
\vspace{-7mm}
\end{center}
\end{figure}

SM $t\bar t$ production at the LHC is dominated by gluon fusion and the
$t$-channel top quark exchange diagram is playing an important role in
this process. As a result, top quarks tend to be produced with a fairly
large rapidity, {\it ie.} with a large $t\bar t$ invariant mass but a
relatively small top quark $p_T$. The steeply falling $p_T(t)$
distribution (see Fig.~\ref{fig:two}b) reflects this behavior. New
physics particles decaying into $t\bar t$, $X\to t\bar t$, manifest
themselves as 
$s$-channel resonances, leading to a Jacobian peak in the top quark
transverse momentum distribution which peaks at $M_X/2$, where $M_X$ is
the $X$ mass. The relatively larger impact the 
$t\bar t\to\ell\nu+2,3$~jet final states have on the SM $p_T(t)$
distribution, therefore, should carry over to the $m(t\bar t)$
distribution in the vicinity of the $X$ resonance, {\it ie.} the $X$
resonance should be significantly more pronounced in lepton+jets events
with 2~or 3~jets. This is borne out
in Fig.~\ref{fig:three}, where we show the $t\bar t$ invariant mass and
top quark $p_T$ distributions for the $t\bar t\to\ell\nu+n$~jet final
states with $n=2,\,3$ in the SM (black solid lines), and for two types
of KK excitations of the gluon. For comparison, we also show the
differential cross sections for $n=4$.
The red curves give predictions for a KK
gluon, $g^*$, with $M_{g^*}=2$~TeV, vector like couplings to quarks and
coupling strength $g_{g^*}=\sqrt{2}g_s$, where $g_s$ is the QCD coupling
constant~\cite{nandi,atlasglu}. The solid and dashed blue lines show the
cross sections for bulk RS KK gluons, $G$, with $M_G=2$~TeV and 3~TeV,
respectively~\cite{Agashe:2006hk,Lillie:2007yh}. Bulk RS KK gluons
have vector-like couplings with strength $g_G=-0.2g_s$ to all quarks
except the
top and bottom quarks for which $g^b_{GL}=g_s$, $g^b_{GR}=-0.2g_s$,
$g^t_{GL}=g_s$ and $g^t_{GR}=4g_s$. The width of the KK 
gluons is taken to be $\Gamma_{g^*,G}=0.17\,M_{g^*,G}$. They do not
couple to gluons at LO. It is obvious
that the $\ell\nu+2$~jet and $\ell\nu+3$~jet final states offer a much
better chance to discover such $t\bar t$ resonances, especially if the mass
of the resonance is larger than 2~TeV. Qualitatively similar results are
obtained for other types of $t\bar t$ resonances. However, if they
couple weakly to top quarks, such as $Z'$ bosons appearing in Little
Higgs or topcolor models, their significance may be very much reduced.

\subsection{NLO QCD corrections to the lepton+jets final state at high
energies} 
\label{sec:twob}

So far, our calculations have been limited to lowest order in
perturbation theory. Since QCD corrections to top pair production are
known to be 
significant, we now study how NLO QCD corrections affect the lepton+jets
final state topologies. QCD corrections may
change the normalization and/or the shape of distributions. In addition,
hard QCD bremsstrahlung may produce additional isolated jets which
complicate the 
reconstruction of the $t\bar t$ invariant mass and the top quark transverse
momentum distribution. 
QCD corrections apply to both the top production and decay processes; 
interference between the two is negligible in the narrow width 
approximation, which we employ.
For very energetic top quarks, most extra jets originate 
from production-stage radiation, {\it i.e.} from QCD corrections for 
$t\bar t$ production. Jets coming from decay-stage radiation,
{\it i.e.} from QCD corrections to $t\to b\ell\nu$ and $t\to bq\bar q'$,
rarely lead
to additional isolated jets, due to the large Lorentz boost for very
energetic top quarks.

We first investigate how NLO QCD corrections to $t \bar t$ production 
modify the
shape and normalization of the $m(t\bar t)$, $p_T(t\to b\ell\nu)$ and
$p_T(t\to bjj)$ distributions for a given number of $t\bar t$ decay
jets. This assumes that the jets from the hadronic top
decay have been correctly identified, for example by imposing an
invariant mass cut on one or several jets. Subsequently, we will then discuss
how QCD corrections affect these distributions for fixed observed jet
multiplicities. 

\begin{figure}[t!] 
\begin{center}
\includegraphics[width=14.2cm]{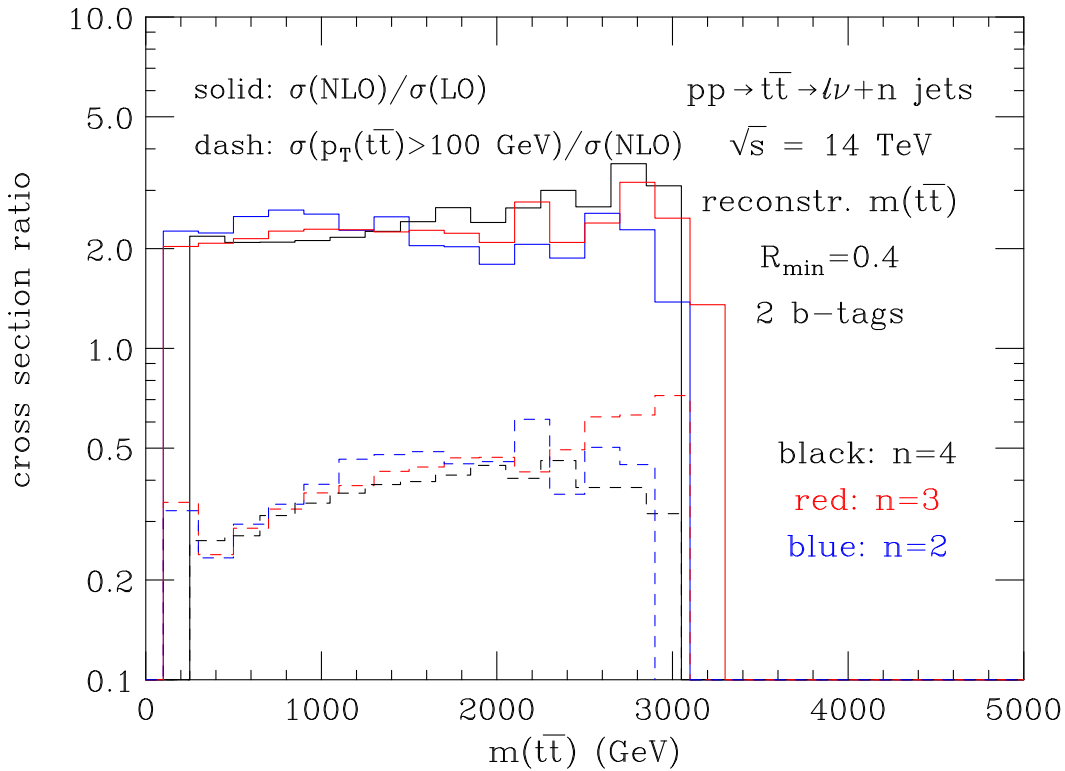} 
\vspace*{2mm}
\caption[]{\label{fig:four} 
The NLO to LO $pp\to t\bar t\to\ell\nu +n$~jets cross section ratio
(solid histograms) at the LHC as a function of the
reconstructed $t\bar t$ invariant mass. Here, $n$ is the number of
$t\bar t$ decay jets. The dashed histograms display
the fraction of the NLO $pp\to t\bar t\to\ell\nu +n$~jets events for
which $p_T(t\bar t)>100$~GeV. Shown are the cross section ratios
for two (blue), three (red) and four (black) $t\bar t$ decay jets. Two
of the jets are assumed to be $b$-tagged. The cuts imposed 
are listed in Eqs.~(\ref{eq:cuts1}) 
--~(\ref{eq:cuts4}). In addition an isolation cut (see
Eq.~(\ref{eq:cuts5})) with $R_{min}=0.4$ is imposed. $t\bar t$ decay
jets with a separation $\Delta R<0.4$ have been merged.} 
\vspace{-7mm}
\end{center}
\end{figure}

The NLO QCD corrections to $t\bar t$ production have been know for more
than 15~years~\cite{Nason:1987xz}. A more recent
calculation~\cite{Bernreuther:2004jv} includes top quark decays and spin
correlations. The NLO QCD corrections to $t\bar t$
production have been interfaced with the HERWIG shower Monte
Carlo\cite{Corcella:2002jc} in the program
MC@NLO~\cite{Frixione:2003ei}. This produces a realistic $t\bar t$
transverse momentum distribution. Furthermore, MC@NLO includes top quark
decay~\cite{Frixione:2007zp} and thus makes it possible to include
acceptance cuts in  
the calculation. Using MC@NLO to compute the $t\bar t$ cross section
including QCD corrections, we show the NLO to LO cross section ratio
($k$-factor) for $pp\to t\bar t\to\ell\nu+n$~jets, $n=2,\,3,\,4$, as a
function of the 
reconstructed $t\bar t$ invariant mass in Fig.~\ref{fig:four} (solid
histograms). Here, and in all other figures in this Section, $n$ is the
number of jets resulting from the decay of 
the $t\bar t$ pair, {\sl not} the number of jets in the
event. Furthermore, we call the cross section obtained with MC@NLO the
``NLO'' cross section, although this is, strictly speaking, not correct:
MC@NLO does take into account multiple gluon radiation in the
leading log approximation. 

For  3~jet and 4~jet final states, Fig.~\ref{fig:four} shows that the
$k$-factor increases slowly with 
$m(t\bar t)$. In the 2~jet case, it rises at low invariant
masses, and then decreases somewhat for $m(t\bar t)>1$~TeV. We also 
show the fraction of NLO $pp\to t\bar t\to\ell\nu +n$~jet events with
$p_T(t\bar t)>100$~GeV in Fig.~\ref{fig:four} (dashed histograms). The
fraction increases from about 25\% at low invariant mass to $40-50\%$ at
$m(t\bar t)\geq 2$~TeV. 

\begin{figure}[th!] 
\begin{center}
\includegraphics[width=12.8cm]{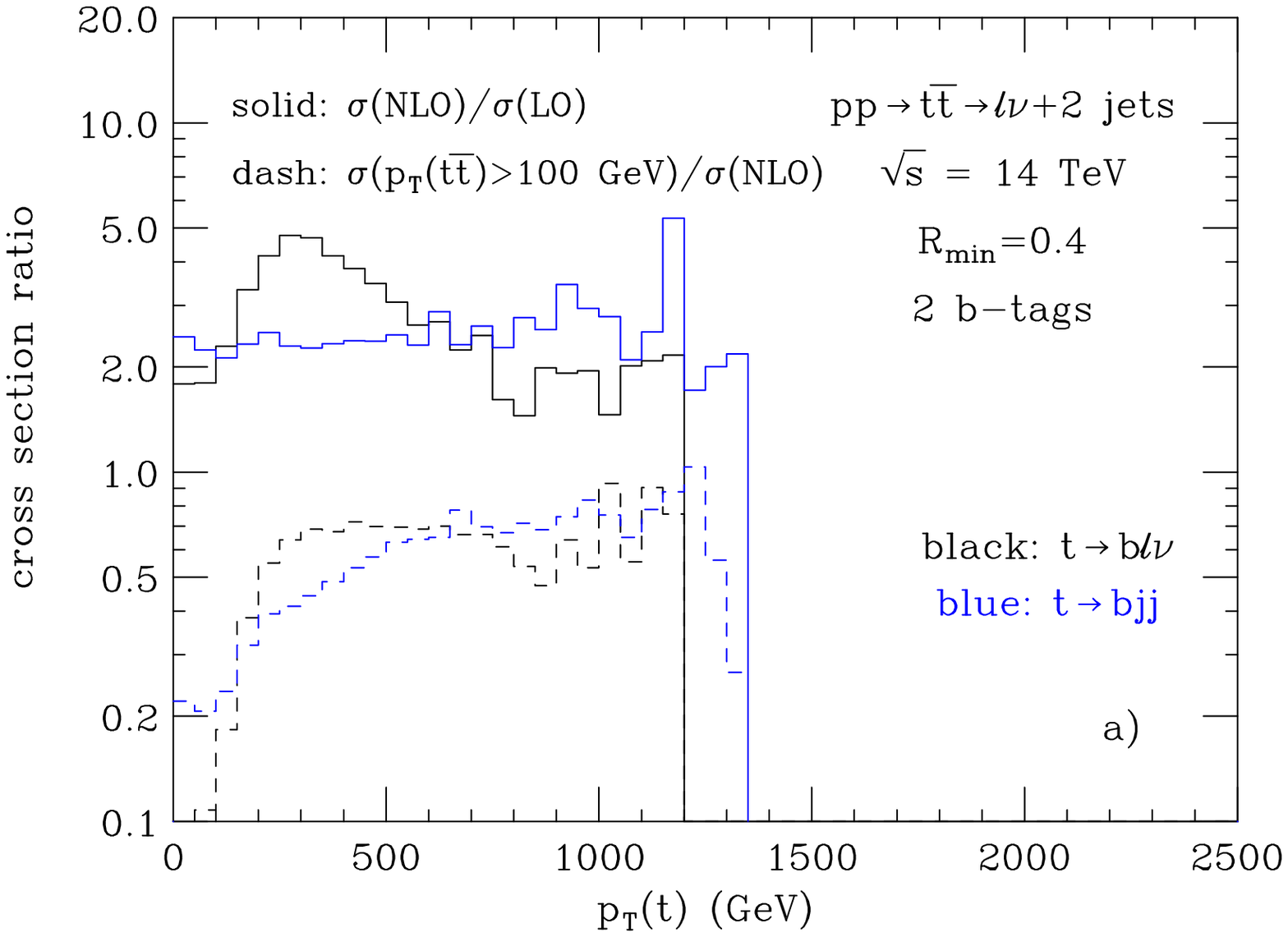} \\[3mm]
\includegraphics[width=12.8cm]{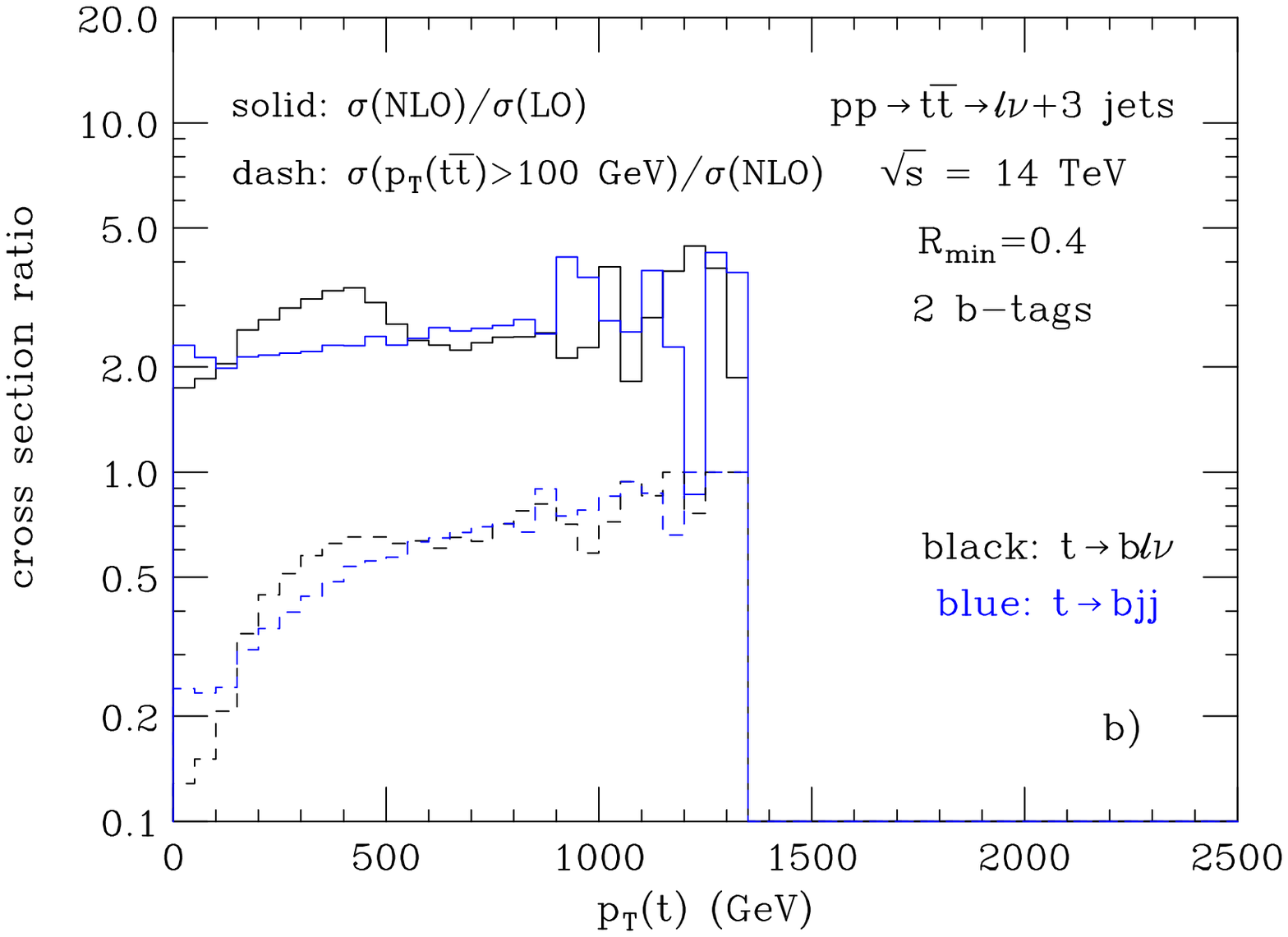}
\vspace*{2mm}
\caption[]{\label{fig:five} 
The NLO to LO cross section ratio (solid histograms) for $pp\to t\bar
t\to\ell\nu +n$~jets at the LHC as a function of the $t\to b\ell\nu$
(black) and $t\to bjj$ (blue) transverse
momentum for a) two ($n=2$) and b) three ($n=3$) $t\bar t$ decay
jets. The dashed histograms display 
the fraction of the NLO $pp\to t\bar t\to\ell\nu +n$~jet events
($n=2,\,3$) for which $p_T(t\bar t)>100$~GeV. Two
of the jets are assumed to be $b$-tagged. The cuts imposed 
are listed in Eqs.~(\ref{eq:cuts1}) 
--~(\ref{eq:cuts4}). In addition an isolation cut (see
Eq.~(\ref{eq:cuts5})) with $R_{min}=0.4$ is imposed. $t\bar t$ decay
jets with a separation $\Delta R<0.4$ have been merged.} 
\vspace{-7mm}
\end{center}
\end{figure}

\begin{figure}[t!] 
\begin{center}
\includegraphics[width=14.2cm]{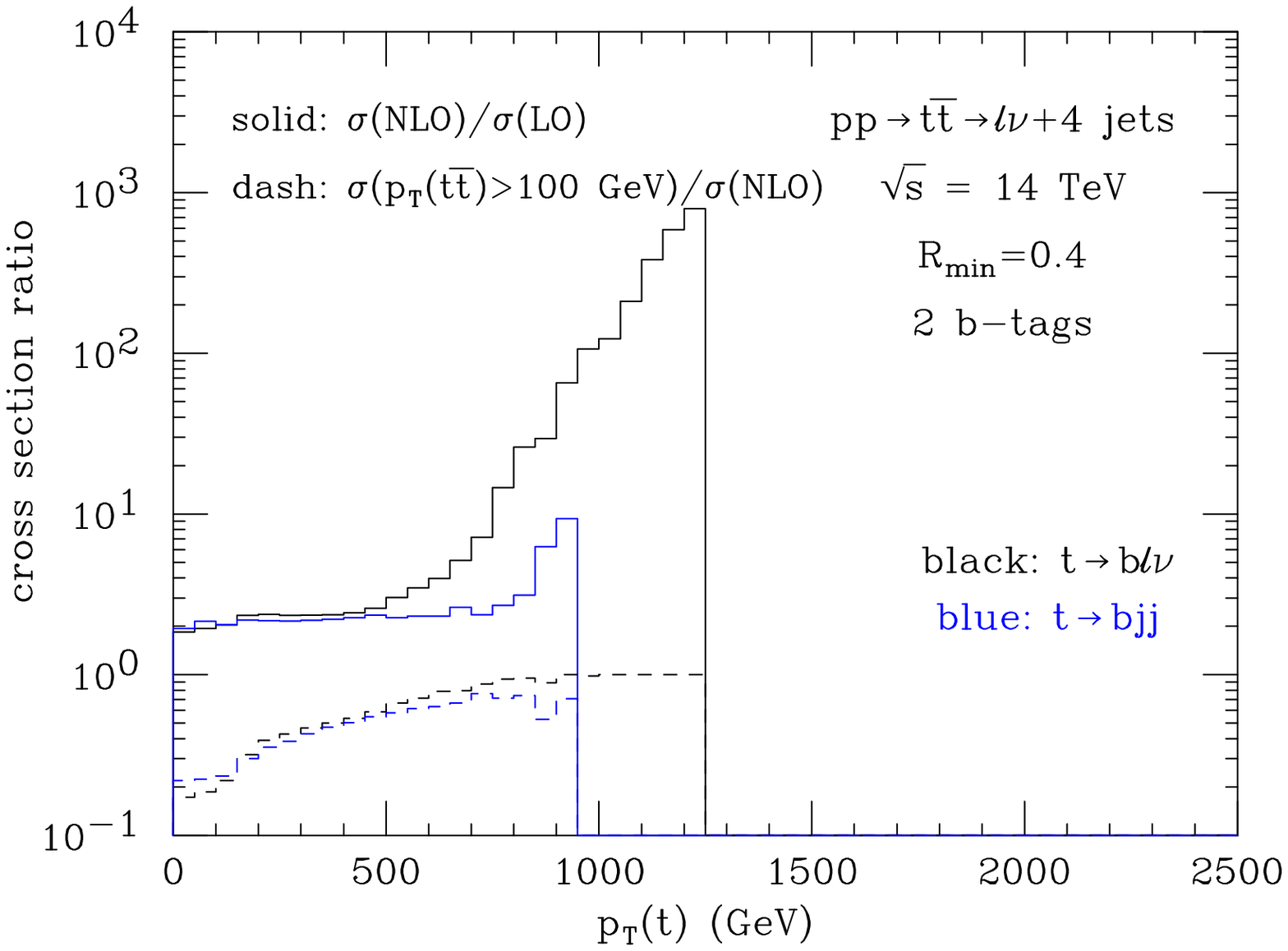} 
\vspace*{2mm}
\caption[]{\label{fig:six} 
The NLO to LO $pp\to t\bar t\to\ell\nu +4$~jets cross section ratio
(solid histograms) at the LHC as a function of the $t\to b\ell\nu$
(black) and $t\to bjj$ (blue) transverse momentum. The dashed histograms
display the fraction of the NLO $pp\to t\bar t\to\ell\nu +4$~jet events
for which $p_T(t\bar t)>100$~GeV. The number of jets here refers to
$t\bar t$ decay jets, and we assume that two of them are $b$-tagged.
The cuts imposed are listed in Eqs.~(\ref{eq:cuts1}) 
--~(\ref{eq:cuts4}). In addition an isolation cut (see
Eq.~(\ref{eq:cuts5})) with $R_{min}=0.4$ is imposed. $t\bar t$ decay
jets with a separation $\Delta R<0.4$ have been merged.} 
\vspace{-7mm}
\end{center}
\end{figure}

Figures~\ref{fig:five} and~\ref{fig:six} show the $k$-factor and the
fraction of events with $p_T(t\bar t)>100$~GeV (at NLO) as a function of the
$p_T$ of the semileptonically and the hadronically decaying top quark. 
As mentioned before, at LO the $p_T$ distributions of the two top quarks
in $t\bar t$ production are identical. This is no longer the case at
NLO. Figures~\ref{fig:five} and~\ref{fig:six} show that the differential
$k$-factors can be quite different for $t\to b\ell\nu$ and $t\to
bjj$. The difference is most pronounced in the 2~jet final state
(Fig.~\ref{fig:five}a). In the region $p_T(t\to bjj)>700$~GeV most
$t\to bjj$ jets merge into a single jet. This favors a kinematical
configuration where $p_T(t\to b\ell\nu)<p_T(t\to bjj)$, {\it ie.} where 
the QCD jet(s) and the semileptonically decaying top quark are in the
same hemisphere. As a result, the fraction of events with $p_T(t\bar
t)>100$~GeV for $p_T(t\to bjj)>700$~GeV is larger than that for $t\to
b\ell\nu$ in the same range. In turn, the fraction of events with $p_T(t\bar
t)>100$~GeV for $p_T(t\to bjj)<700$~GeV is smaller than that for $t\to
b\ell\nu$. Below a $p_T$ of about 250~GeV, a new effect comes into play.
If $p_T(t\bar 
t)>100$~GeV and $p_T(t\to b\ell\nu)$ is small, the hadronically decaying
top quark has to carry a transverse momentum of ${\cal O}(100$~GeV).
This makes it likely that one of the light
quark jets originating from $t\to bjj$ satisfies the jet acceptance
cuts. On the other hand, the top quark transverse momentum is not high
enough for jet merging. For small $p_T(t\to b\ell\nu)$, events with a
large transverse momentum of the $t\bar t$ system thus are very unlikely
(black dashed histogram). 

The evolution of the event fraction with $p_T(t\bar t)>100$~GeV as a
function of $p_T(t\to b\ell\nu)$ is directly reflected in the
corresponding 
$k$-factor. The preference for events with $p_T(t\to b\ell\nu)<p_T(t\to
bjj)$ for $p_T(t\to bjj)>700$~GeV leads to a very large $k$-factor for
$250~{\rm GeV}<p_T(t\to b\ell\nu)<700$~GeV. 
The suppression of events with high $t\bar t$ transverse momentum at
small $p_T(t\to b\ell\nu)$ then causes the $k$-factor to sharply drop for
$p_T(t\to b\ell\nu)<250$~GeV (solid black histogram). While the
$k$-factor varies significantly with $p_T(t\to b\ell\nu)$, it is
essentially uniform for $p_T(t\to bjj)<800$~GeV. 

The $k$-factor and the fraction of events with $p_T(t\bar t)>100$~GeV
for the 3~jet and the 2~jet final state show the same qualitative
behavior. In the 3~jet final state, an even larger fraction of events
has a $t\bar t$ transverse momentum larger than 100~GeV, especially for
very large top quark $p_T$ (dashed histograms in
Fig.~\ref{fig:five}b). At low $p_T(t\to b\ell\nu)$, the suppression of
events with high $p_T(t\bar t)$ is less pronounced than in the 2~jet final
state. Consequently, the variation of the $k$-factor with $p_T(t\to
b\ell\nu)$ is smaller than for $t\bar t\to\ell\nu+2$~jets. The
$k$-factor slowly increases with $p_T(t\to bjj)$ over the entire $p_T$
range. 

In the 4~jet final state, the $k$-factor is uniform for $p_T(t\to
b\ell\nu)<500$~GeV (see Fig.~\ref{fig:six}). For higher values, it
becomes very large. For very 
large $p_T(t\to b\ell\nu)$, the $t\bar t$ transverse momentum
of essentially all events exceeds 100~GeV (dashed black histogram in
Fig.~\ref{fig:six}). In contrast, the $k$-factor stays uniform up to
transverse momenta of about 800~GeV for the hadronically decaying top
quark. The extremely large $k$-factor for $p_T(t\to b\ell\nu)>500$~GeV
is a consequence of the separation cut which affects $t\to bjj$ much
more than $t\to b\ell\nu$ at large $p_T$. As a result, events which
contain one or more hard QCD jets in the hemisphere opposite to that of the
$t\to b\ell\nu$ decay products are kinematically favored. For the same
reason, $t\bar tW$ and $t\bar tZ$ production becomes important for large
values of $p_T(t\to b\ell\nu)$~\cite{Baur:2006sn}.

\begin{figure}[t!] 
\begin{center}
\includegraphics[width=14.2cm]{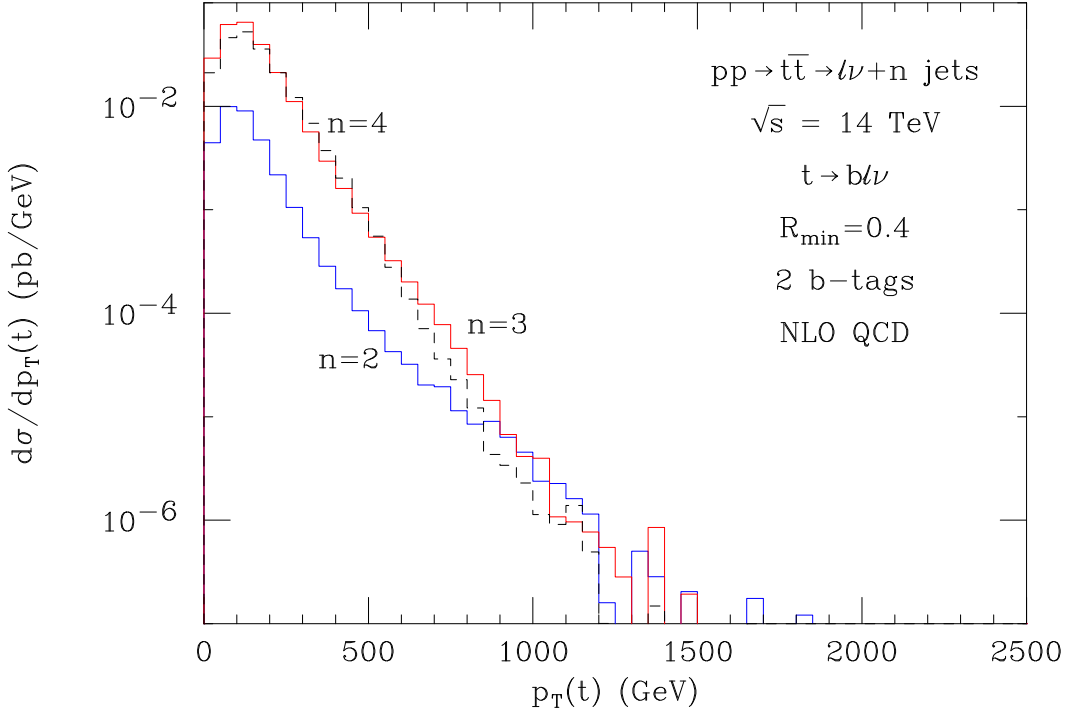} 
\vspace*{2mm}
\caption[]{\label{fig:seven} 
The NLO $pp\to t\bar t\to\ell\nu +n$~jets cross section 
at the LHC as a function of the $t\to b\ell\nu$ transverse momentum. $n$
is the number of $t\bar t$ decay jets in the event, and two
of these jets are assumed to be $b$-tagged. The solid blue and red
histograms show the cross sections for $n=2$ and $n=3$, respectively. The
dashed histogram represents the $n=4$ cross section. The cuts imposed 
are listed in Eqs.~(\ref{eq:cuts1}) 
--~(\ref{eq:cuts4}). In addition an isolation cut (see
Eq.~(\ref{eq:cuts5})) with $R_{min}=0.4$ is imposed. $t\bar t$ decay
jets with a separation $\Delta R<0.4$ have been merged.} 
\vspace{-7mm}
\end{center}
\end{figure}

NLO QCD corrections mostly change the normalization of the $m(t\bar t)$
and $p_T(t\to bjj)$ distributions. For these distributions, the
cross section hierarchy for 2, 3 and 4~$t\bar t$ decay jets shown in
Fig.~\ref{fig:two}  remains unchanged. With the extremely large
$k$-factor for the 4~jet final state, this is not obvious for the
$p_T(t\to b\ell\nu)$ distribution. Figure~\ref{fig:seven} shows that for
$p_T(t\to b\ell\nu)>600$~GeV ($p_T(t\to b\ell\nu)>900$~GeV) the 
cross section for the final state with 3~(2) jets from
$t\bar t$ decays still exceeds that of
the channel with 4~$t\bar t$ decay jets when NLO QCD corrections are
taken into account. 

In phase space regions where most top pair events have a large
transverse momentum, the NLO $t\bar t$ cross section is dominated by the
tree level process $pp\to t\bar tj$. As a result, the NLO cross section
depends significantly on the choice of the factorization and
renormalization scale in these regions.

So far we have classified events by
the number of $t\bar t$ decay jets. The number of extra jets from QCD 
radiation in $t \bar t$ production 
was not specified. At large top quark transverse momentum, we found that
most events have $p_T(t\bar t)>100$~GeV and thus have one or more extra
hard jets. These extra jets introduce a combinatorial
background. Considering final states with a fixed jet multiplicity of 2,
3 or 4~jets, and requiring that the invariant mass of the jet(s) excluding
the $b$-jet with the smaller separation from the charged lepton is
consistent with $m_t$ is expected to suppress hard extra QCD
jets in the event, and thus the combinatorial background. As we shall
demonstrate in Secs.~\ref{sec:three} and~\ref{sec:four}, such a cut will
also be helpful in reducing the background to an acceptable level. If
no hard QCD jets are produced, the $t\bar t$ transverse momentum will be
small. A rough estimate of the QCD corrections to 
$pp\to t\bar t\to\ell\nu +n$~jets with no hard 
extra QCD
jets can be obtained from the $k$-factor for events with $p_T(t\bar
t)<100$~GeV, $k_{<100}$. $k_{<100}$ can be calculated from the
inclusive $k$-factor, $k_{incl.}$, and the ratio of the NLO cross
section with $p_T(t\bar t)>100$~GeV and the inclusive NLO rate, $r$,
\begin{equation}
k_{<100}=k_{incl.}(1-r),
\end{equation}
which are both shown in Figs.~\ref{fig:four} --~\ref{fig:six}. The
$k_{<100}$ distributions as a function of $m(t\bar t)$, $p_T(t\to
b\ell\nu)$, and $p_T(t\to bjj)$ are qualitatively very similar.
$k_{<100}$ is found to decrease smoothly from $k_{<100}\approx 1.5-1.7$
at low $m(t\bar t)$ and $p_T(t)$ to $k_{<100}\approx 0.7-0.8$ at large
values. QCD corrections for top quark pairs with small transverse
momentum thus are smaller than in the inclusive case, and result
in somewhat steeper falling $m(t\bar t)$ and $p_T(t)$
distributions than at LO.

Quantitative results of course depend on the jet $p_T$ threshold
considered; for a lower (higher) threshold of $p_T(t\bar t)$, a smaller
(larger) $k$-factor is found. More detailed simulations, which are
beyond the scope of our 
paper, are required to develop a better understanding how QCD
corrections affect the $m(t\bar t)$ and $p_T(t)$ distributions for fixed
jet multiplicities when an invariant mass cut on one or more jets is
imposed. 

\subsection{$b$-tagging for very energetic top quarks}
\label{sec:twoc}

So far, in our analysis, we have assumed that both $b$-quarks in
$pp\to t\bar t\to\ell\nu b\bar bq\bar q'$ are tagged with an efficiency
of $\epsilon_b=60\%$ each. However, for very energetic top quarks, the
$b$-tagging efficiency is expected to degrade~\cite{Lillie:2007yh}. This
is easy to understand. The energy of the $b$-quark in $t\to b\ell\nu$ or
$t\to bjj$ is on average about 1/3 of that of the parent top quark. The
higher the energy of the $b$-quark, the more collimated the $b$ decay
products are. Due to the finite angular resolution of the LHC detectors,
this will increase the uncertainty in the position of the
reconstructed secondary vertex, and thus decrease the tagging
efficiency. In hadronic top decays where two or more jets merge, the
overlapping of the $b$-jet with one or several light quark jets may
additionally complicate the reconstruction of the secondary vertex which
is expected to result in a further decrease in $\epsilon_b$. The
increased decay length of very energetic $b$-quarks, which makes it
easier to tag $b$-quarks, is not expected to compensate these effects. 

Although detailed simulations of $b$-tagging for very energetic top
quarks do not exist yet, preliminary studies~\cite{atlaslh,atlasglu}
indicate that $\epsilon_b$ may decrease by a factor $2-3$ in the TeV
region. Simultaneously, the probability for misidentifying a light
quark, gluon or $c$-jet as a $b$-jet may increase by up to a factor~3.

A decrease of $\epsilon_b$ by a factor $2-3$ in the high energy regime
results in a reduction of the observable $t\bar t$
cross section by up to a factor~10. However, the efficiency for tagging
only one $b$-quark in a $t\bar t$ event,
\begin{equation}
\epsilon(1~{\rm tag})=2\epsilon_b(1-\epsilon_b),
\end{equation}
is much less sensitive to $\epsilon_b$ than $\epsilon(2~{\rm
tags})=\epsilon^2_b$. For $\epsilon_b=0.2-0.6$, $\epsilon(1~{\rm
tag})=0.32-0.48$, {\it ie.} it varies by less than a factor of two. For
small $\epsilon_b$, the cross section of the lepton+jets final state
with one $b$-tag is much larger than that for two $b$-tags. For example,
for $\epsilon_b=0.2$, $\epsilon(1~{\rm tag})/\epsilon(2~{\rm tags})=8$.
This, and the relative stability of the one tag cross section to
variations of $\epsilon(b)$, make
the lepton+jets final state with one $b$-tag an attractive channel in
the search for resonances in the $t\bar t$ channel. 

The increase in rate in the lepton+jets channel with one $b$-tag comes
at the price of a potentially much larger background. The background for
both one and two $b$-tags in the lepton+jets channel will be examined in
detail in Sec.~\ref{sec:three}.

\subsection{The di-lepton+jets and all-hadronic final states}
\label{sec:twod}

Our discussion, so far, has been focused on the lepton+jets final
state. In this 
section we investigate whether the $t\bar t$ di-lepton and all-hadronic
final states can significantly increase the range in $m(t\bar t)$ and/or
$p_T(t)$ which can be accessed at the LHC. 

In the di-lepton channel, $pp\to t\bar
t\to\ell^\pm\nu_\ell{\ell'}^\mp\nu_{\ell'}b\bar b$
($\ell,\,\ell'=e,\,\mu$), one requires two 
isolated leptons with opposite charge, two jets with at least one
$b$-tag, and a substantial amount of $p\llap/_T$. The main disadvantages
of the di-lepton final state are its small branching ratio of $\approx
4.7\%$, and the two neutrinos in the final state which make it
impossible to reconstruct the $t\bar t$ invariant mass or the $p_T$ of
the individual top quarks. However, the $\ell\ell'p\llap/_T+2$~jets
final state is much less sensitive to the isolation cut at high energies
than $pp\to t\bar t\to\ell\nu+4$~jets which makes it worthwhile to
investigate. The main background in this channel comes from $Z+$jets and
$WWb\bar b$ production~\cite{cmstdr}.

\begin{figure}[t!] 
\begin{center}
\includegraphics[width=14.2cm]{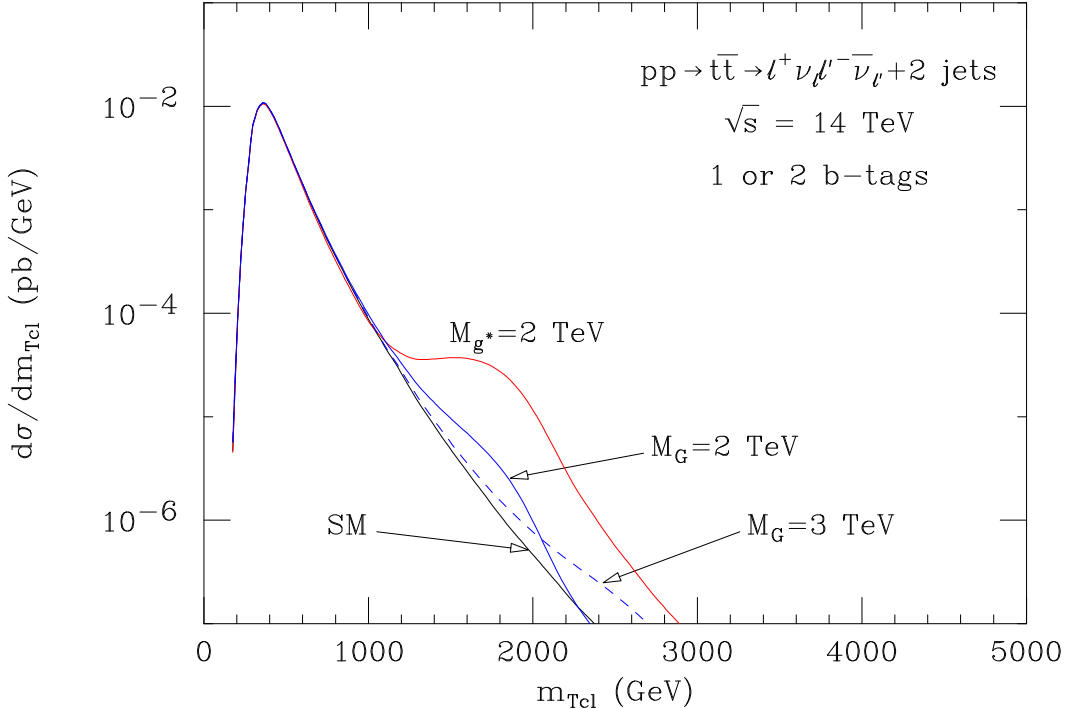} 
\vspace*{2mm}
\caption[]{\label{fig:eight} 
The LO $pp\to t\bar t\to\ell^\pm\nu_\ell{\ell'}^\mp\nu_{\ell'}b\bar b$
differential cross section at the LHC as a function of the cluster
transverse mass (see Eq.~(\ref{eq:mtcl})). At least one of the
$b$-quarks is assumed to be tagged. Shown are the cross sections 
in the SM (black lines) and for two types of KK gluon
excitations (see Sec.~\ref{sec:twoa}). The cuts imposed are listed
in Eqs.~(\ref{eq:cuts1}) --~(\ref{eq:cuts4}). In addition an isolation
cut (see Eq.~(\ref{eq:cuts5})) with $R_{min}=0.4$ is imposed.} 
\vspace{-7mm}
\end{center}
\end{figure}

Since the $t\bar t$ invariant mass cannot be reconstructed in the
di-lepton final state, one has to use the $\ell\ell'b\bar b$ cluster
transverse mass,
\begin{equation}
\label{eq:mtcl}
m^2_{Tcl}=\left(\sqrt{p_T^2(\ell\ell'b\bar b)+m^2(\ell\ell'b\bar b)}
+ p\llap/_T\right)^2-\left(\vec{p}_T(\ell\ell'b\bar b)+
\vec{p\llap/}_T\right)^2\, ,
\end{equation}
where $p_T(\ell\ell'b\bar b)$ and $m(\ell\ell'b\bar b)$ are the transverse
momentum and invariant mass of the $\ell\ell'b\bar b$ system,
respectively, to search for resonances. The cluster transverse mass
distribution in the SM and for the KK gluon states discussed in
Sec.~\ref{sec:twoa} is shown in Fig.~\ref{fig:eight}.
We require at least one $b$-quark to be tagged and
impose the same cuts as in Secs.~\ref{sec:twoa} and~\ref{sec:twob}. The
cluster transverse mass distribution is seen to fall much more rapidly
than that of the $t\bar t$ invariant mass. The di-lepton final state
therefore will not be competitive with the lepton+jets final state when
searching for $t\bar t$ resonances, and we will not discuss it further
in this paper.

The all-hadronic final state, $t\bar t\to bq_1\bar q_2\bar bq_3\bar q_4$
has the largest branching ratio ($\approx 46\%$) but also the largest
background. In order to reduce the QCD multi-jet background to an
acceptable level, two $b$-tags have to be required. Imposing a standard
$R_{min}=0.4$ isolation cut on the $b$-jets and the four light quark
jets, the main background originates from QCD $b\bar b+4$~jet
production, which is approximately one order of magnitude larger than
the signal~\cite{atlastdr,cmstdr}. 

As in the lepton+jets mode, the isolation cut strongly reduces the
observable cross section in the all-hadronic final state for very
energetic top quarks. In this region of phase space, some or all of the
jets originating from top quark decay may merge, {\it ie.} the LO $t\bar t$
signal is spread out over the $b\bar b+n$~jets final states with $0\leq
n\leq 4$. The QCD background for $0\leq n\leq 3$ is expected to be
of the same size or larger than that in the $b\bar b+4$~jet channel. 

The large background makes it difficult to utilize the all-hadronic
final state 
in a search for $t\bar t$ resonances. In the following, we therefore
concentrate on the lepton+jets final state with one or two $b$-tags.

\section{Background Calculations for the lepton+jets Final State}
\label{sec:three}

In Sec.~\ref{sec:two}, we have shown that extending the search criteria
for the lepton+jets final state to include topologies with less than
4~jets and/or one $b$-tag may considerably increase
the number of $t\bar t$ candidate events. These final states,
however, will be useful in a search for $t\bar t$ resonances only if the
backgrounds are sufficiently small. It is well
known~\cite{atlastdr,cmstdr,Hubaut:2005er} that the background is indeed
small in the 4~jet case with two $b$-tags. Here we calculate the
backgrounds contributing to the 2~jet 
and 3~jet final states and compare it with that obtained in the 4~jet
case. We consider final states with one or two $b$-tags. Backgrounds
where one or two light quark, gluon or $c$-quark jets are misidentified
as a $b$-jet are included in our calculation. Numerical results are
presented for $\epsilon_b=0.6$ and 
misidentification probabilities of $P_{q,g\to b}=P_{j\to b}=1/100$
($q=u,\,d\,s$) and $P_{c\to b}=1/10$~\cite{atlastdr,cmstdr}. Since these
numbers may well be considerably higher for very energetic top
quarks~\cite{atlaslh,atlasglu} we comment wherever appropriate on how
our results change if $P_{j\to b}$ is increased by a factor~3 and the
$b$-tagging efficiency, $\epsilon_b$, is decreased by a factor~3.

The main background processes contributing to the $\ell\nu+n$~jet final
states with $n=2,\,3,\,4$ are $Wb\bar b+m$~jets, $(Wb+W\bar b)j+m$~jets,
and $Wjj+m$~jets production, $(t\bar b+\bar tb)+m$~jets, $(t+\bar
t)j+m$~jets production with $t\to b\ell\nu$, and $Wbt$, $Wt$ and $Wjt$
production with $t\to bjj$. For each process,
$m=0,\,1,\,2$, and $j$ represents a light quark or gluon jet, or a
$c$-jet. $Wt$ production only contributes to the 2~jet and 3~jet final
states. The $(Wb+W\bar b)j+m$~jets ($(t+\bar t)j+m$~jets) background 
originates from $Wb\bar b+m$~jets ($(t\bar b+\bar
tb)+m$~jets) production where one of the $b$-quarks is not detected. We
calculate these processes in the $b$-quark structure function
approximation. We have verified that, for $m=0$, the differential cross
sections for $pp\to Wbj$ ($(t+\bar
t)j$) and $pp\to Wb\bar bj$ ($(t\bar b+\bar tb)j$) where one $b$-jet is not 
detected are very similar. For the 2~jet final state ($m=0$), the NLO QCD
corrections for all background processes except $Wbt$ and $Wjt$
production are
known~\cite{Ellis:1998fv,Campbell:2002tg,Stelzer:1997ns,smith,Sullivan:2004ie,Campbell:2006cu,Campbell:2005bb}.
The background processes relevant for the 3~jet and 4~jet final states,
however, are only known at LO. We therefore calculate all background cross
sections consistently at LO, and comment wherever appropriate how NLO QCD
corrections modify our results. To calculate $pp\to Wb\bar b+m$~jets and
$pp\to Wjj+m$~jets we use {\tt ALPGEN}~\cite{Mangano:2002ea}. All other
background processes are calculated using {\tt
MadEvent}~\cite{Maltoni:2002qb}. 

$b\bar b+m$~jets production where one $b$-quark decays semileptonically
also contributes to the background. 
Once a lepton isolation cut has been imposed, this background is known to
be small for standard lepton+jets cuts~\cite{Hubaut:2005er}. For $b\bar
b+m$~jets events to mimic a $t\bar t$ production with very energetic top
quarks, the $b$-quarks also have to be very energetic. This will make
the lepton isolation cut even more efficient. We therefore ignore the
$b\bar b+m$~jets background here. 

$Wjj+m$~jets production in {\tt ALPGEN}
includes $c$-jets in the final state. Since $P_{c\to b}\approx 10
\cdot P_{q,g\to b}$, this underestimates the background from $W+$~charm
production. However, the cross section of $W+$~charm final states is
only a tiny fraction of the full $Wjj+m$~jets rate, resulting in an
error which is much smaller than the uncertainty on the background from
other sources. One can also estimate the $W+$~charm cross section
from that of $pp\to Wb\bar b+m$~jets and $pp\to (Wb+W\bar
b)+m$~jets. For the phase space cuts imposed, quark mass effects are
irrelevant. The $(Wc+W\bar c)j+m$~jets ($Wc\bar c+m$~jets) cross section
thus is about a factor 10 (100) smaller than the $(Wb+W\bar b)j+m$~jets
($Wb\bar b+m$~jets) rate. 

In the following we impose the standard acceptance cuts of
Eqs.~(\ref{eq:cuts1}) --~(\ref{eq:cuts5}) with $R_{min}=0.4$ and
reconstruct the $t\bar t$ invariant mass using the
procedure described in Sec.~\ref{sec:twoa}. The reconstruction of
$p_T(t\to b\ell\nu)$ depends on the number of $b$-tags and is discussed
in more detail below. For the background
processes, the $m(t\bar t)$ distribution is replaced by the
reconstructed $Wb\bar b+m$~jets and $Wbj+m$~jets invariant mass
distribution. The
renormalization and factorization scales of background processes
involving top quarks are set to $m_t$; for all other background
processes we choose the $W$ mass. Since our calculations are performed
at tree level, the cross section of many background processes exhibits a
considerable scale dependence. However, uncertainties on the current
$b$-tagging efficiencies and the light jet mistag probability at high
energies introduce an even larger uncertainty.

\subsection{Background for lepton+jets events with two $b$-tags}

\begin{figure}[th!] 
\begin{center}
\includegraphics[width=13.8cm]{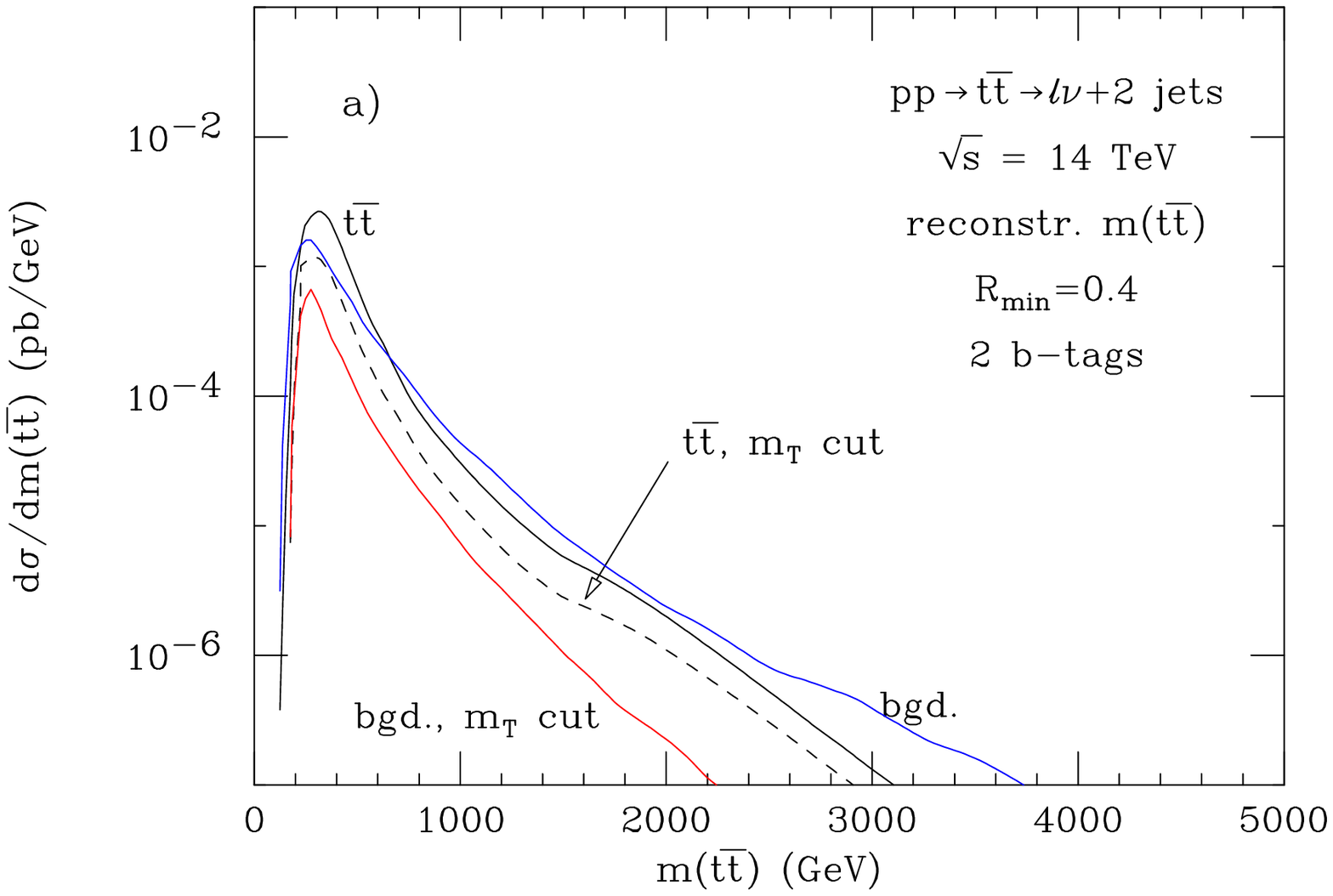} \\[3mm]
\includegraphics[width=13.8cm]{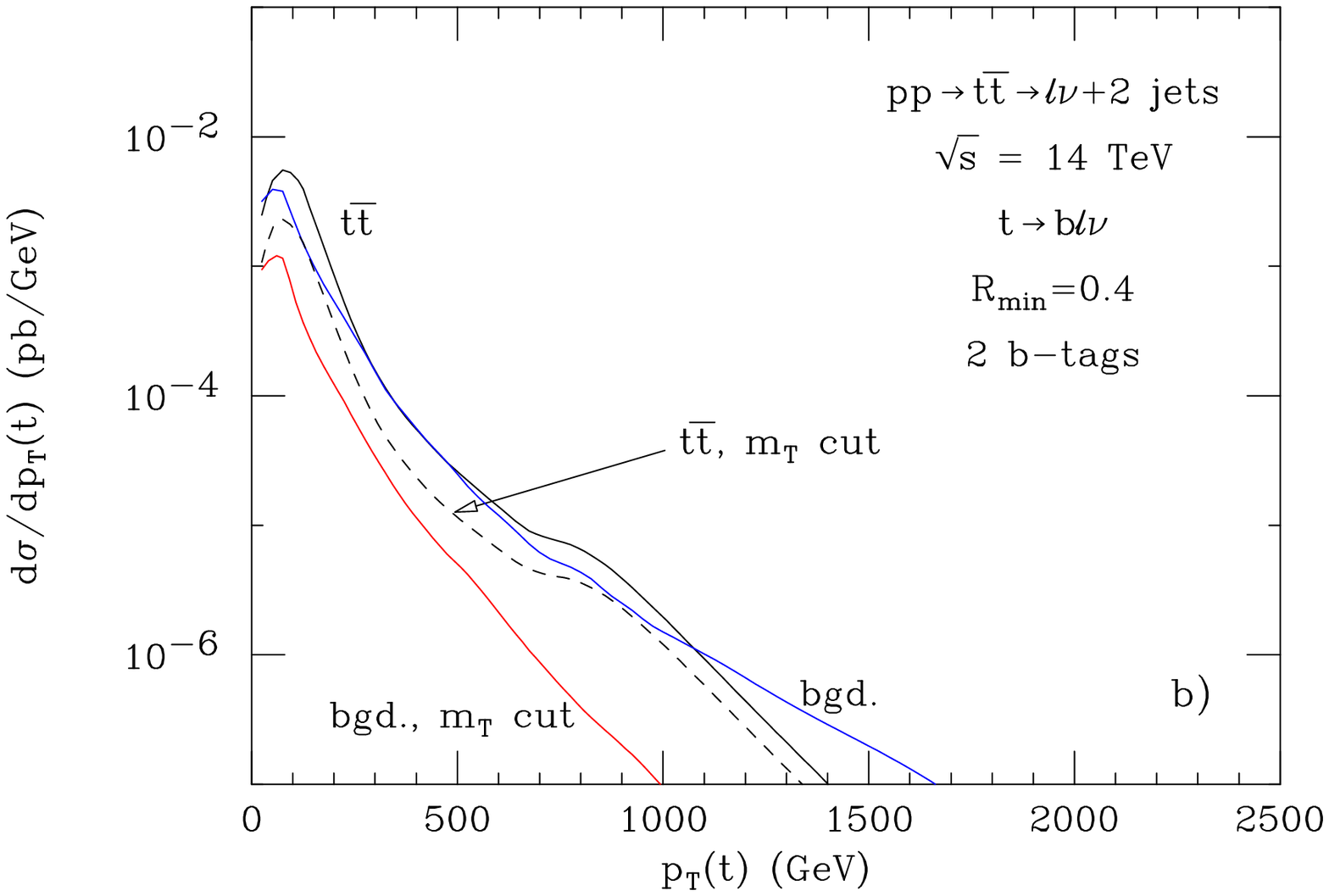}
\vspace*{2mm}
\caption[]{\label{fig:nine} 
The LO differential cross section of the
SM $t\bar t\to\ell\nu+2$~jets signal and the combined background as a
function of a) the reconstructed $t\bar t$ invariant mass and b) the
$t\to b\ell\nu$ transverse momentum at the LHC. 
The two jets are assumed to be $b$-tagged. The black and blue curves
show the signal and background, respectively, imposing standard cuts
(Eqs.~(\ref{eq:cuts1}) --~(\ref{eq:cuts5})) with $R_{min}=0.4$). The
dashed and red curves show signal and background if in addition a
cluster transverse mass cut is imposed on the $b\ell\nu$ system (see
Eq.~(\ref{eq:mtcl1})).} 
\vspace{-7mm}
\end{center}
\end{figure}

The differential cross section of the
SM $t\bar t\to\ell\nu+2$~jets signal and the combined background
from the processes discussed above as a function of the reconstructed
$m(t\bar t)$ and $p_T(t\to b\ell\nu)$ is shown in Fig.~\ref{fig:nine}.
To reconstruct $p_T(t\to b\ell\nu)$ for signal and background
$\ell\nu+n$~jet final states with two $b$-tags, we use the method
discussed in Sec.~\ref{sec:twoa}. 
Imposing the standard cuts of Eqs.~(\ref{eq:cuts1}) --~(\ref{eq:cuts5})
with $R_{min}=0.4$, signal and background are seen to be about
equal. Only for $m(t\bar t)> 2.5$~TeV and $p_T(t)>1.2$~TeV does the
background dominate. 
The main background source is $pp\to Wb\bar b$, except at
very high $m(t\bar t)$ and $p_T(t)$ where $Wbt$ production dominates.

The signal to background ratio can be significantly improved by imposing
a cut
\begin{equation}
\label{eq:mt}
|m_T(b_{min}\ell)-m_t|< 20~{\rm GeV}
\end{equation}
on the cluster transverse mass, $m_T$, of the $b\ell\nu$ system which is
defined by
\begin{equation}
\label{eq:mtcl1}
m^2_{T}(b_{min}\ell)=\left(\sqrt{p_T^2(b_{min}\ell)+m^2(b_{min}\ell)}
+ p\llap/_T\right)^2-\left(\vec{p}_T(b_{min}\ell)+
\vec{p\llap/}_T\right)^2\, ,
\end{equation}
where $p_T(b_{min}\ell)$ and $m(b_{min}\ell)$ are the transverse
momentum and invariant mass of the $b_{min}\ell$ system, respectively,
and $b_{min}$ is the 
$b$- or $\bar b$-quark which has the smaller separation from the charged
lepton. $m_T$ sharply peaks at the top mass. The cluster transverse mass
reduces the signal by about a factor~2 (1.5) at small (large) invariant
masses and $p_T$'s (dashed curves). The background, on the other hand,
decreases by a factor $5-10$. At large $m(t\bar t)$ and $p_T(t\to
b\ell\nu)$, $(t\bar b+\bar tb)$ production is the dominant contribution
to the background after the $m_T$ cut has been imposed. At low energies,
$pp\to Wb\bar b$ is still the largest background source.

The $t\bar t$ signal in the $\ell\nu+2$~jets final state with two
$b$-tags necessarily requires a small transverse momentum of the $t\bar
t$ system. As discussed in Sec.~\ref{sec:twob}, the NLO $m(t\bar t)$
and $p_T(t)$ distributions fall somewhat faster than those at LO if the
transverse momentum of the $t\bar t$ system is small. It is
important to know whether the corresponding distributions of the dominant
background processes, $pp\to Wb\bar b$ and $pp\to (t\bar b+\bar tb)$,
show the same behavior, or whether QCD corrections worsen the signal to
background ratio. Calculating the NLO corrections to these
processes using the program {\tt MCFM}~\cite{mcfm} and imposing a veto
on additional hard jets, we find that they 
have a similar effect on the signal and the background
distributions. QCD corrections thus will not change the signal to
background ratio significantly.

\begin{figure}[th!] 
\begin{center}
\includegraphics[width=13.5cm]{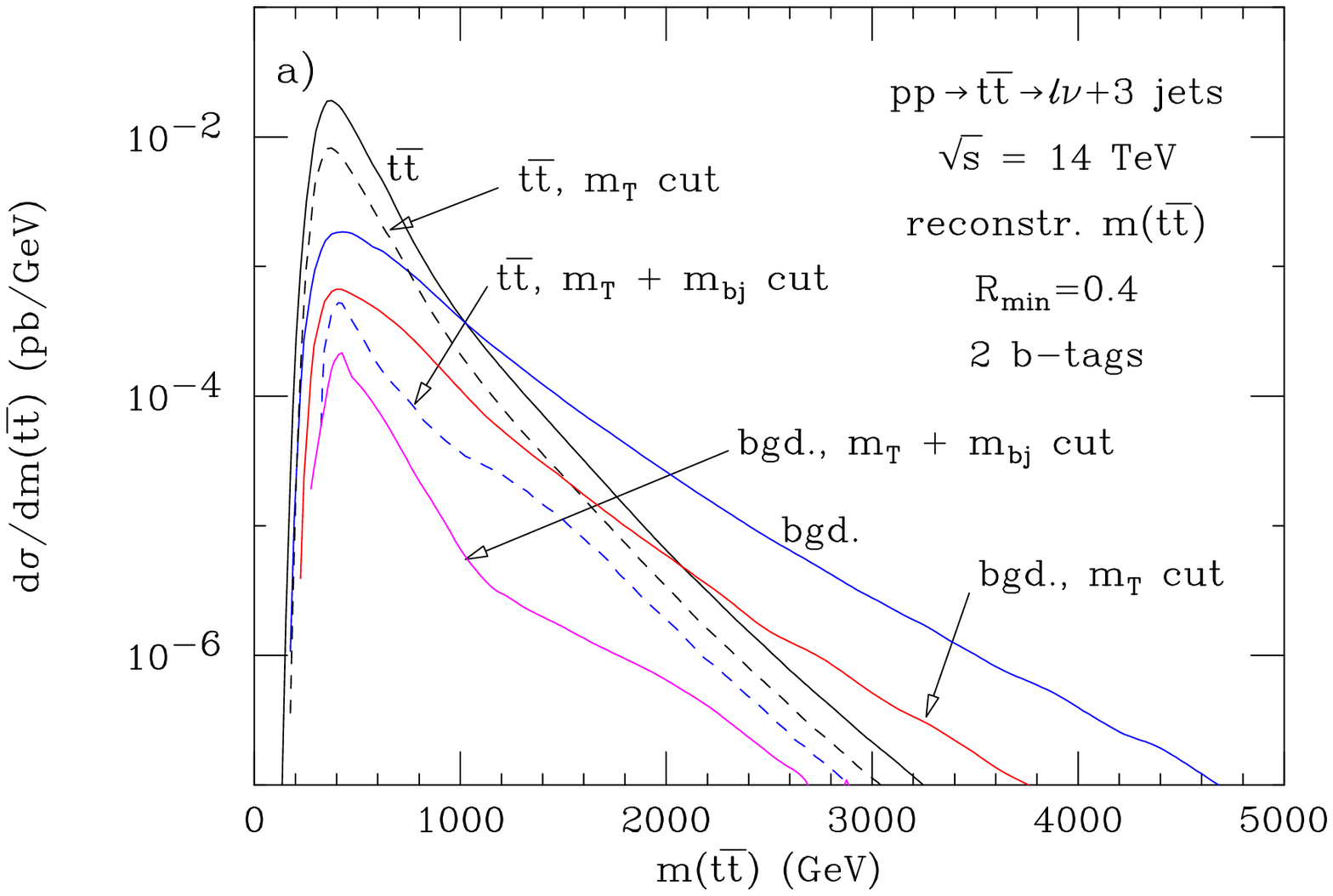} \\[3mm]
\includegraphics[width=13.5cm]{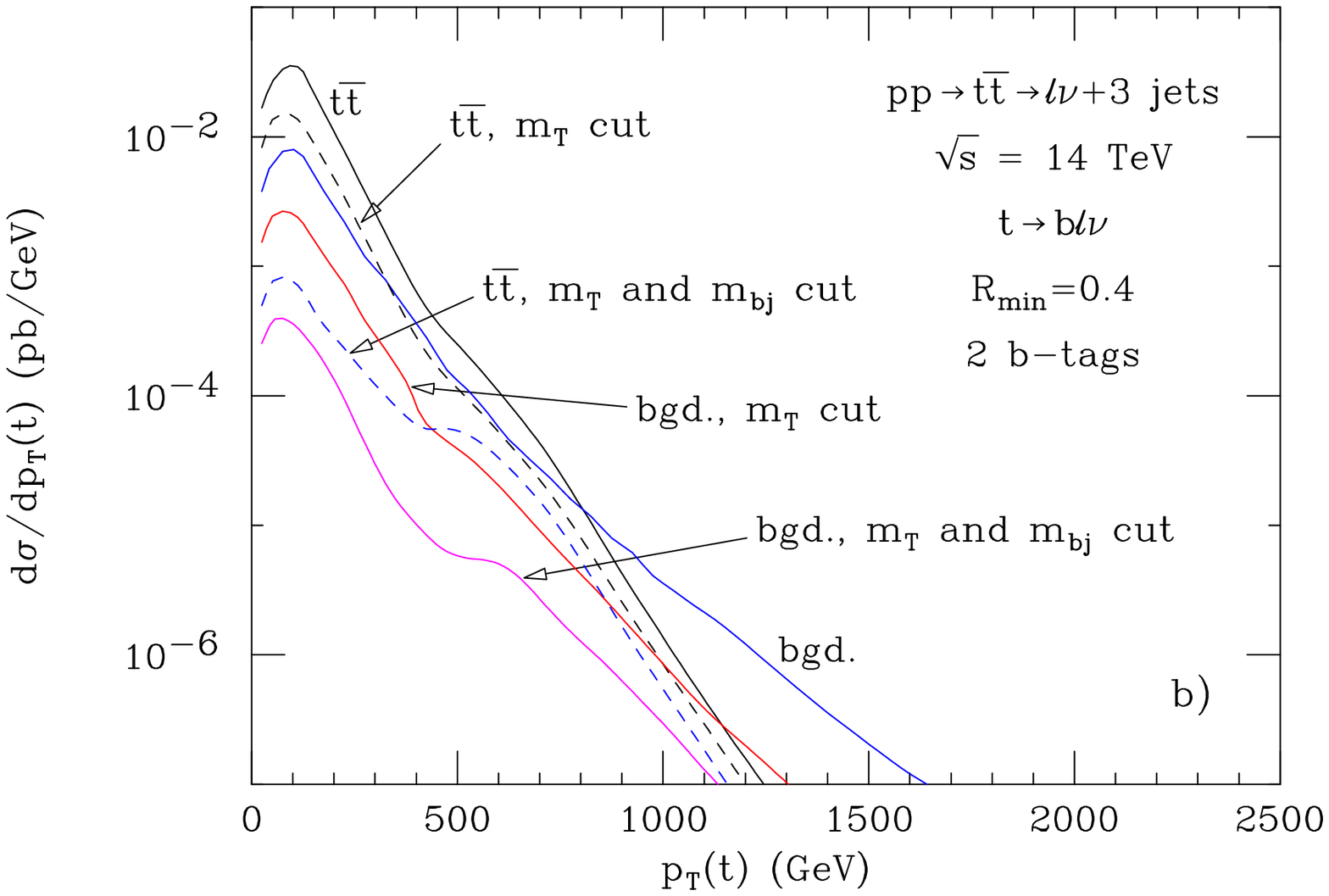}
\vspace*{2mm}
\caption[]{\label{fig:ten} 
The LO differential cross section of the
SM $t\bar t\to\ell\nu+3$~jets signal and the combined background as a
function of a) the reconstructed $t\bar t$ invariant mass and b) the
$t\to b\ell\nu$ transverse momentum at the LHC. 
Two jets are assumed to be $b$-tagged. The black and blue curves
show the signal and background, respectively, imposing standard cuts
(Eqs.~(\ref{eq:cuts1}) --~(\ref{eq:cuts5})) with $R_{min}=0.4$). The
dashed and red curves show signal and background if in addition a
cluster transverse mass cut is imposed on the $b\ell\nu$ system (see
Eq.~(\ref{eq:mtcl1})). The blue dashed and magenta curves, finally,
show signal and background if in addition the invariant mass cut of
Eq.~(\ref{eq:invm}) is imposed.} 
\vspace{-7mm}
\end{center}
\end{figure}

In Fig.~\ref{fig:ten}, we show the $t\bar t$ signal and the combined
background for the 3~jet final state.
Imposing the standard cuts of Eqs.~(\ref{eq:cuts1}) --~(\ref{eq:cuts5})
only, the background is small at low values of $m(t\bar t)$ and
$p_T(t\to b\ell\nu)$ but dominates over the signal in the TeV
region. Imposing a $m_T$ cut (see Eq.~(\ref{eq:mt})) improves the
situation. However, the background is still larger than the signal for
$m(t\bar t)>1.8$~TeV and $p_T(t)>1$~TeV. To 
further improve the signal to background ratio one can impose an
invariant mass cut on the $b_{max}j$ system,
\begin{equation}
\label{eq:invm}
|m(b_{max}j)-m_t|<20~{\rm GeV},
\end{equation}
where $j$ is the non-tagged jet and $b_{max}$ is the $b$-quark with the
larger separation from the charged lepton. The $m(b_{max}j)$ cut
suppresses the background by an additional factor $3-10$ for
$m(t\bar t)>1.8$~TeV (magenta line), while it has a much smaller effect
on the signal in this region (dashed blue curve). Note that the
$m(b_{max}j)$ cut does reduce the signal by up to a factor~10 at small
$t\bar t$ invariant masses. In this region most $t\bar
t\to\ell\nu+3$~jets events are the result of one jet not
satisfying the $p_T$ and pseudo-rapidity cuts, and not of the merging of
$t\to bjj$ jets. If a jet fails the acceptance cuts, the $b_{max}j$
system usually will not be in the vicinity of the top quark mass.

Once a $m(b_{max}j)$ cut has been imposed, the background is smaller
than the signal over the entire $p_T$ and invariant mass range of
interest. The main background sources in this case are $(t\bar b+\bar
tb)j$ and $Wbt$ production. Without the $m(b_{max}j)$ cut, 
the main contributions to the background in the 3~jet final state
originate from $Wb\bar bj$ and $(t\bar b+\bar tb)j$ production.

\begin{figure}[th!] 
\begin{center}
\includegraphics[width=13.5cm]{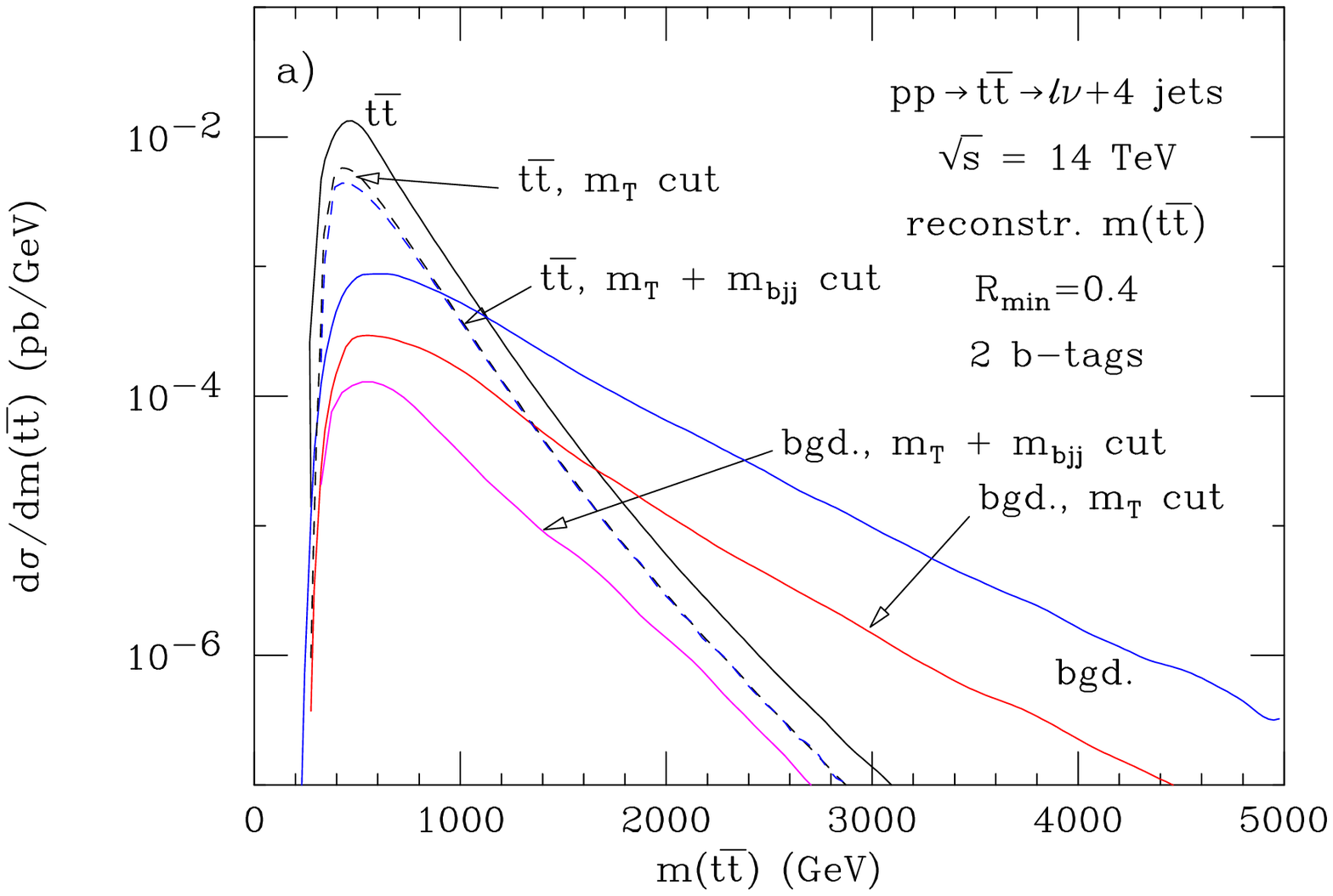} \\[3mm]
\includegraphics[width=13.5cm]{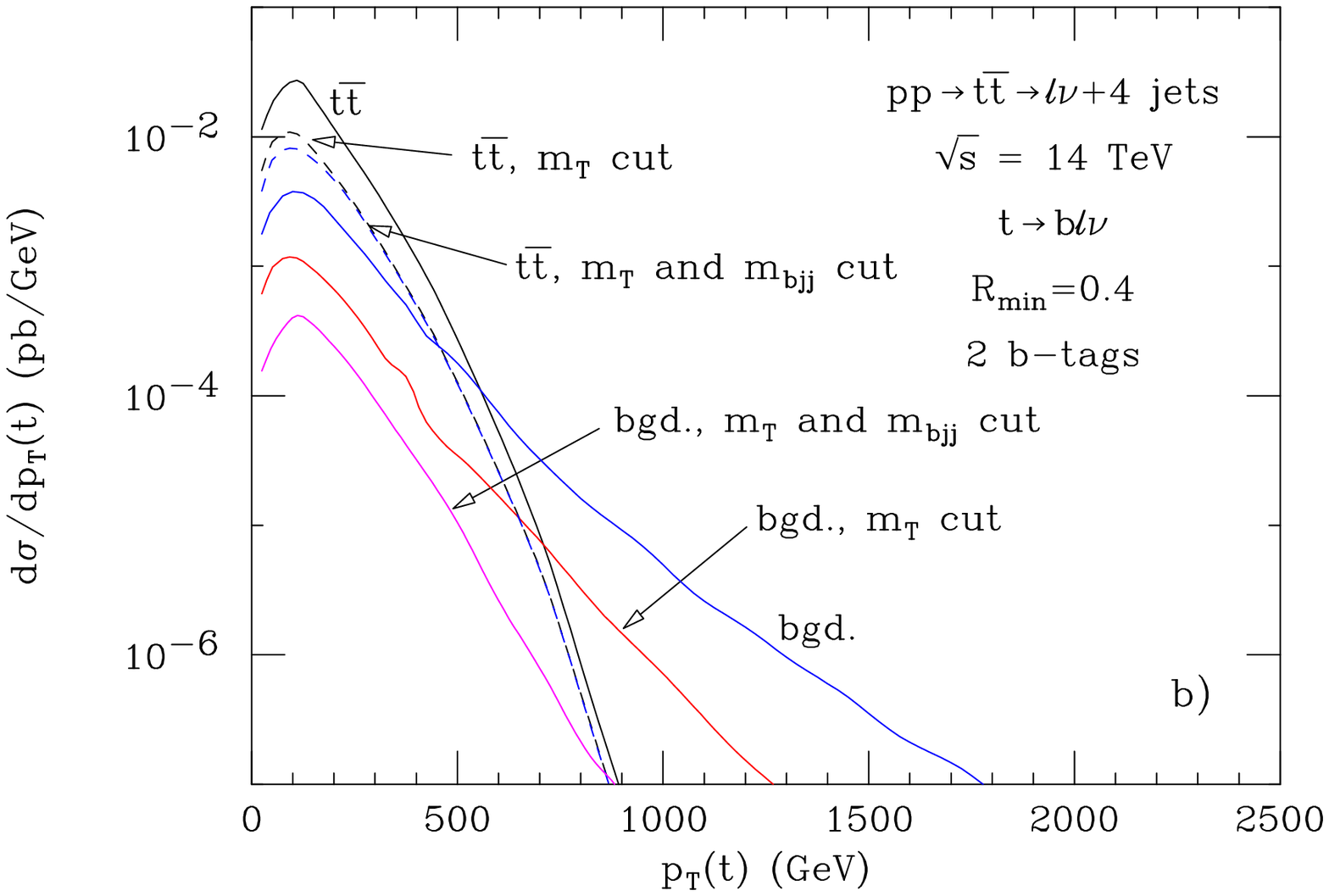}
\vspace*{2mm}
\caption[]{\label{fig:eleven} 
The LO differential cross section of the
SM $t\bar t\to\ell\nu+4$~jets signal and the combined background as a
function of a) the reconstructed $t\bar t$ invariant mass and b) the
$t\to b\ell\nu$ transverse momentum at the LHC. 
Two jets are assumed to be $b$-tagged. The black and blue curves
show the signal and background, respectively, imposing standard cuts
(Eqs.~(\ref{eq:cuts1}) --~(\ref{eq:cuts5})) with $R_{min}=0.4$). The
dashed and red curves show signal and background if in addition a
cluster transverse mass cut is imposed on the $b\ell\nu$ system (see
Eq.~(\ref{eq:mtcl1})). The blue dashed and magenta curves, finally,
show signal and background if in addition the invariant mass cut of
Eq.~(\ref{eq:invm1}) is imposed.} 
\vspace{-7mm}
\end{center}
\end{figure}

For completeness, we show the $t\bar t$ invariant mass and top quark
$p_T$ distribution for the 4~jet final state in Fig.~\ref{fig:eleven}.
Once a $m_T$ and a 
\begin{equation}
\label{eq:invm1}
|m(b_{max}jj)-m_t|<20~{\rm GeV},
\end{equation}
cut have been imposed, the background is below the signal for all top quark
transverse momenta and $t\bar t$ invariant masses of interest. 
The $m(b_{max}jj)$ cut has essentially no
effect on the signal for $m(t\bar t)>600$~GeV and $p_T(t)>200$~GeV.
The main background source in the 4~jet channel with (without) a
$m(b_{max}jj)$ cut is $Wbt$ (single resonant $(t\bar b+\bar tb)jj$ and
$Wb\bar bjj$) production. Note that we have not imposed a cut on the
invariant mass of the two light quark jets, $m(jj)$. It peaks near $M_W$
for both the signal and the $Wbt$, $t\to bjj$, background. Once a
$m(b_{max}jj)$ cut has been imposed, a $m(jj)$ cut thus will have little
effect on the signal to background ratio.

As we have mentioned before, the $b$-tagging efficiency at high
invariant masses and transverse momenta may be up to a factor~3 smaller,
and the misidentification probability for light quark and gluon jets may
be up to factor~3 larger, than at small $m(t\bar t)$ and $p_T(t)$. If
this is indeed the case, $Wjj$ production becomes the largest
background source in $pp\to t\bar t\to\ell\nu +2$~jets, and exceeds the
signal by about a factor~2 in the $m(t\bar t)$ distribution, even if a
$m_T$ cut is imposed. To further 
improve the signal to background ratio in this channel, 
a cut on the invariant mass of the ``$t$-jet'' which originates from the
$t\to bjj$ jet merging may be useful. This will be discussed in more
detail in Sec.~\ref{sec:four}. In the $p_T(t)$ distribution, the
background remains smaller than the signal for $p_T(t)>700$~GeV.
For the 3~jet and 4~jet final states,
$Wbt$ production remains the most important background for very
energetic top quarks, and the signal is still larger than the combined
background once a cluster transverse mass and $m(b_{max}j(j))$ cut have
been imposed. 

\subsection{Background for lepton+jets events with one $b$-tag}

As discussed in Sec.~\ref{sec:twoc}, the cross section for $pp\to t\bar
t\to\ell\nu+n$~jets with one $b$-tag may be significantly larger than
that of final states with two $b$-tags if $\epsilon_b$ is small. In this
Section, we consider the background to the lepton+jets mode with one
$b$-tag. As before, we assume $\epsilon_b=0.6$ and $P_{j\to b}=1/100$ in our
simulations and comment on how the signal to background ratio changes if
the $b$-tagging efficiency decreases, and $P_{j\to b}$ increases, by a
factor of~3. 

\begin{figure}[t!] 
\begin{center}
\includegraphics[width=14.2cm]{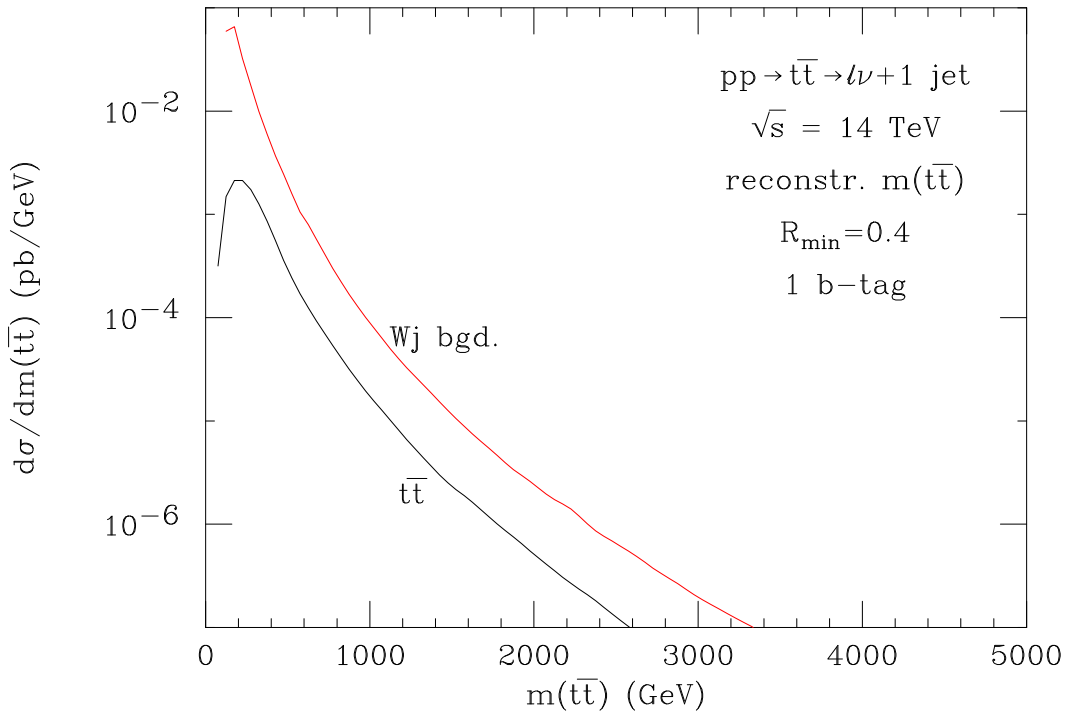} 
\vspace*{2mm}
\caption[]{\label{fig:twelve} 
The LO differential cross section of the
SM $t\bar t\to\ell\nu+1$~jet signal (black) and the $Wj$ background
(red) as a function of the reconstructed $t\bar t$ ($Wb$) invariant mass
at the 
LHC. The jet is assumed to be $b$-tagged. The cuts imposed are listed in
Eqs.~(\ref{eq:cuts1}) --~(\ref{eq:cuts4}). In addition an isolation cut
(see Eq.~(\ref{eq:cuts5})) with $R_{min}=0.4$ is imposed.} 
\vspace{-7mm}
\end{center}
\end{figure}

In addition to final states with 2, 3, or 4~jets, $t\bar t$ production
can also contribute to the $\ell\nu+1$~jet channel if only one $b$-tag
is required. Top pair events where the $b$-quark in $t\to b\ell\nu$ is
not detected, and the two light quark jets in $t\to bjj$ are either
missed or are merged with the tagged $b$-quark are the dominant source
for signal $\ell\nu b$ events. The $Wb$ invariant mass distribution thus
may still carry useful information on heavy $t\bar t$
resonances. However, as shown in
Fig.~\ref{fig:twelve}, the background from $W+1$~jet 
production where the jet is misidentified as a $b$-quark is much larger
than the signal.
Since the $b$-jet from $t\to b\ell\nu$ is usually lost, a $m_T$ cut is
ineffective. Furthermore, it prevents reconstruction of the top quark
transverse momentum. A cut on the jet invariant mass may help to reduce
the background (see Sec.~\ref{sec:four}). However, any gain in
the signal to background ratio from such a cut is at least partially
canceled by a
reduced $b$-tagging efficiency and an enhanced light jet mistagging
probability at large invariant masses. Furthermore, the $\ell\nu+1$~jet
cross section is significantly smaller
than that in the 2, 3, and 4~jet channels (see below) at
large values of the reconstructed $t\bar t$ invariant mass. 
We will not consider the $\ell\nu+1$~jet final state further here. 

\begin{figure}[th!] 
\begin{center}
\includegraphics[width=13.8cm]{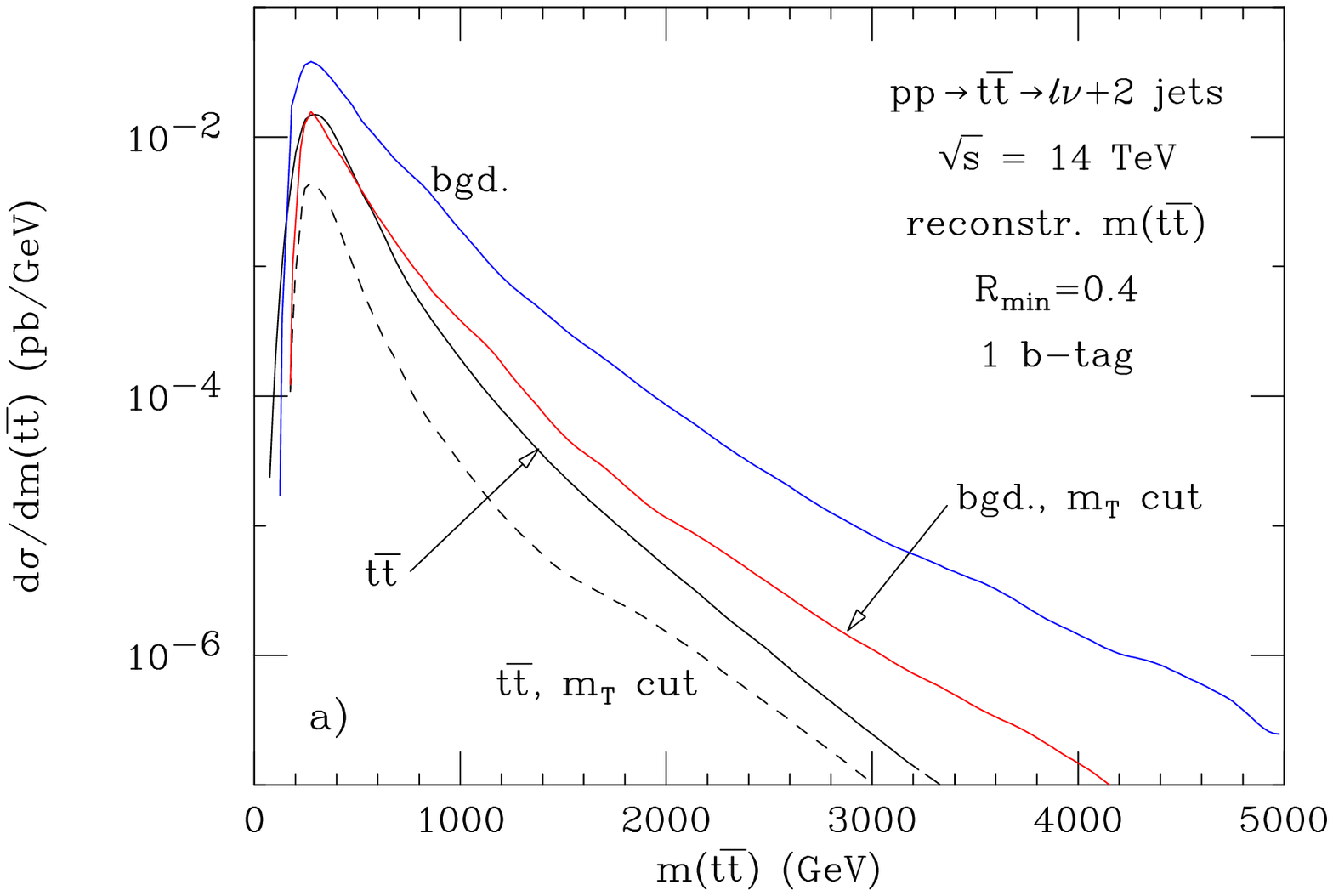} \\[3mm]
\includegraphics[width=13.8cm]{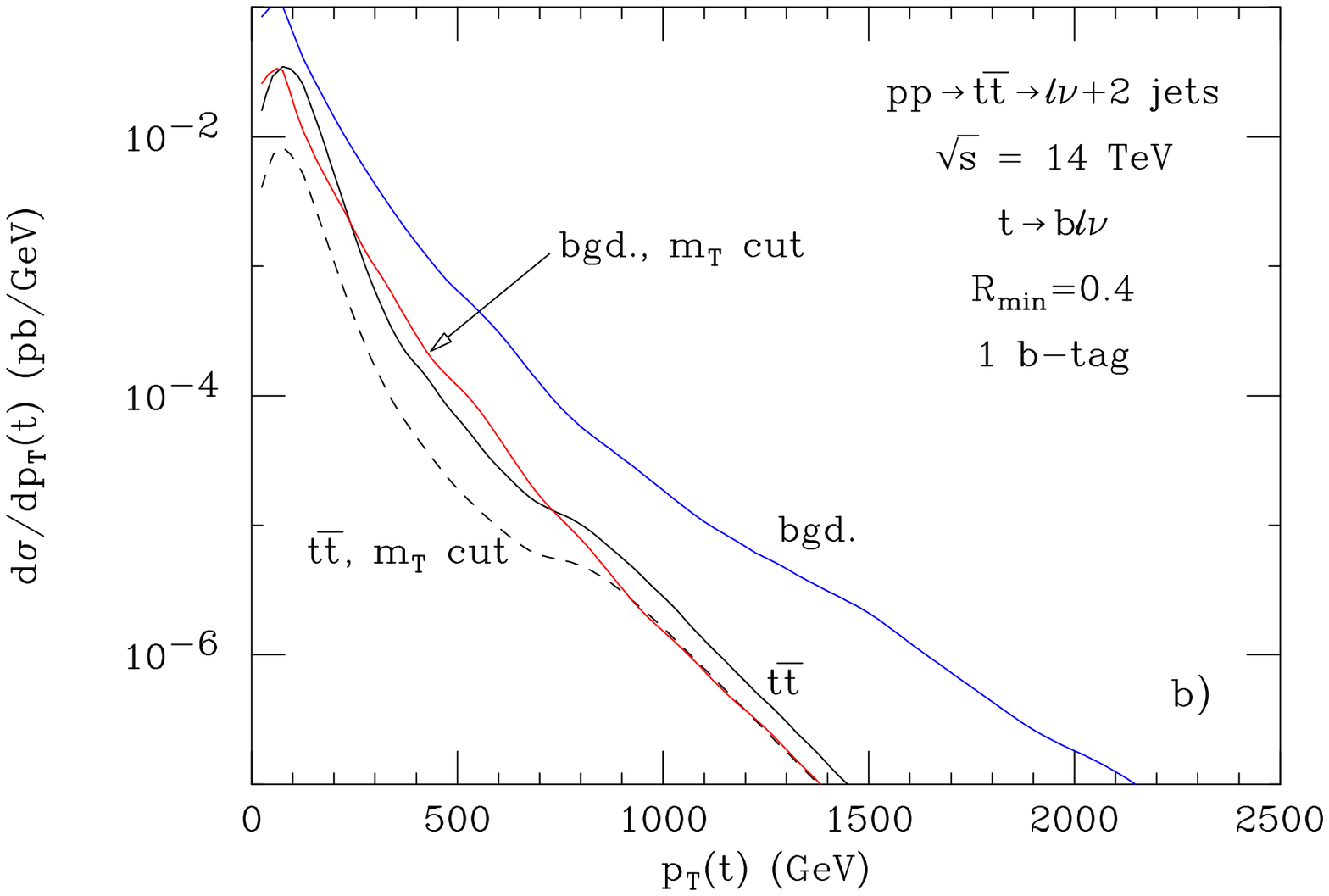}
\vspace*{2mm}
\caption[]{\label{fig:thirteen} 
The LO differential cross section of the
SM $t\bar t\to\ell\nu+2$~jets signal and the combined background as a
function of a) the reconstructed $t\bar t$ invariant mass and b) the
$t\to b\ell\nu$ transverse momentum at the LHC. 
One of the jets is assumed to be $b$-tagged. The black and blue curves
show the signal and background, respectively, imposing standard cuts
(Eqs.~(\ref{eq:cuts1}) --~(\ref{eq:cuts5})) with $R_{min}=0.4$). The
dashed and red curves show signal and background if in addition a
cluster transverse mass cut is imposed on the $b\ell\nu$ system (see
Eq.~(\ref{eq:mtcl2})).} 
\vspace{-7mm}
\end{center}
\end{figure}

The $m(t\bar t)$ and $p_T(t\to b\ell\nu)$ distributions of the signal
and the combined background in the $\ell\nu+2$~jets final state with one
$b$-tag are shown in Fig.~\ref{fig:thirteen}.
Even when a $m_T$ cut is imposed in addition to the standard transverse
momentum, rapidity and separation cuts, the background is still
considerably larger than the signal in the $t\bar t$ invariant mass
distribution. In the $p_T(t)$ distribution, the signal to background
ratio is more favorable. Requiring $|m_T-m_t|<20$~GeV, signal and
background are approximately equal at large $p_T$. 

In final states with two $b$-tags we used the $b$-jet with the smaller
separation from the charged lepton to reconstruct $p_T(t\to b\ell\nu)$ and
to compute $m_T$. Now, with only one $b$-tag, we use the jet, $j_{min}$,
whether it is tagged or not, which is closest to the charged lepton in
$\Delta R$, {\it ie.} we require
\begin{equation}
\label{eq:mtcl2}
|m_T(j_{min}\ell)-m_t|<20~{\rm GeV}.
\end{equation}

If only the standard cuts of Eqs.~(\ref{eq:cuts1}) --~(\ref{eq:cuts5})
are imposed, $Wjj$ production is the dominant background source for the
$\ell\nu+2$~jet final state with one $b$-tag. If we require in addition
that the $j_{min}\ell\nu$ cluster transverse mass satisfies
Eq.~(\ref{eq:mtcl2}), $pp\to Wjj$ and $pp\to tj$ each contribute about
one half of the total background.

To improve the signal to background ratio, one may consider a cut on the
invariant mass of the jet with the larger separation from the charged
lepton. This will be discussed in more detail in Sec.~\ref{sec:four}. 

\begin{figure}[th!] 
\begin{center}
\includegraphics[width=13.5cm]{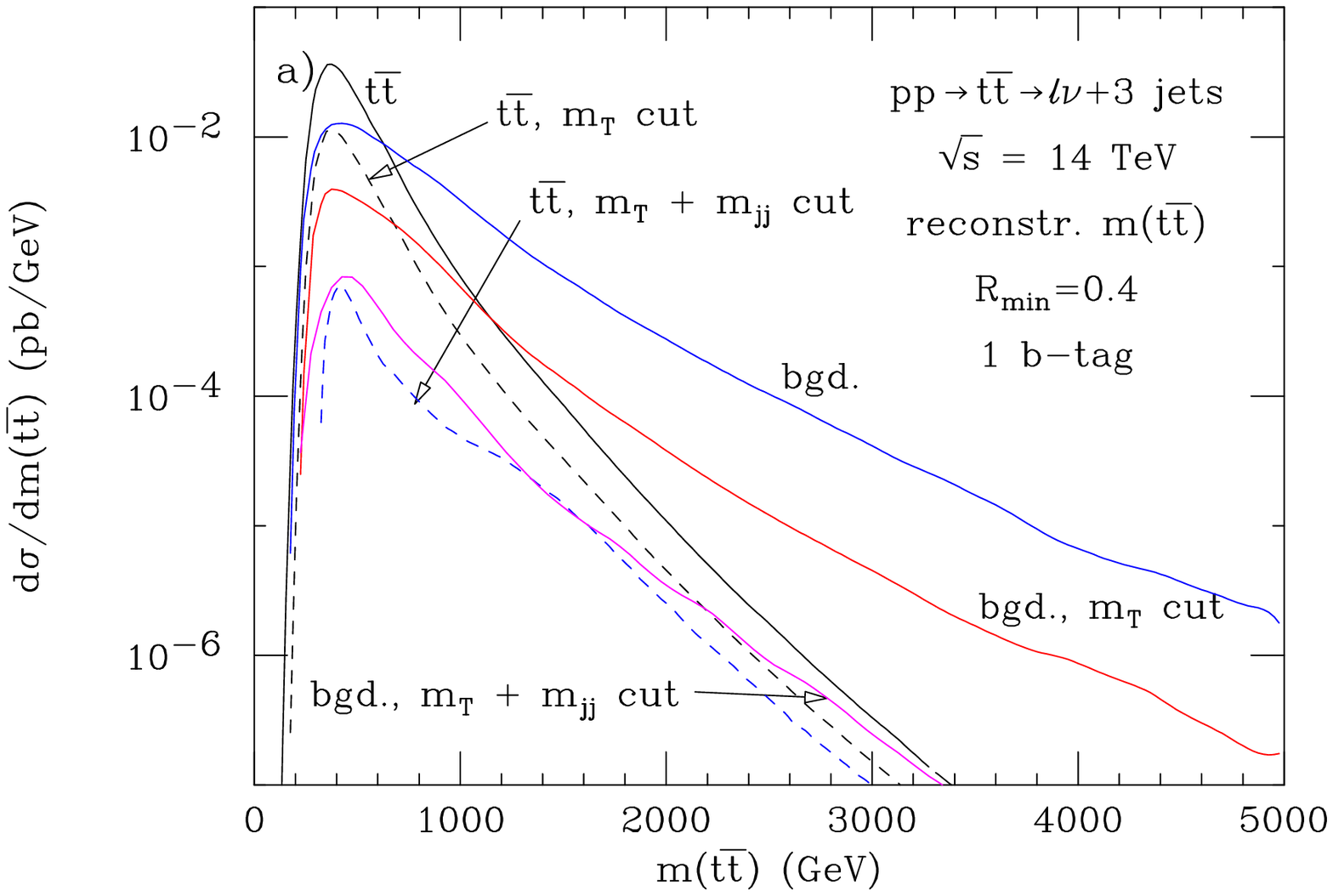} \\[3mm]
\includegraphics[width=13.5cm]{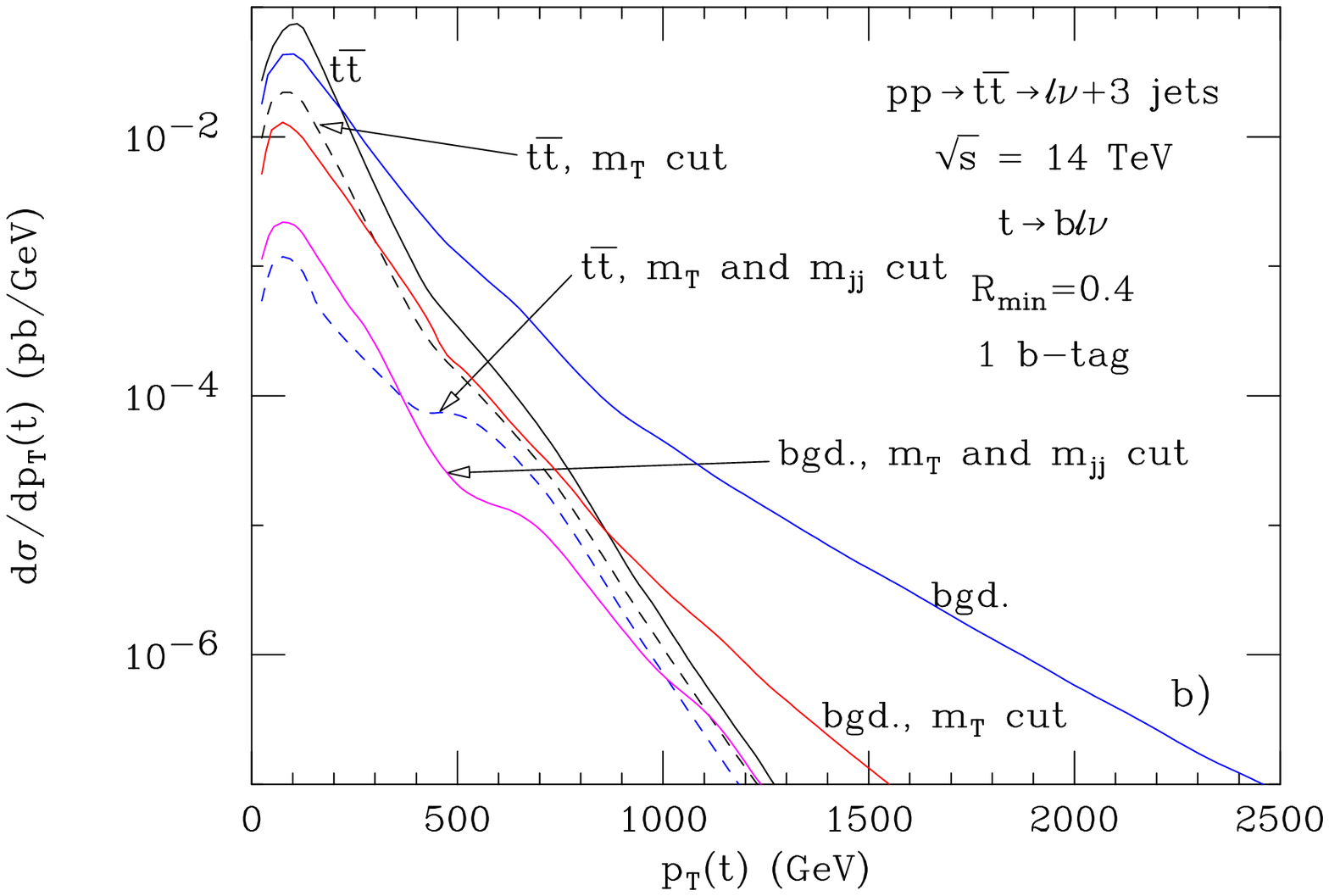}
\vspace*{2mm}
\caption[]{\label{fig:fourteen} 
The LO differential cross section of the
SM $t\bar t\to\ell\nu+3$~jets signal and the combined background as a
function of a) the reconstructed $t\bar t$ invariant mass and b) the
$t\to b\ell\nu$ transverse momentum at the LHC. 
One of the jets is assumed to be $b$-tagged. The black and blue curves
show the signal and background, respectively, imposing standard cuts
(Eqs.~(\ref{eq:cuts1}) --~(\ref{eq:cuts5})) with $R_{min}=0.4$). The
dashed and red curves show signal and background if in addition a
cluster transverse mass cut is imposed on the $b\ell\nu$ system (see
Eq.~(\ref{eq:mtcl2})). The blue dashed and magenta curves, finally,
show signal and background if in addition the invariant mass cut of
Eq.~(\ref{eq:invm2}) is imposed.} 
\vspace{-7mm}
\end{center}
\end{figure}

In Fig.~\ref{fig:fourteen} we show signal and background predictions for
the 3~jet final state.
%
Even when we require  
\begin{equation}
\label{eq:invm2}
|m(j_2j_3)-m_t|<20~{\rm GeV}
\end{equation}
in addition to the $m_T$ cut, the background is
found to be somewhat larger than the signal in the $m(t\bar t)$
distribution. $j_2$ and $j_3$ in Eq.~(\ref{eq:invm2}) are the jets with
the larger separations from the charged lepton. The signal to background
ratio is slightly better in the $p_T(t)$ distribution. Without the
$m(j_2j_3)$ cut, the background is dominated by $W+3$~jet production. If
the invariant mass cut of Eq.~(\ref{eq:invm2}) is imposed, $W+3$~jet,
$Wjt$ and $Wbt$ production all contribute about equally to the background.

\begin{figure}[th!] 
\begin{center}
\includegraphics[width=13.5cm]{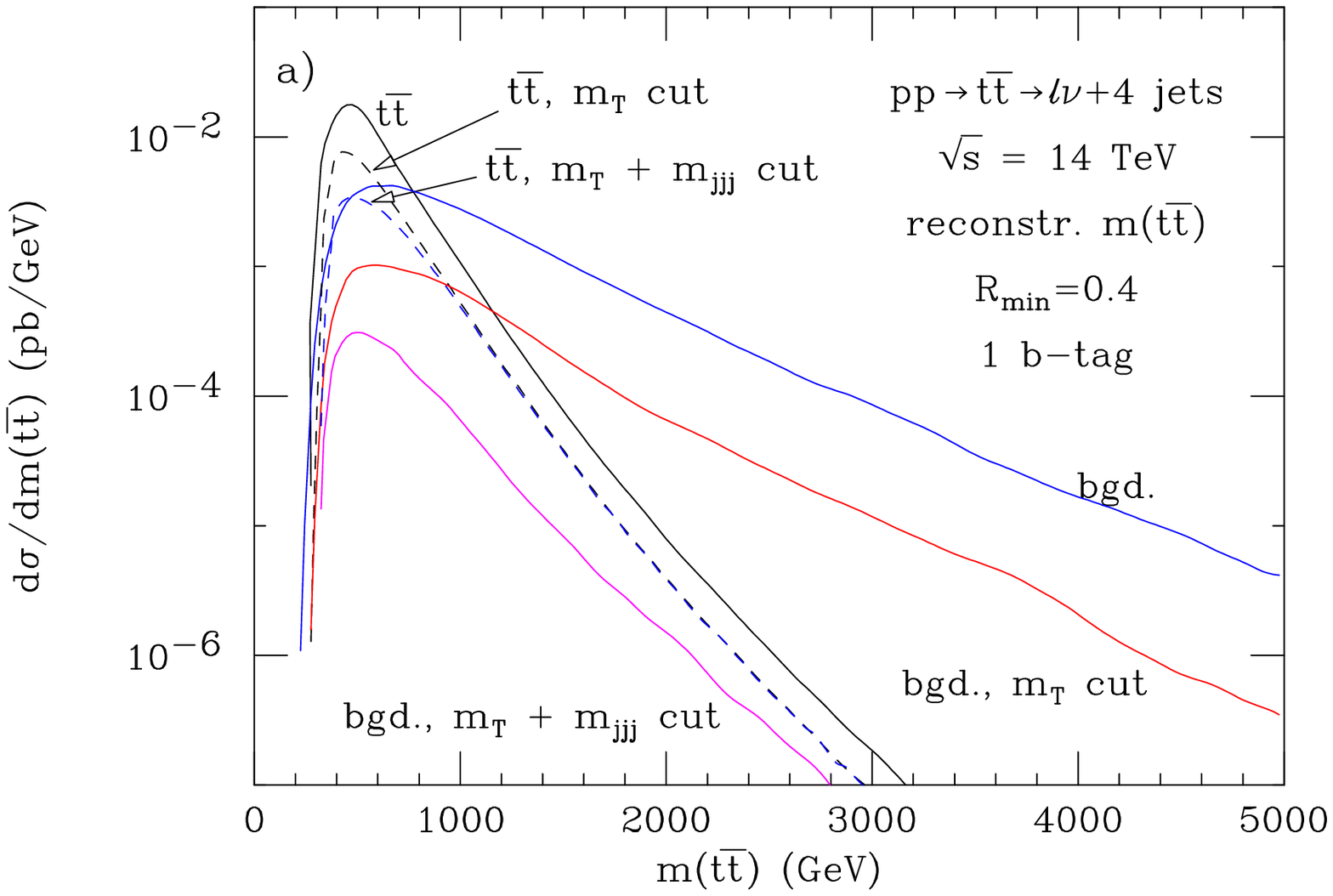} \\[3mm]
\includegraphics[width=13.5cm]{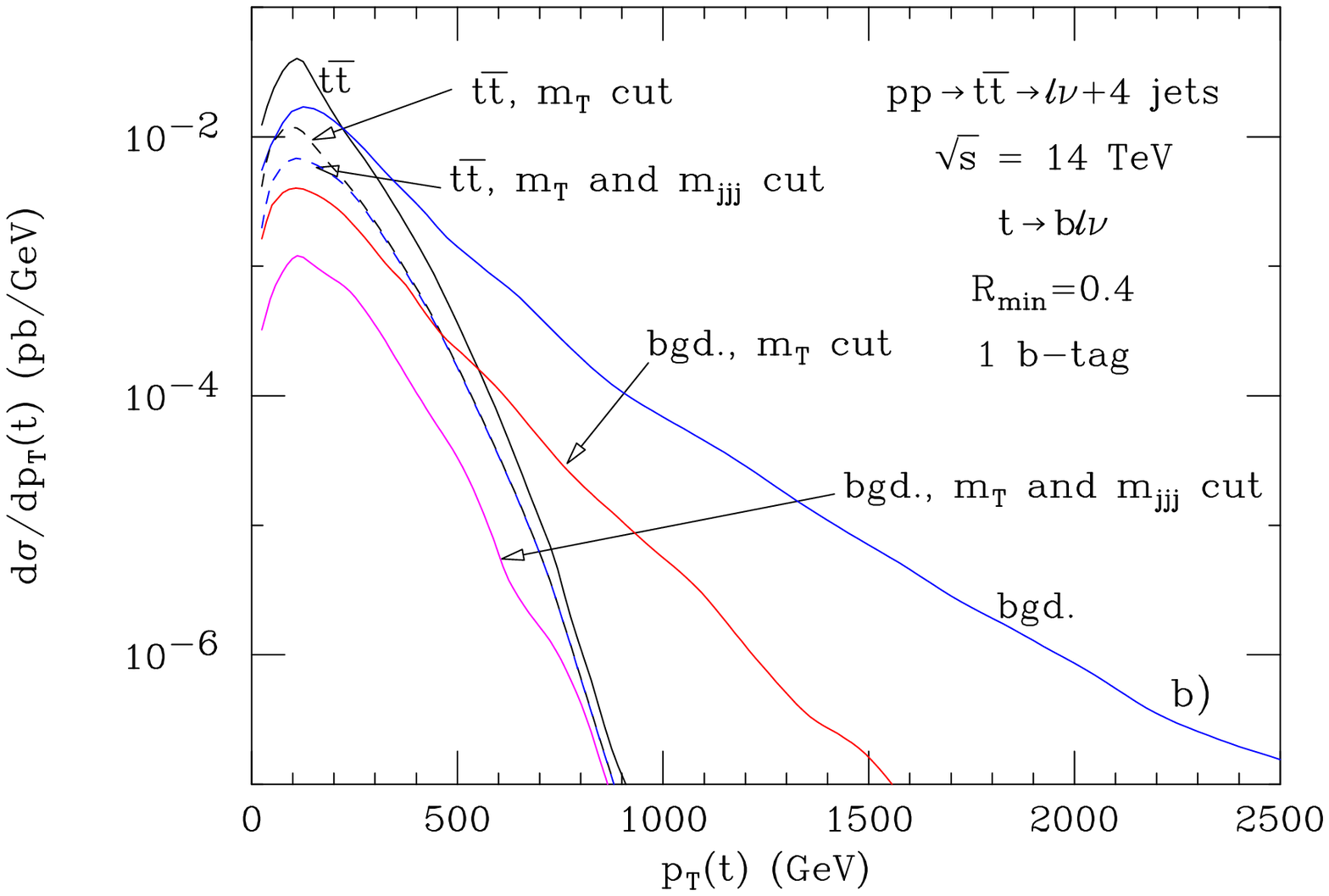}
\vspace*{2mm}
\caption[]{\label{fig:fifteen} 
The LO differential cross section of the
SM $t\bar t\to\ell\nu+4$~jets signal and the combined background as a
function of a) the reconstructed $t\bar t$ invariant mass and b) the
$t\to b\ell\nu$ transverse momentum at the LHC. 
One of the jets is assumed to be $b$-tagged. The black and blue curves
show the signal and background, respectively, imposing standard cuts
(Eqs.~(\ref{eq:cuts1}) --~(\ref{eq:cuts5})) with $R_{min}=0.4$). The
dashed and red curves show signal and background if in addition a
cluster transverse mass cut is imposed on the $b\ell\nu$ system (see
Eq.~(\ref{eq:mtcl2})). The blue dashed and magenta curves, finally,
show signal and background if in addition the invariant mass cut of
Eq.~(\ref{eq:invm3}) is imposed.} 
\vspace{-7mm}
\end{center}
\end{figure}

In Fig.~\ref{fig:fifteen}, finally, we show signal and background for
the 4~jet final state. 
Once a $m_T$ and an 
\begin{equation}
\label{eq:invm3}
|m(j_2j_3j_4)-m_t|<20~{\rm GeV}
\end{equation}
invariant mass cut have been imposed, the background is considerably
smaller than the 
signal. $j_2$, $j_3$ and $j_4$ in Eq.~(\ref{eq:invm3}) are the three
jets with the larger separations from the charged lepton. Without the
$m(j_2j_3j_4)$ cut, the background is dominated by $W+4$~jet
production. If Eq.~(\ref{eq:invm3}) is imposed, $W+4$~jet,
$Wjt$ and $Wbt$ production all contribute about equally to the
background. 

For $\epsilon_b=0.2$ and $P_{j\to b}=1/30$, as suggested in
Refs.~\cite{atlaslh,atlasglu}, the signal to background ratio in the TeV
region would be about a factor~2 (factor~5) worse than the results
shown in Figs.~\ref{fig:thirteen} --~\ref{fig:fifteen}
(Fig.~\ref{fig:twelve}). 

\section{A jet invariant mass cut}
\label{sec:four}

In Sec.~\ref{sec:three} we have seen that a cut on the $j\ell\nu$
cluster transverse mass is not sufficient to suppress the background to
an acceptable level in the $\ell\nu+2$~jets final state with one
$b$-tag. At large $t\bar t$ invariant mass, events where the jets from
$t\to bjj$ all merge into a single ``$t$-jet'' dominate the signal. The
invariant mass of such a ``$t$-jet'' is consistent with the top quark
mass. The ``$t$-jet'' usually is the jet with the larger separation from
the charged lepton. The main background for the $\ell\nu+2$~jets final
state with one $b$-tag originates from $Wjj$ and
$(t+\bar t)j$, $t\to b\ell\nu$, ($t$-channel single top) production. In
both processes, the jet which is further away from the charged lepton in
$\eta-\phi$ space is a light quark or gluon jet. At LO, such jets are
(almost) massless. Once higher order QCD corrections are taken into
account, light quark or gluon jets acquire a mass which depends on the
jet algorithm and the distance in $\Delta R$ space used to cluster
particles. The average invariant mass of such a jet is expected to
be~\cite{Owens:1979bt} 
\begin{equation}
\langle m(j)\rangle\propto\sqrt{\alpha_s}\,p_T(j).
\end{equation}
This suggests that a cut on the invariant mass of the jet with the
larger separation from the charged lepton may be helpful in suppressing
the background in the $\ell\nu+2$~jets final state with one $b$-tag. As
discussed in Sec.~\ref{sec:twob}, such a cut would also be helpful in
reducing the combinatorial background from extra QCD
jets. 

For both $Wjj$ and $(t+\bar t)j$ production, the NLO QCD corrections
are known~\cite{Campbell:2002tg,Sullivan:2004ie}. However,
non-perturbative QCD effects may significantly contribute to the
invariant mass of a light quark or gluon jet. Calculations which only
include NLO QCD corrections may therefore not be reliable in predicting
how a well a jet invariant mass cut reduces the background in the
$\ell\nu+2$~jets final state with one $b$-tag. 

In order to {\sl estimate} the effect of a jet invariant mass cut on the
$Wjj$ and $(t+\bar t)j$ background, we convolute the differential cross
sections obtained from {\tt ALPGEN} and {\tt MadEvent} 
with ${\cal P}(m(j_{top}),p_T(j_{top}))$ where $j_{top}$ is the jet with
the larger separation from the charged lepton ({\it ie.} the ``t-jet''
candidate). A cut on $m(j_{top})$ is then imposed (see below). ${\cal
P}(m(j),p_T(j))$ is the two-dimensional probability density that a jet
with transverse momentum $p_T(j)$ has an 
invariant mass $m(j)$. We calculate ${\cal P}(m(j),p_T(j))$ by
generating $10^5$ $W+$~jets events in {\tt
PYTHIA}~\cite{Sjostrand:2006za} and passing them through {\tt
PGS4}~\cite{conway}, which simulates the response of a generic
high-energy physics collider detector with a tracking system,
electromagnetic and hadronic calorimetry, and muon system. Jets are
required to have $p_T(j)>30$~GeV. The rapidity coverage of the tracking
system is assumed to be $|\eta|<3$. Jets are reconstructed in the
cone~\cite{cone} and $k_T$ algorithms~\cite{Ellis:1993tq} as implemented
in {\tt PGS4}, 
using a cone size ($D$ parameter) of $R=0.5$ ($D=0.5$) in the cone
($k_T$) algorithm. Since it is infrared safe, the $k_T$ algorithm is the
theoretically preferred algorithm. For a discussion of the advantages
and disadvantages of the two algorithms at hadron colliders, see
Ref.~\cite{Blazey:2000qt}. The cone size ($D$ parameter) is deliberately
chosen to be slightly larger than in our parton level studies to avoid
drawing conclusions which are too optimistic.

\begin{figure}[t!] 
\begin{center}
\includegraphics[width=14.2cm]{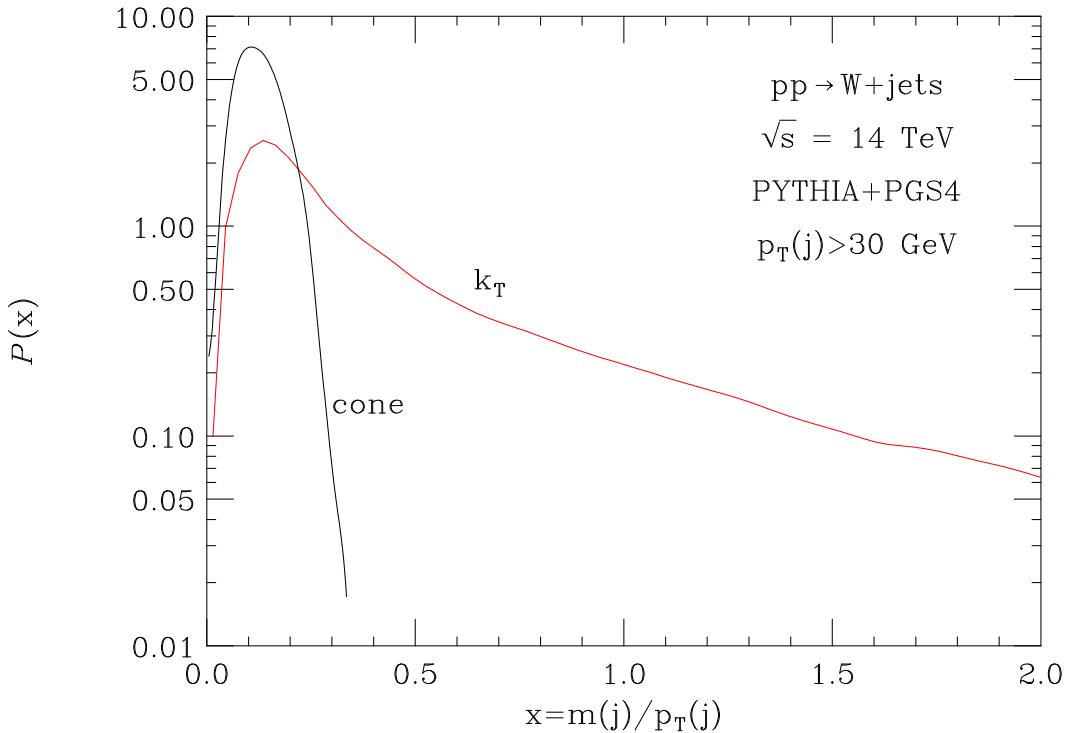} 
\vspace*{2mm}
\caption[]{\label{fig:sixteen} 
The probability density $P(x)$ versus $x=m(j)/p_T(j)$ for a jet with
transverse momentum $p_T(j)$ to have an invariant mass $m(j)$ in 
$pp\to W+$jets at the LHC. Shown are results for the cone (black) and the
$k_T$ jet algorithm (red). The $W+$jets events are generated with {\tt
PYTHIA} requiring $p_T(j)>30$~GeV and then processed into physics
objects using {\tt PGS4}. The {\tt PGS4} parameters used are described
in the text. } 
\vspace{-7mm}
\end{center}
\end{figure}

We find that the probability density function, ${\cal P}$, for the
$k_T$ algorithm has 
a much longer tail at large jet invariant masses than for the cone
algorithm. This is illustrated in Fig.~\ref{fig:sixteen}, where we show
the one-dimensional probability density $P(x)$, $x=m(j)/p_T(j)$, for
$W+$jets events with $p_T(j)>30$~GeV. Very similar results are obtained
for jets in other processes, such as $t$-channel single top production.
$P(x)$ peaks at $x\approx 0.1-0.13$ for both algorithms. In the cone
algorithm, a jet has a fixed size in $\eta - \phi$ space. This limits
the jet invariant mass. For the parameters chosen here, it is very
difficult for a jet in the cone algorithm to have an invariant mass
larger than about $0.3\cdot p_T(j)$. In contrast, the $k_T$ algorithm
has a tendency to ``vacuum up'' contributions from the underlying event,
and $k_T$ jets therefore do not have a well defined size. As a result,
the $k_T$ jet mass tail extends to $x=1$ and beyond and is sensitive to
how the underlying event is simulated. 

Figure~\ref{fig:sixteen} shows $P(x)$ for jets with $p_T(j)>30$~GeV. For
a higher $p_T$ threshold, the $x$ value for which the distribution peaks
remains almost constant. However, $P(x)$ falls considerably faster with
$x$ for larger $p_T(j)$ values when using the $k_T$ algorithm. Removing
the contributions from the underlying event, or decreasing the $D$
parameter, leads to a similar result. 

\begin{figure}[th!] 
\begin{center}
\includegraphics[width=13.5cm]{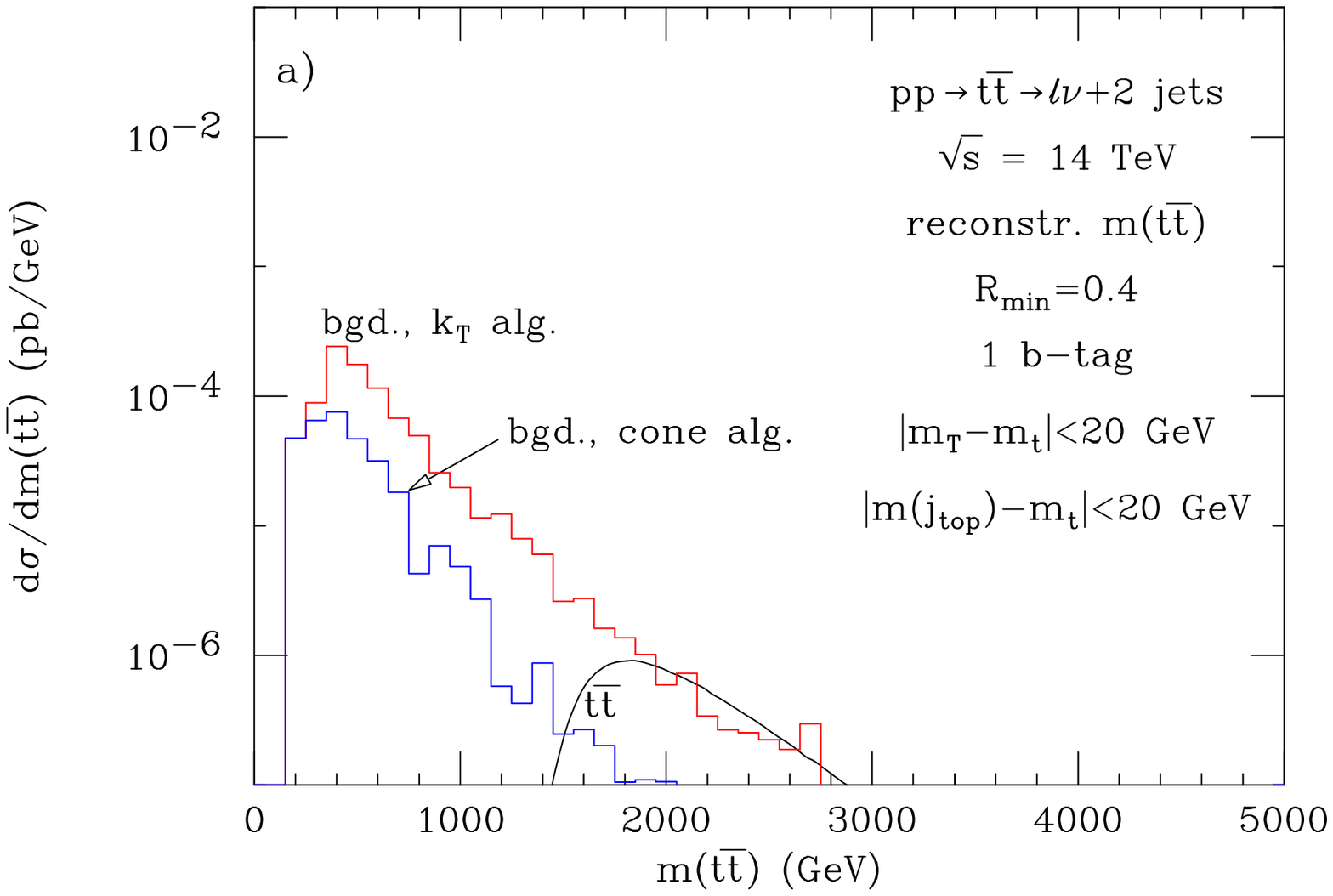} \\[3mm]
\includegraphics[width=13.5cm]{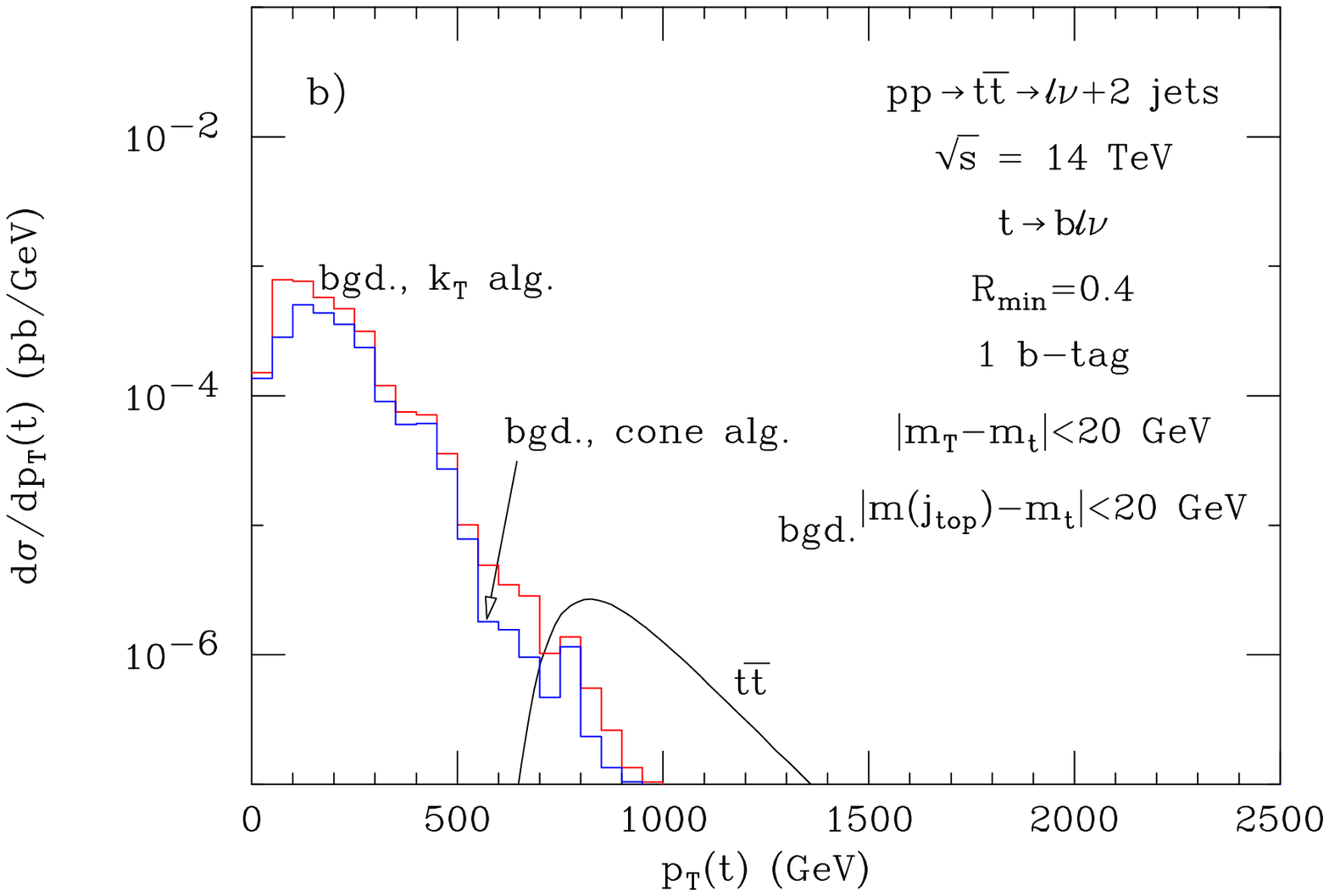}
\vspace*{2mm}
\caption[]{\label{fig:seventeen} 
The LO differential cross section of the
SM $t\bar t\to\ell\nu+2$~jets signal and the combined background as a
function of a) the reconstructed $t\bar t$ invariant mass and b) the
$t\to b\ell\nu$ transverse momentum at the LHC. 
One of the jets is assumed to be $b$-tagged. In addition to the standard
cuts (Eqs.~(\ref{eq:cuts1}) --~(\ref{eq:cuts5})) with $R_{min}=0.4$) a
$|m_T-m_t|<20$~GeV (see Eq.~(\ref{eq:mtcl2})) and a
$|m(j_{top})-m_t|<20$~GeV cut are imposed.  
The black curve shows the $t\bar t$ signal cross section. The red and
blue histograms are the predictions for the combined background using
the $k_T$ and cone algorithms, respectively. As before, we assume
$\epsilon_b=0.6$ and $P_{j\to b}=1/100$.} 
\vspace{-7mm}
\end{center}
\end{figure}

The differences between ${\cal P}(m(j),p_T(j))$ in the $k_T$ and the
cone algorithm have a significant impact on the background in
the $m(t\bar t)$ distribution of the $\ell\nu+2$~jet final state with
one $b$-tag when a 
\begin{equation}
\label{eq:mjet}
|m(j_{top})-m_t|<20~{\rm GeV}
\end{equation}
cut is imposed. This is shown in Fig.~\ref{fig:seventeen}a, where, in 
addition to the jet invariant mass cut, we also impose the cluster
transverse mass cut of Eq.~(\ref{eq:mtcl2}). 
$j_{top}$ in the figure denotes the ``$t$-jet'' resulting from $t\to
bjj$. The key in understanding the significant dependence of the
background on the jet algorithm in the $m(t\bar t)$ distribution is the
observation that a large fraction of the $\ell\nu+2$~jet background
events at high invariant mass contains a jet with 
a relatively small $p_T$. 
If the cone algorithm is used, a jet transverse momentum of at least
450~GeV is needed to satisfy Eq.~(\ref{eq:mjet}). This is not
the case for the $k_T$ algorithm, where jets with a $p_T$ as low as
75~GeV may pass the jet invariant mass cut. In the cone algorithm, the
$Wjj$ and $tj$ backgrounds, which dominate if no $m(j_{top})$ cut is
imposed, thus are negligible compared with the $Wbt$ and $Wjt$, $t\to bjj$,
background. On the other hand, if the $k_T$ algorithm is used, $pp\to
Wjj$ and $pp\to tj$ are still the dominant background sources. Note that
the $m(j_{top})$ invariant mass cut (Eq.~(\ref{eq:mjet})) has no effect
on the $Wbt$ and $Wjt$ background. 

In either case, the jet invariant mass cut strongly suppresses the
background. It also decimates the signal for $m(t\bar
t)<1.6$~TeV, eliminating all events where one or two jets are not
passing the acceptance cuts of Eq.~(\ref{eq:cuts2}). For larger $t\bar
t$ invariant masses, the $m(j_{top})$ cut reduces the
signal by at most a factor~2. In this region the signal to background
ratio is ${\cal O}(1)$ or better, an improvement of a factor~10 or more
compared with the result without such a cut (see
Fig.~\ref{fig:thirteen}a). If one were able to eliminate the 
effect of the underlying event on the jet invariant mass in the $k_T$
algorithm, the background could be reduced by an additional factor~2.

Since the $Wjj$ and $tj$ contributions to the background at large
$p_T(t)$ are smaller than at large $m(t\bar t)$ once the jet invariant
mass cut has been imposed, 
the background $p_T(t)$ distribution, which is shown in
Fig.~\ref{fig:seventeen}b, is found to be much less sensitive to the
details of the jet algorithm. For $p_T(t)>700$~GeV, the $t\bar t$ signal
dominates over the background. $Wjt$ production is the largest
background source in the $p_T(j)$ distribution in the 2~jet final state
with one $b$-tag when a cluster transverse mass cut and a jet invariant
mass cut are imposed. Comparing the results of Fig.~\ref{fig:seventeen}b
with those of Figs.~\ref{fig:fourteen}b and~\ref{fig:fifteen}b shows
that, once an invariant mass cut on the jet(s) originating from the
hadronic top decay has been imposed, the 2~jet final state is by far the
largest source for $t\bar t$ events with $p_T(t)>900$~GeV. Furthermore,
the signal to background ratio in the 2~jet channel in this region is
considerably better than that for the 3~jet and 4~jet final states. 

As before, we have used $\epsilon_b=0.6$ and
$R_{j\to b}=1/100$ in the numerical results presented in this
Section. If the $b$-tagging efficiency is reduced by a factor~3 in the
TeV region, and $R_{j\to b}$ simultaneously increases by a factor~3, 
the signal to background ratio worsens by about a factor~$1.5-3$.

From the calculations presented in this Section, we conclude that a jet
invariant mass cut may be useful in improving the signal to background
ratio in the $\ell\nu+2$~jet final state with one $b$-tag. Details,
however, may depend on the jet algorithm used. Our calculation uses a
rather crude approximation to model the invariant mass of a light quark
or gluon jet. Our results thus should be viewed as order of magnitude
estimates. The main disadvantage of the jet invariant mass cut clearly
is that only few signal events will survive it.

\section{Summary and Conclusions}
\label{sec:five}

Many New Physics models predict the existence of $t\bar t$ resonances
with masses in the TeV region. New particles with mass $M$ which decay
into a $t\bar t$ pair lead to a 
peak in the $t\bar t$ invariant mass distribution located at $m(t\bar
t)=M$ and to a Jacobian peak at $p_T(t)=M/2$ in the top quark transverse
momentum distribution. The lepton+jets final state offers a good
opportunity to search for such particles. The di-lepton channel suffers
from a small branching ratio and the inability to reconstruct the $t\bar
t$ invariant mass and the top quark transverse momentum. The
all-hadronic final state is subject to a very large QCD background which
limits the sensitivity to new physics.

The search for resonances in the $t\bar t$ channel with masses in the
TeV region requires the reconstruction of very energetic top quarks. 
These are strongly
boosted, and their decay products are highly collimated. This leads to
overlapping and merging jets from hadronically decaying top quarks. As a
result, the standard $t\bar t$ identification requirements for the
lepton+jets final state -- an isolated charged lepton, missing transverse
momentum and four isolated jets with two $b$-tags -- strongly reduce the
observable $t\bar t$ cross section. In addition, the tagging efficiency
for $b$-quarks in $t\bar t$ events with very energetic top quarks may be
up to a factor~3 smaller than at low energies. This further reduces the
number of $t\bar t$ events which can be identified using standard $t\bar
t$ identification requirements. 

In this paper we investigated in detail how the efficiency for detecting
very energetic top quarks in the lepton+jets channel can be
improved. We found that the $\ell\nu+n$~jets ($\ell=e,\,\mu$) final states
with $n=2$ and $n=3$, and one or two $b$-tags, 
increase the observable rate significantly. For $s$-channel resonances,
the gain is even larger. In particular, $\ell\nu+n$~jets events
with only one $b$-tag offer a crucial advantage 
in the TeV region, provided that the background can be controlled. 

Using {\tt MC@NLO}, we investigated how NLO QCD corrections affect the
lepton+jets final states at very high energies. 
We found that QCD corrections in this 
channel are large, often produce extra hard jets from radiation in the 
$t \bar t$ production process, 
and may substantially modify the $p_T$ distribution of top
quarks in the high transverse momentum region. Details depend on how
many jets from the hadronic top decay are observed. When final state
topologies with a fixed number of jets are considered and invariant mass
cuts on the observed $t\to bjj$ jets are imposed, 
the radiation of extra quarks and gluons is strongly reduced, 
and QCD
corrections are moderate and decrease with increasing $m(t\bar t)$ and
$p_T(t)$. In the TeV region, the ratio of NLO to LO differential cross
sections typically is $0.7-0.8$ if one requires $p_T(t\bar t)<100$~GeV.

We also presented a comprehensive analysis of the background processes
which contribute to the $\ell\nu+n$~jets final states with one or two
$b$-tags. All relevant $W+$~jets and single top background processes
were considered. In the 2~jet and 3~jet final states, the background
processes occur at a lower order in perturbation theory, and thus are
potentially more worrisome, than in the 4~jet channel, in particular if
only one $b$-quark is tagged. We found that a cluster transverse mass
cut on the $t\to b\ell\nu$ system is sufficient to adequately
suppress the background in the $\ell\nu+2$~jet channel with two
$b$-tags. In the 3~jet and 4~jet final states with two tagged $b$-jets,
an additional cut on the invariant mass of the observed $t\to bjj$ jets 
is needed in order to
achieve a signal to background ratio better than one over the entire
kinematic al region of interest. 

As expected, the background is significantly higher if only one
$b$-quark is tagged. In the 2~jet channel, an invariant mass cut on the
``$t$-jet'' which results from $t\to bjj$ is needed to suppress the
background sufficiently. We simulated the effect of such a cut on the
background by convoluting the two-dimensional probability density that a
light quark or gluon jet with a given $p_T$ has a jet mass $m(j)$ with
the differential cross section of the relevant background processes. The
probability density was determined for both the $k_T$ and the cone
algorithm using {\tt PYTHIA} and {\tt PGS4}. For the $t\bar t$ invariant
mass distribution we found that the background differential cross
section significantly depends on the jet algorithm used. For
the 3~jet final state, the background in the $m(t\bar t)$ distribution
is somewhat larger than the signal even after imposing a cluster transverse
mass cut and a cut on the invariant mass of the jets originating from
$t\to bjj$ (see Fig.~\ref{fig:fourteen}a). The signal to background
ratio in the $p_T(t)$ distribution in all cases is better than in the
$t\bar t$ invariant mass distribution. Although we presented all
numerical results for a $b$-tagging efficiency of $\epsilon_b=60\%$ and
a light jet mistagging probability of $P_{j\to b}=1\%$, we commented on
how our results change if $\epsilon_b$ decreases, and $P_{j\to b}$
simultaneously increases, by a factor~3 as indicated by ATLAS
simulations~\cite{atlaslh,atlasglu}. 

\begin{figure}[th!] 
\begin{center}
\includegraphics[width=13.1cm]{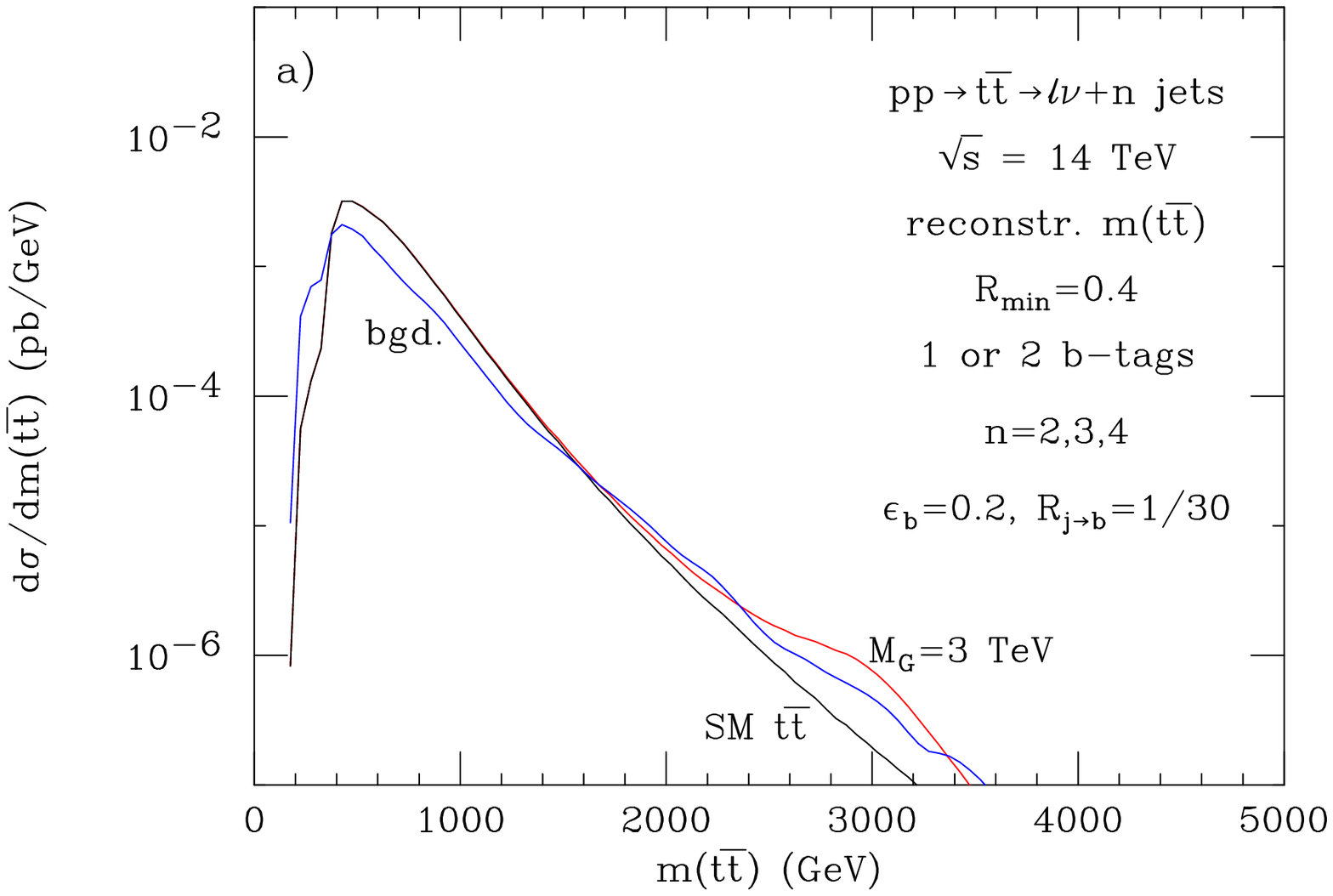} \\[3mm]
\includegraphics[width=13.1cm]{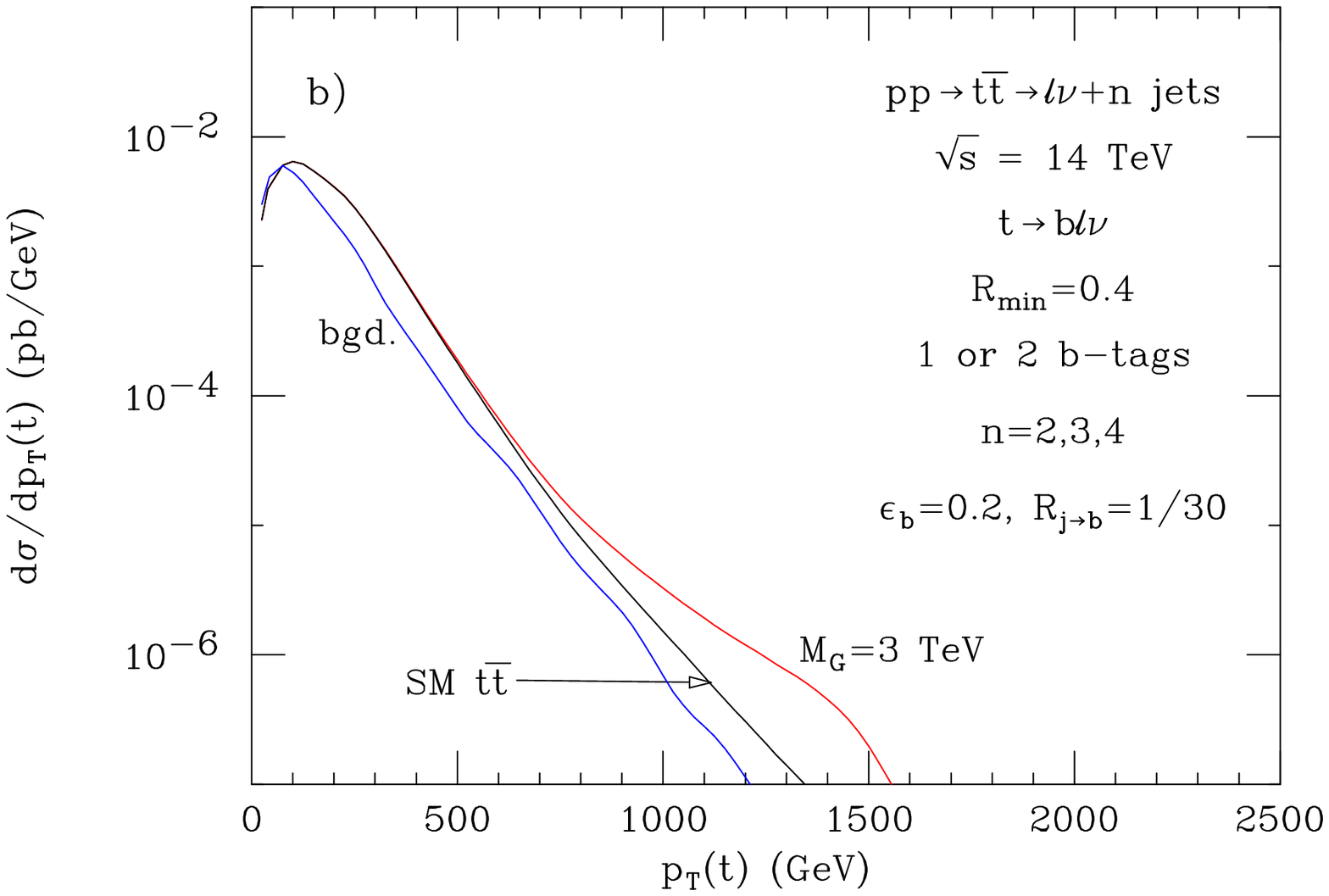}
\vspace*{2mm}
\caption[]{\label{fig:eightteen} 
The LO differential cross section of the combined 
SM $t\bar t\to\ell\nu+n$~jets ($n=2,\,3,\,4$) signal (black line), a
bulk RS KK gluon, $G$, with $M_G=3$~TeV (red line), and the combined
background (blue line) as a 
function of a) the reconstructed $t\bar t$ invariant mass and b) the
$t\to b\ell\nu$ transverse momentum at the LHC. 
One or two of the jets are assumed to be $b$-tagged. In addition to the
standard 
cuts (Eqs.~(\ref{eq:cuts1}) --~(\ref{eq:cuts5})) with $R_{min}=0.4$) a
$|m_T-m_t|<20$~GeV cut, and a cut on the invariant mass, $m$, of the jet(s)
from $t\to bjj$ of $|m-m_t|<20$~GeV are imposed in all cases except for
$n=2$ and two tagged $b$-jets where the $m$-cut has not been imposed. 
We assume $\epsilon_b=0.2$ and $P_{j\to b}=1/30$.} 
\vspace{-7mm}
\end{center}
\end{figure}

From the results presented in Secs.~\ref{sec:three} and~\ref{sec:four}
it may not be clear how good or bad the signal to background ratio is, and
whether a $t\bar t$ resonance with a mass in the TeV region can be
observed, for the worst-case scenario of $\epsilon_b=0.2$, $P_{j\to
b}=1/30$, if one or two jets are $b$-tagged, and the 2~jet, 3~jet
and 4~jet final states are combined. To answer this question, we show
the LO $m(t\bar t)$ and $p_T(t)$ distributions for these parameters in
Fig.~\ref{fig:eightteen}. 
As before, we have imposed the standard acceptance cuts of
Eqs.~(\ref{eq:cuts1}) --~(\ref{eq:cuts5})) with $R_{min}=0.4$. In
addition, events are assumed to satisfy the cluster transverse mass
cuts, Eqs.~(\ref{eq:mt}) and~(\ref{eq:mtcl2}), and the invariant mass
cuts of Eqs.~(\ref{eq:invm}), (\ref{eq:invm1}), (\ref{eq:invm2})
and~(\ref{eq:invm3}). In order to be conservative, we have used the
background obtained with the $k_T$ algorithm for the contribution from
the 2~jet final state with one $b$-tag. We emphasize that, for $t\bar t$
invariant masses below about 2~TeV, and top quark transverse momenta
smaller than 600~GeV, our results are probably overly pessimistic as the
$b$-tagging efficiency in this region should be significantly higher,
and the light jet mistagging probability considerably lower, than what
we have assumed in Fig.~\ref{fig:eightteen}.

The black lines in Fig.~\ref{fig:eightteen} represent the SM $t\bar t$
signal. The red lines show 
the prediction for a bulk RS KK gluon with mass $M_G=3$~TeV. The blue
curves, finally, represent the combined background. For an integrated
luminosity of 100~fb$^{-1}$, a $M_G=3$~TeV bulk RS KK gluon leads to
about 100 (130)~signal events for $m(t\bar t)\geq 2$~TeV ($p_T(t)\geq
700$~GeV) on a total background (SM $t\bar t$ production
(red lines) and other backgrounds (blue lines) combined) of about
400 (350)~events. This
corresponds to about a $5\sigma$ signal in the invariant mass
distribution, and to about a $7\sigma$ effect in the $p_T(t)$
differential cross 
section, reflecting the smaller background in the top quark transverse
momentum distribution. For comparison, if one were to consider the 4~jet
final state with two $b$-tags only, less than one signal event would be
expected. 

The significances given here are for illustration purposes only. All
calculations have been performed at LO, and thus retain a significant
dependence on the factorization and renormalization scales. This could
easily change the signal to background ratio by a
factor~2. Nevertheless, it is clear that lepton+jets topologies with
less than 4~jets and/or only one $b$-tag will be able to significantly
enhance the capabilities of the LHC experiments in discovering resonances
in the $t\bar t$ channel, even if the $b$-tagging
efficiency and the light jet mistagging probability in the TeV region
are much worse than at low energies.

Figure~\ref{fig:eightteen} also demonstrates that the cluster transverse
mass and invariant mass cuts imposed using the techniques described in
Sec.~\ref{sec:three} work well in suppressing the $W+$~jets and single
top background at large energies. As noted before, these cuts also 
suppress 
the radiation of extra QCD jets 
especially at high energies. For example, in an event where 
all $t\to bjj$ jets merge into a single jet and where there are one or two
extra QCD jets, the invariant mass of the two or three jets in the final
state will usually not be in the vicinity of $m_t$. However, for a more
quantitative statement on how well a $t\to bjj$ jet(s) invariant mass
cut suppresses extra QCD radiation, detailed simulations are needed
which are beyond the scope of this paper.

While the non-$t\bar t$ background in the invariant mass distribution
does not seriously impact the search for a strongly coupled resonance in
the $t\bar t$ channel such as a bulk RS KK gluon, it will considerably
limit the search for weakly coupled resonances. In order to further
reduce the background, one can try to make use of the substructure of
jets in $t\bar t$ events in the 2~jet and 3~jet final states. In the
2~jet final state, the single ``$t$-jet'' from $t\to bjj$ contains two
light quark jets which form a $W$ boson. In the 3~jet channel, the
$b$-jet receives contributions from merging with one of the light quark
jets, while the second (non-taggable) jet has a component originating
from the merging of the two jets from $W$ decay. The invariant mass
distribution of these jets thus should differ from that of QCD
jets. Other possibilities to discriminate $t\bar t$ and background
events include track isolation~\cite{strassler}, and identifying the
substructure of jets using methods such as those proposed in
Refs.~\cite{Butterworth:2007ke} and~\cite{Butterworth:2002tt}. 

We also reemphasize the importance of more accurate
estimates of the $b$-tagging efficiency $\epsilon_b$ for top events at
very high energies. Existing calculations~\cite{atlaslh,atlasglu}
indicate that $\epsilon_b$ may be significantly smaller in this region
than at low energies. Overlapping and merging jets in lepton+jets events
with 2 or 3 jets are likely to further complicate the reconstruction of 
secondary
vertices.

\acknowledgements
We would like to thank B.~Acharya, A.~Belyaev, M.A.~Dufour, B.~Knuteson,
T.~LeCompte, B.~Lillie, S.~Mrenna, and B.~Vachon 
for useful discussions. We are also grateful to J.~Huston for a private
tutorial on jet algorithms. One of us would like to thank the
Fermilab Theory Group and the High Energy Physics Group, McGill
University, where part of this work was done, for
their generous hospitality. This research was supported in part by the
National Science Foundation under grant No.~PHY-0456681 and the
Department of Energy under grant DE-FG02-91ER40685.


\bibliographystyle{plain}

\end{document}